\documentclass[preprint,3p,times,authoryear]{elsarticle}

\usepackage{graphicx}
\usepackage{amsmath}
\usepackage{amssymb}
\usepackage{amsfonts}
\usepackage{mathrsfs}
\usepackage{adjustbox}
\newcommand{\q}[1]{``#1''}
\usepackage{mdframed}
\usepackage[figuresright]{rotating}
\usepackage{courier} 
\usepackage[final]{changes}

\usepackage{setspace}

\usepackage{dutchcal}

\graphicspath{ {figures/} }

\usepackage[utf8x]{inputenc} 

\usepackage{nomencl}
\usepackage{etoolbox}
\usepackage{wrapfig}
\usepackage{stfloats} 
\usepackage{framed} 
\usepackage[skins,breakable, most]{tcolorbox}

\renewcommand*\nompreamble{\begin{multicols}{2}} 
\usepackage{epstopdf} 
\usepackage{todonotes}
\usepackage{setspace}
\usepackage{siunitx}

\usepackage[caption = false]{subfig}
\usepackage[hyphens]{url}
\usepackage{graphicx}
\usepackage{booktabs}

\usepackage{lmodern,textcomp}
\usepackage{rotating} 
\usepackage{multicol} 
\usepackage{multirow} 
\usepackage{array} 
\usepackage{mathtools, cuted}
\usepackage{lscape}
\usepackage{tabularx} 
\usepackage{longtable}  
\usepackage[flushleft]{threeparttable}
\usepackage{threeparttablex} 
\usepackage{tabulary}
\usepackage{amssymb}
\usepackage{pifont}
\RequirePackage[ngerman=ngerman-x-latest]{hyphsubst}
\usepackage{microtype}
\usepackage[american]{babel} 
\usepackage{hyperref}
\usepackage{float}
\usepackage[normalem]{ulem}
\useunder{\uline}{\ul}{}

\usepackage{natbib}
\biboptions{square,comma,sort&compress}

\setcitestyle{numbers}
\journal{}
\makeatletter
\def\ps@pprintTitle{%
 \let\@oddhead\@empty
 \let\@evenhead\@empty
 \def\@oddfoot{}%
 \let\@evenfoot\@oddfoot}
\makeatother

\begin{document}
\begin{frontmatter}

\title{A meta-analysis of residential PV adoption: the important role of perceived benefits, intentions and antecedents in solar energy acceptance}

\author[IIRM]{Emily Schulte\corref{cor1}}
\ead{schulte@wifa.uni-leipzig.de}
\author[IIRM,DTU]{Fabian Scheller}
\author[K]{Daniel Sloot}
\author[IIRM]{Thomas Bruckner}
\cortext[cor1]{Corresponding author}

\address[IIRM]{Chair of Energy Management and Sustainability, Institute for Infrastructure and Resources Management (IIRM), Leipzig University}
\address[DTU]{Energy Economics and System Analysis, Division of Sustainability, Department of Technology, Management and Economics, Technical University of Denmark (DTU)}
\address[K]{Chair of Energy Economics, Institute for Industrial Production (IIP), Karlsruhe Institute of Technology (KIT)}

\begin{abstract}
The adoption of residential photovoltaic systems (PV) is seen as an important part of the sustainable energy transition. To facilitate this process, it is crucial to identify the determinants of solar adoption. This paper follows a meta-analytical structural equation modeling approach, presenting a meta-analysis of studies on residential PV adoption intention, and assessing four behavioral models based on the theory of planned behavior to advance theory development. Of 653 initially identified studies, 110 remained for full-text screening. Only eight studies were sufficiently homogeneous, provided bivariate correlations, and could thus be integrated into the meta-analysis.The pooled correlations across primary studies revealed medium to large correlations between environmental concern, novelty seeking, perceived benefits, subjective norm and intention to adopt a residential PV system, whereas socio-demographic variables were uncorrelated with intention. Meta-analytical structural equation modeling revealed a model (N = 1,714) in which adoption intention was predicted by benefits and perceived behavioral control, and benefits in turn could be explained by environmental concern, novelty seeking, and subjective norm. Our results imply that measures should primarily focus on enhancing the perception of benefits. Based on obstacles we encountered within the analysis, we suggest guidelines to facilitate the future aggregation of scientific evidence, such as the systematic inclusion of key variables and reporting of bivariate correlations.

\end{abstract}




\end{frontmatter}https://www.overleaf.com/project/6177acf780624b68f1ee2277

\section*{Highlights}
\begin{itemize}
\item Meta-analytical structural equation modeling (MASEM) approach
\item Heterogeneity among studies and inconsistent use of predictors hampers analysis
\item Perceived benefits are the strongest predictor of residential PV adoption intention
\item Benefits can be explained by environmental concern, novelty seeking and social norm
\item Consistent predictors and reporting standards mandatory for future meta-analyses
\end{itemize}

\section{Introduction}
\label{S:1}
According to the World Energy Outlook 2020, \q{solar becomes the new king of electricity} in the near future \citep{IEA.2020}, being the main driver of global growth in the share of renewable energies in all scenarios considered. Particularly rooftop photovoltaic systems (PV) are widely accepted by the population (e.g. \citep{BundesministeriumfurUmwelt.}). 
Considering that the decarbonization of the residential sector, which largely depends on individual decision-making to adopt low-carbon energy sources, is central to reaching global goals to reduce carbon emissions to net-zero by 2050 (c.f. \citep{Mercure.2014, Geels.2018, Axsen.2012}), the antecedents of residential adoption of PV systems have been researched widely to help to accelerate diffusion \citep{Alipour.2020, Karakaya.2015b}.

Residential uptake of PV systems has been approached from various perspectives. Drawing on the relative novelty of the product, the Diffusion of Innovation Theory (DOI) \citep{Rogers.2003} has been applied to identify differences between adopters and non-adopters concerning their individual characteristics and their perception of product traits, e.g. \citep{Labay.1981, Sawyer.1982, Faiers.2006}. Based on the interpretation of PV adoption as a pro-environmental behavior, theories of moral decision-making such as the Value-Belief-Norm Theory (VBN) \citep{Stern.2000} have been used, e.g. \citep{Wolske.2017}. Furthermore, the adoption of a PV system is a consumer behavior, making the Theory of Planned Behavior (TPB) \citep{Ajzen.1991} applicable to explain adoption intention and behavior using the product-specific constructs attitude, subjective norm and perceived behavioral control (PBC), e.g. \citep{Irfan.2020, Abreu.2019, Sun.2020}. Typically, the general structure of the TPB is extended with variables relating to DOI and VBN to suit the specific application case, which has been done in multiple areas of environmental science \citep{Si.2019}. Due to the different theoretical perspectives and the large number of explanatory variables employed, questions remain regarding the role of different predictors for PV adoption.

In their recent review of 173 studies related to residential PV adoption, Alipour et al. \cite{Alipour.2020} identify a total of 333 predictors. They find that \q{the quality of research varies widely, depending on the availability of data, depth and tools used in analysis, models and theories, and particularly the quality and appropriate use of [...] predictors.} \citep{Alipour.2020}. Investigating the use of predictors in the literature over time, no trend towards an emerging, comprehensive list of predictors could be identified, and results suggest that the most frequently applied predictors (age, income, financial knowledge) are less correlated with behavior than attitudinal traits \citep{Alipour.2020}.

Aside from using a plethora of different predictors, there are inconsistencies in the use of predictors whose measurement is not as straightforward as age and gender. On the one hand, predictors relating to the same underlying concept have various names. For example, the effect of expectations of the social network on decision-making has been referred to as normative beliefs \citep{Wolske.2017}, subjective norm \citep{Tan.2017, Korcaj.2015}, injunctive norm \citep{Litvine.2011}, social influence \citep{Ozaki.2011}, social endorsement \citep{Gerpott.2010}, social norm \citep{Ek.2008}, and passive and active peer effect \cite{Bollinger.2012, Rai.2013, Palm.2017, Rode.2020, scheller2021active}. On the other hand, semantically similar predictors are not operationalized similarly. For example, whilst Sun et al. \cite{Sun.2020} assess environmental concern with three items out of the New Environmental Paradigm scale (NEP) \citep{Dunlap.2008, Hawcroft.2010}, Robinson and Rai \cite{Rai.2015b} operationalize environmental concern with \q{In general, I am concerned about environmental issues} and \q{I am concerned about air pollution}. These differences raise questions concerning construct reliability and validity \citep{Hunter.2004}.

Furthermore, comparing the results of different studies is complicated by different sampling procedures and contexts. For example, Sun et al. \cite{Sun.2020} and Parkins et al. \cite{Parkins.2018} include homeowners only, but the latter includes only those where PV is technically feasible. Conversely, Arroyo and Carrete \cite{Arroyo.2019} screen for single-family residential homes, but not for homeownership. Yet, it is likely that intentions to adopt a PV system are lower for renters, and that the technical feasibility is a major determinant for adoption intentions \citep{Galvin.2020}.
In addition, the literature body on PV adoption covers different countries with unique policy regimes, technical systems, geographical constraints and general living conditions (e.g., Norway \citep{Cherry.2020}, Israel \citep{Bashiri.2018}, Japan \citep{Mukai.2011}, and Germany \citep{Scheller.2020}).
Lastly, one should remember that due to the limited sample size, sampling error is present in any study, hampering the informative value of single studies \citep{Hunter.2004, Schulze.2004, Borenstein.2009}.
Consequently, although a large number of studies has examined the determinants of residential PV adoption, general insights into the roles of the various predictors remain unclear.\\

In this research paper, we aim to advance theory development concerning residential PV adoption. To this end, our analysis follows the meta-analytical structural equation modeling (MASEM) approach as outlined by Bergh et al. \cite{Bergh.2016} and applied by e.g. Kloeckner \cite{Klockner.2013} and Bamberg and Moeser \cite{Bamberg.2007}. MASEM combines traditional meta-analysis, analyzing direction and significance of bivariate relationships, with structural equation modeling, allowing researchers to \q{draw on accumulated findings to test the explanatory value of a theorized model against one or more competing models} \cite{Bergh.2016}.
With this approach, we aim to (1) determine point estimates of relationships between socio-demographic variables, the typical TPB constructs attitude, subjective norm and perceived behavioral control, and additional attitudinal variables related to environmental motivation and innovativeness, and adoption intention, (2) assess the suitability of the original TPB to replicate the extracted data and (3) extend the theory to match the specific application case better.

Contrary to narrative reviews as in \cite{Karakaya.2015b} and \cite{Kastner.2015}, and the semi-quantitative systematic literature review in \cite{Alipour.2020}, a statistical meta-analysis necessitates more rigid inclusion criteria, since only such studies working on the same research question can be meaningfully combined \citep{Hunter.2004, Schulze.2004, Borenstein.2009}. For this reason, only studies analyzing adoption intentions are assessed, as such intentions are a central predictor of actual behavior \citep{Ajzen.1991}.\\

This research paper contributes to the scientific debate in three substantial ways. On the one hand, it is the first attempt to quantitatively summarize the results of empirical studies in the rapidly growing research area of residential PV adoption. Furthermore, we advance theory development by testing the explanatory value of TPB and three extended versions tailored to residential PV adoption, providing a fundamental analytical structure for future studies. Lastly, we encountered several obstacles in selecting and coding the literature, which caused us to include only eight studies in the meta-analysis. Moreover, the studies used overlap only to a limited extent in the use of variables. This issue is not unprecedented (c.f. \cite{Klockner.2013, Bamberg.2003}), and Stamm and Schwarb state that in the past, \q{major impulses of meta-analysis came not so much from the actual results, but rather from side effects such as the evidence of moderators, quality control, standard setting for single-studies, and indication of research gaps} \cite{Stamm.1995}. Consequently, the last contribution of this research paper are guidelines for future empirical studies to improve the aggregability of scientific evidence.\\

Section \ref{S:2} provides insights into the current research on residential PV adoption and recent findings, and the four models assessed in the main part are presented. Thereafter, the methodology is described (Section \ref{S:3}). Results are provided in Section \ref{S:4}. In the discussion, results are put in perspective with the literature body, and recommendations for future studies are given (Section \ref{S:5}). The paper closes with a brief conclusion and policy implications.

\section{Background}
\label{S:2}
\subsection{Research on the residential adoption of PV systems}
\label{S:2.1}
First studies on the residential adoption of PV systems were conducted in the early 1980s (e.g. \citep{Labay.1981, Sawyer.1982}), when the DOI \citep{Rogers.2003} was applied to explain the motives of early adopters. Since then, a large literature body emerged following different conceptual and methodological approaches. Data is typically generated from qualitative interviews \citep{Palm.2018, Koch.2018, Karjalainen.2019, Scheller.2021}, quantitative surveys \citep{Korcaj.2015, Engelken.2018, Zander.2019}, choice experiments \citep{Galassi.2014, Scarpa.2010, Willis.2011}, or panel data at the individual \citep{Best.2019, Nair.2010} or spatial level \citep{Dharshing.2017, Groote.2016, BaltaOzkan.2015}. The conceptualization of adoption ranges from interest \citep{Setyawati.2020, Leenheer.2011}, intention \citep{Cherry.2020, Aggarwal.2019, Korcaj.2015}, willingness to pay \citep{Cherry.2020, Scarpa.2010, Abdullah.2017} to actual adoption behavior \citep{Simpson.2015, Schelly.2014, Rai.2013b}. Data is assessed through methods like content analysis or semi-quantitative methods such as vote counting in qualitative studies \cite{Galvin.2020,Karjalainen.2019,Palm.2016}. In quantitative studies, bivariate methods are applied, to, for example, compare groups \cite{Axsen.2012,Petrovich.2019,scheller2021active}, and linear \cite{Wolske.2017,Curtius.2018} or logistic regression analysis \cite{Bashiri.2018,Jan.2020} or structural equation modeling \cite{Parsad.2020,Kapoor.2020,Aggarwal.2019} is performed. As aforementioned, behavioral models such as TPB \citep{Abreu.2019, Irfan.2020}, VBN \citep{Wolske.2017} and DOI \citep{Kapoor.2020, Wolske.2020, Alrashoud.2019} are used on their own or in combination, and tailored to the specific application case. 
In single studies, individual subsets of the explanatory variables that affect individual decision-making are analyzed.\\

In 2011, the meta-analysis of Arts et al. \cite{Arts.2011} revealed that socio-demographic predictors can generally only weakly explain individual innovation adoption decisions. In line with this, the review of Kastner and Stern \cite{Kastner.2015} deems such variables unimportant for energy-related investment decisions, suggesting that the use of socio-demographic measures in surveys might be related to the ease of measurement as compared to attitudinal variables.
However, the recent review of Alipour et al. \cite{Alipour.2020} shows that socio-demographic variables are the most commonly used predictors of PV adoption, although in this study too, they appear not to be the best predictor variables.

The analyses of Niamir et al. \cite{Niamir.2020} and Jan et al. \cite{Jan.2020} reveal a key role of education for household's investments in, and social acceptability of solar panels, respectively. This could be because a certain level of education is needed to understand the financial costs and benefits of PV systems, as 85\% of respondents of Simpson and Clifton \cite{Simpson.2017} indicate. Also Nair et al. \cite{Nair.2010} find effects of education on homeowners preferences concerning energy efficiency measures, with income and age also playing a role. The findings of Karytsas et al. \cite{Karytsas.2019} do not indicate an influence of education for the installation of microgeneration systems, however, they show effects of gender, age, and income. Particularly the effect of income is broadly confirmed in the literature body.

Several studies identify effects of income or wealth on adoption intention, e.g. \citep{Best.2019, Jan.2020}. In this vein, Jacksohn et al. \cite{Jacksohn.2019} find that high upfront costs hinder adoption to a much larger extent than revenues drive adoption. Several studies draw on the fact that despite costs decreased and revenues increased, high-income households remain more likely to adopt than low and medium income (LMI) households \citep{OShaughnessy.2020} and aim to understand better how uptake among LMI households can be enhanced \citep{Reames.2020, Wolske.2020, OShaughnessy.2020, Chapman.2019}. Results indicate that the elimination of the constraining factor income through LMI-specific policies, typically reducing upfront costs, shifts deployment to LMI markets \citep{OShaughnessy.2020, Chapman.2019}. Moreover, current adopters among LMI and high-income households appear to be \q{more alike than different} concerning their psychographic image \citep{Wolske.2020}.

Oppositely, Zander \cite{Zander.2020},  Balta-Ozkan et al. \cite{BaltaOzkan.2015} and Anugwom et al. \cite{Anugwom.2020} cannot find significant effects of income on adoption. The latter instead find that adoption is primarily influenced by \q{issues like the reliability of the systems, perceived problems in access to grid connections, and limited scope of energy generated}. However, the study took place in the context of rural Nigeria and might not translate to Western countries.\\

The TPB has been proven useful in explaining many types of environmental behavior \citep{Si.2019}.
In the realm of residential PV adoption, Abreu et al. show that norms and attitudes have strong- (norms) and medium-sized (attitudes) effects on intentions to adopt \cite{Abreu.2019}. In their analysis, PBC does not explain intentions. This is surprising, given that several barriers related to the socio-technical system, management and economic factors, and unfavourable policy regimes have been previously identified \citep{Karakaya.2015b}. Recent studies indicate that, for example, \q{roof orientation causes a degree of self-selection among would-be prosumers} \citep{Galvin.2020}, and that \q{capital cost, lack of information, and maintenance requirements} \citep{Alsabbagh.2019} are keeping interested Bahrain citizens from adoption. The studies of Korcaj et al. \cite{Korcaj.2015} and Engelken et al. \cite{Engelken.2018} both find effects of medium strength for attitudes and subjective norm, and a small \citep{Korcaj.2015} and medium \citep{Engelken.2018} effect of PBC on purchase intention.

Whereas subjective norms appear to be measured in a rather consistent way, typically asking respondents about perceived descriptive and injunctive norms, there appears to be disagreement concerning the operationalization of attitudes and PBC among the three studies.
Abreu et al. \cite{Abreu.2019} investigate paths between attitudes, subjective norm, PBC, and purchase intention, and followed indications of a factor analysis to cluster positive statements under attitudes, and negative statements under PBC. 
Korcaj et al. \cite{Korcaj.2015} and Engelken et al. \cite{Engelken.2018} measure attitudes with simple, global statements such as \q{I find a PV system gives me a good feeling} \citep{Korcaj.2015}, and add variables to explain the formation of attitudes in more detail. Korcaj et al. include economic, social, financial and autarky benefits, and perceived costs, and Engelken et al. account for environmental awareness and technology affinity, financial and autarky benefits and perceived costs. For PBC, Engelken et al. employ two items, asking respondents whether they had the power to decide for or against a PV system. Aside from decision power, Korcaj et al. examine technical feasibility, the ability to finance the system and the ability to get a permit. Going even further, Abreu et al. incorporate amongst others statements about maintenance and the ability to find a contractor \cite{Abreu.2019}.

In line with the factor analysis of Abreu et al.\cite{Abreu.2019}, Claudy et al. point towards a meaningful distinction between attitudes, arising exclusively from positive statements, and PBC, accumulating all kinds of barriers \cite{Claudy.2013}.
Their results imply that whereas benefits have a strong and significant positive relationship with attitudes (determined by simple, global statements), but not with intention, barriers have a medium significant negative relationship with intention, but not with attitudes. Attitudes have a strong and significant relationship with intention, which is stronger than the relationship between barriers and intention.\\

A large number of studies shows positive relations between general personal motivations and PV adoption.
Variables designed to measure the effect of environmental attitudes on behavior include those that measure general attitudes toward the environment (e.g. environmental awareness \citep{Mundaca.2020}, interest in addressing global warming \citep{Alrashoud.2020}) as well as variables that measure prior pro-environmental behavior (e.g. environmental behavior \citep{Karytsas.2019}) or both (e.g. environmental preferences and related behavior \citep{Best.2019}). The variables have been found to influence positively intention \citep{Mundaca.2020}, likelihood to install \citep{Best.2019}, social acceptance \citep{Alrashoud.2020} and installation \citep{Karytsas.2019}. Evaluating the influence of environmental behavior over time, Karytsas et al. reveal that it had a consistent effect on installation \cite{Karytsas.2019}. Schelly and Letzelter even find that \q{environmental motivations are slightly more important than economics} for adopters \cite{Schelly.2020}. Oppositely, Zander et al. reveal that economic considerations are of higher importance than environmental motivations \cite{Zander.2019}.

Wolske shows that irrespective of income differences, current PV adopters in the US are characterized by strong pro-environmental norms and a strong propensity to novel goods \cite{Wolske.2020}, which was true for the early adopters in Australia too, where adoption is nowadays driven more by economic considerations \cite{Simpson.2017}. In Sweden, the role of financial incentives increased over time \citep{Palm.2018}, however, in line with Karytsas et al \cite{Karytsas.2019}, environmental motivations appear to drive adopters consistently \citep{Palm.2018}. Interestingly, also Karjalainen and Ahvenniemi describe the early adopters in Finland as people who enjoy effortlessly producing pollution-free energy, \q{and being able to deliver information about clean energy production to others through their own installations} \citep{Karjalainen.2019}, evoking the typical early adopters of DOI. It appears that particularly in the early stages of the diffusion process, general personal motivations of the decision-maker are useful explanatory variables.

Contrary to the findings of Simpson and Clifton \cite{Simpson.2017} and Palm \cite{Palm.2018}, the analysis of Jacksohn et al. suggests that economic factors (costs and revenues) are the most important factors for explaining adoption. Environmental concern had comparatively low relevance \citep{Jacksohn.2019}.\\

Despite the fact that there appears to be consensus about the general directions of some effects, e.g. negative effects of perceived barriers \citep{Claudy.2013, Korcaj.2015, Karakaya.2015b} and low wealth or income \citep{Engelken.2018, Best.2019, OShaughnessy.2020}, as well as positive effects of perceived benefits \citep{Wolske.2017, AzizN.S..2017, Labay.1981}, and general personal motivations such as environmental attitudes and innovativeness \citep{Wolske.2020, Karjalainen.2019}, a number of uncertainties remain. Amongst others, this is related to the inconsistent use of predictors, different operationalizations of predictors, differences in dependent variables assessing PV adoption in the form of interest, intention, willingness to pay, acceptability and different contextual environments including, e.g. subsidy programs and different market maturity states.

Attempts to accumulate knowledge from multiple studies are surprisingly scarce. Karakaya et al. \cite{Karakaya.2015b} and Kastner and Stern \cite{Kastner.2015} perform narrative reviews focusing on barriers for PV adoption and explanatory variables for household energy investments, respectively. Alipour et al. \cite{Alipour.2020} conduct a systematic literature review and employ semi-quantitative methods like vote counting and correlational analysis of effect signs (negative/positive) and significance levels. The mentioned reviews help in identifying research gaps and provide up-to-date knowledge, however, they involve subjectivity, and cannot offer clear conclusions regarding specific relationships such as point estimates of effect sizes \citep{Hunter.2004, Schulze.2004, Borenstein.2009}.

\subsection{Four models to analyze residential PV adoption}
\label{S:2.2}
\begin{figure}[ht]
    \centering
    \includegraphics[width=0.5\textwidth]{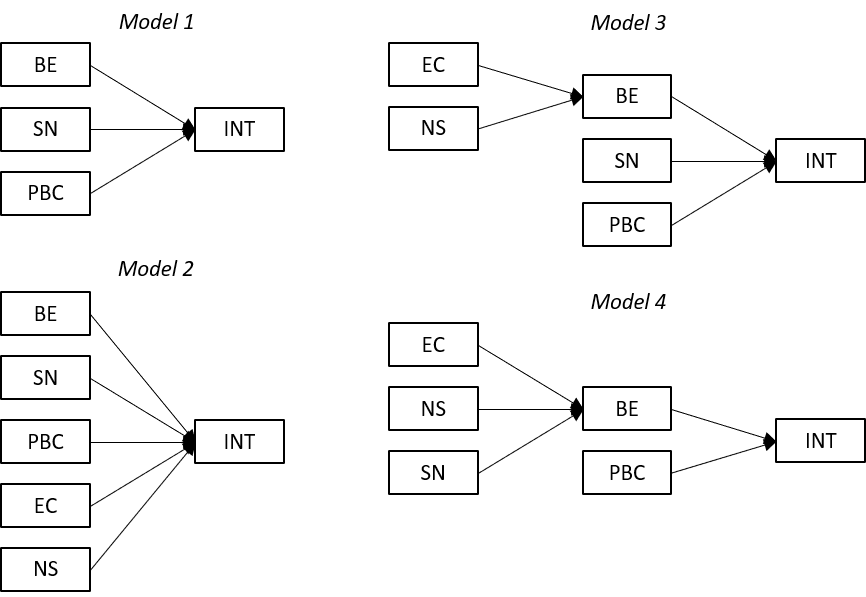}
    \caption{The Theory of Planned Behavior (Model 1) and three alternative models. Model 2 includes additional general personal motivations as antecedents of intention. Model 3 proposes the general personal motivations to be predictors of benefits. Model 4 suggests subjective norm to be a third predictor of benefits. 
    INT Intention; EC Environmental concern; NS Novelty Seeking; PBC Perceived behavioral control; BE Benefits; SN subjective norm}
    \label{fig:sems_structure}
\end{figure}

In Figure \ref{fig:sems_structure}, the four models that are analyzed in the last step of the MASEM analysis are presented. 
The first model is closely related to TPB and is used to assess the suitability of the original TPB to replicate the extracted data. Model 1 deviates from TPB in that attitudes are replaced with benefits. A distinction between positive (benefits) and negative (PBC) perceptions provides a more simple structure to cluster variables from primary studies in the first analytical step, the meta-analysis. For example, variables such as environmental, financial and autarky benefits can be subsumed under benefits, and cost, maintenance and technical barriers under PBC, in line with Abreu et al. \cite{Abreu.2019} and Claudy et al. \cite{Claudy.2013}.

Models 2-4 are alternative models to predict adoption intention that are based on the TPB, but extended based on the literature discussed above. They are thus used to assess whether models tailored to the specific application case better match the extracted data.
Model 2 is a simple integration of the general personal motivations, namely environmental concern and novelty seeking, as antecedents of intention. A structure in which all variables directly predict the dependent variable has amongst others been analyzed by \cite{Mundaca.2020, Jacksohn.2019, Aggarwal.2019}.
In Model 3, we take into account the distinction between general personal motivations and specific beliefs about PV and its attributes, and the logic \q{that broad dispositions may influence how an innovation is perceived} \citep{Wolske.2017}. We thus integrate the general personal motivations as antecedents of perceived benefits. This approach is in line with the integrated framework for predicting interest in pursuing PV of \cite{Wolske.2017}, and the analytical approach of e.g. \cite{Engelken.2018, Sun.2020}. 
In Model 4, subjective norm is proposed as an additional antecedent of perceived benefits. The relationship between diffusion stage and product perception is a central aspect of DOI, where increasing diffusion levels lead to more favorable evaluations due to an increase in familiarity \citep{Rogers.2003}.

\section{Methods}
\label{S:3}

\subsection{Literature selection, data extraction and processing}
\label{S:3.1}

\begin{figure}[ht]
    \centering
    \includegraphics[width=0.75\textwidth]{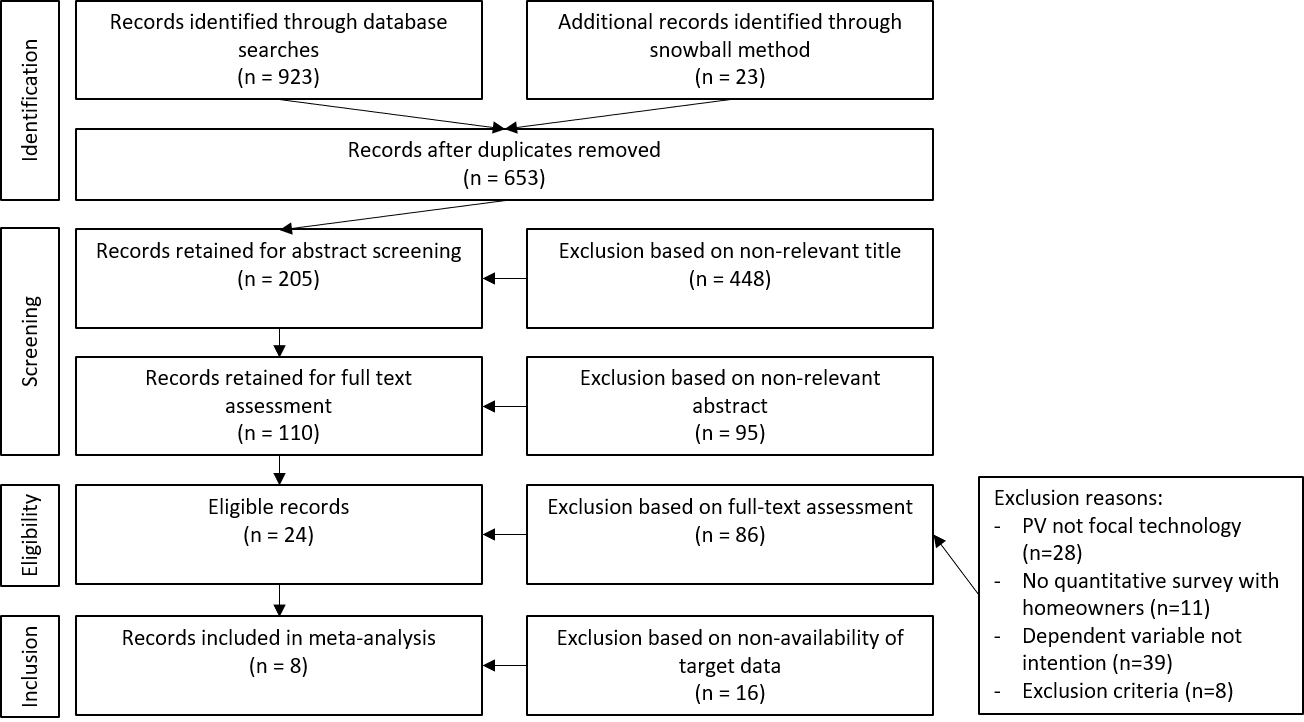}
    \caption{Flow-chart of the literature selection process. The literature identified by a database and snowball search was screened by two independent researchers, investigating titles, abstracts, and full-texts. 24 papers were suitable content-wise, however, due to lacking data, only 8 could be included in the analysis.}
    \label{fig:flow_chart}
\end{figure}

To advance theory development concerning residential PV adoption, we conducted a MASEM analysis of existing empirical studies, consisting of a traditional meta-analysis and structural equation modeling (c.f. \cite{Bergh.2016}). 
Meta-analysis is a method to combine effect sizes from different primary studies and provide a quantitative estimate of the effect. Thereby, meta-analysis is able to provide more accurate estimates with greater statistical power compared to primary studies \citep{Schulze.2004, Borenstein.2009, Hunter.2004}. In addition, MASEM uses the results of a traditional meta-analysis to advance theory integration and development by modeling the relationships between effect sizes from different theories or perspectives (see e.g. \cite{Klockner.2013, Bamberg.2007, vanZomeren.2008}).\\

To ensure adequate rigor and transparency in the process, we followed recommendations of \cite{Brocke.2009}, \cite{Stanley.2013}, and \cite{Aytug.2012}, and the Preferred Reporting Items for Systematic Reviews and Meta-Analyses (PRISMA) \citep{Shamseer.2015}.
Since only such studies working on the same research question can be meaningfully combined \citep{Schulze.2004, Borenstein.2009, Hunter.2004}, we defined inclusion criteria to ensure coherence on the conceptual level. Studies had to be peer-reviewed scientific articles written in English, and include results of a quantitative survey on residential PV adoption intention. As additional exclusion criteria, we defined (1) the inclusion of off-grid households in the primary study, because motivations to adopt a PV system deviate in case of no electricity access via the grid and (2) choice experiments. Choice experiments were excluded because the methodology has no direct measure of adoption intention, and effects of single predictors are not measured separately. Target statistics were bivariate correlations and sample sizes.

A literature search with a fixed set of keywords\footnote{"photovoltaic" OR "solar" OR "PV" OR "own power") AND ("consumer*" OR “household*” OR "homeowner*" OR "residential" OR "private" OR "domestic") AND ("adopt*" OR "decision*" OR "behavio\$r" OR "intent*" OR "push*" OR "hinder*" OR "reason*") AND ("*survey*" OR "question\$a*" OR "empiric*" OR "quant*"} was conducted in August 2020 using the databases Web of Science (518 results), PsychINFO (31 results) and Scopus (374 results)\footnote{Because we recognized overlaps with topics such as "partner violence" in the first two databases, we set additional filters by subject areas on Scopus and excluded Medicine (203), Chemical Engineering (63), Chemistry (0), Biochemistry, Genetics and Molecular Biology (61), Pharmacology, Toxicology and Pharmaceutics (22), Immunology and Microbiology (18), Neuroscience (12), Nursing (11), Veterinary (10)}. Using \cite{Alipour.2020} and 5 quantitative surveys on PV adoption \citep{Zander.2019, Sun.2020, Aggarwal.2019, Rai.2015b, Korcaj.2015}, an additional snowball search was conducted. The search yielded 946 results in total, of which 653 remained after doublets removal. 
Studies were published between 1981 \citep{Labay.1981} and 2020, with exponentially increasing numbers of studies per year starting in 2008. The most frequent journals are Energy Policy, Renewable Energy and Applied Energy. The screening procedure (Figure \ref{fig:flow_chart}) was conducted by two independent researchers and encompassed title screening, abstract screening, and full-text screening (Intercoder agreement: 89\%). Intercoder disagreements were resolved through discussion. Doublet removal and the first two steps were conducted with the free screening software Rayyan\footnote{https://rayyan.qcri.org/}. In the Appendix, a list of the 110 papers eligible for full-text screening with a documentation of the screening results (Table \ref{tab:fulltext}), and an overview of the 24 eligible records (Table \ref{tab:lit_select}) are provided. Exclusion based on full-text assessment was caused in 28 cases by not restricting the analysis to residential PV, in 11 cases by the lack of results of a quantitative survey with residential decision-makers, in 39 cases by not analyzing intention as dependent variable, and in 8 cases by previously determined exclusion criteria. More precisely, \cite{Lay.2013, Smith.2014, Rahut.2018, Opiyo.2019} and \cite{Anugwom.2020} include residents without grid access, and \cite{Zander.2019} and \cite{Bao.2020} perform choice experiments. \cite{Liang.2020} was not accessible as full-text article. From the 24 eligible papers, 5 included the target statistic bivariate correlations \citep{Sun.2020, AzizN.S..2017, Wolske.2017, Chen.2014, Claudy.2013}. This issue is not unparalleled. Bamberg and Moeser lost 31 out of 81 identified studies due to lacking data \cite{Bamberg.2007}, and Delmas et al. had to exclude studies because they did not report effect sizes in a format that enables conflation \cite{Delmas.2013}. After contacting the remainder via e-mail, 3 more datasets were gathered \citep{Rai.2015b, Arroyo.2019, Parkins.2018}. For detailed information on the final selection of 8 papers, please refer to Table \ref{tab:lit_final}. The studies were conducted between 2013 and 2020, and countries from the western and northern hemisphere are in the majority. Study samples covered from 72 up to 2065 respondents, with a mean of 566 participants.

\begin{landscape}
\begin{table}[htbp]
    \centering
    \footnotesize
    \renewcommand{\arraystretch}{1.0} 
    \setlength{\tabcolsep}{4pt}
\begin{threeparttable}
  \caption{Description of 8 primary studies included in the meta-analysis.}
  \label{tab:lit_final}
\begin{tabularx}{\linewidth}{p{1.5em}p{9em}p{8em}p{8em}p{4em}p{2em}p{2em}p{15.5em}p{14.5em}}
    \toprule
    \textbf{Ref.} & \textbf{Intention variable} & \textbf{Conceptual background} & \textbf{Survey Method} & \textbf{Region} & {\textbf{Year}} & {\textbf{N}} & \textbf{Screener} & \textbf{Statistical methods} \\
    \midrule
    \cite{Sun.2020}  & Intention for rooftop PV installation & TPB, self-determination theory & Online questionnaire & Taiwan & 2020 & 300 & Employees of "one of the top-50 national companies" & Confirmatory factor analysis, SEM (AMOS 18) \\
    \midrule
    \cite{Claudy.2013}   & Intention & Behavioral reasoning theory & Computer assisted telephone survey & Ireland & 2013 & 254 & Homeowners, aware of PV, decision power in financial issues, quota sampling based on region, gender, age & Confirmatory factor analysis, SEM (AMOS 18) \\
    \midrule
    \cite{Rai.2015b}  & Intention to call installer & TPB & Post card survey & Texas & 2015 & 522 & Homeowners of single family houses & Logit model (considering a solar installation), ordered logit model (likelihood to call a solar installer) \\
    \midrule
    \cite{Chen.2014}   & Intention &  & Online and paper-and-pen survey & Taiwan & 2014 & 203 & Students and staff of a University & Exploratory and confirmatory factor analysis, SEM \\
    \midrule
    \cite{Arroyo.2019}   & Purchase intention 1 year & Goal framing theory & Face-to-face survey & Mexico & 2019 & 72  & Equal distribution male - female, medium - high socioeconomic level & Descriptive statistics, group comparison (Mann-Whitney test), analysis of covariance \\
    \midrule
    \cite{Parkins.2018}   & Solar adoption intention & VBN & Online survey & Canada & 2018 & 2065 & Homeowner, PV installation applicable, quotas for age, gender, language, income, education, regional distribution & Descriptive statistics, ordered probit model (MIMIC model) \\
    \midrule
    \cite{AzizN.S..2017}   & Purchase intention & TRA, DOI & Online survey & Malaysia & 2017 & 211 & White collar employees, non PV users & Descriptive statistics, reliability, correlational and regression analysis \\
    \midrule
    \cite{Wolske.2017}   & Interest in talking to a PV installer & TPB, VBN, DOI & Online survey & USA & 2017 & 904 & Homeowners, not spoken to solar company, quotas for age and education & SEM (Stata 14), nested sequence of OLS multiple regressions  \\
    \bottomrule
\end{tabularx}
\begin{tablenotes}
    \scriptsize
        \item TPB: Theory of Planned Behavior; VBN: Value-Belief-Norm Theory; DOI: Diffusion of Innovation Theory; TRA: Theory of Reasoned Action; SEM: Structural Equation Modeling
\end{tablenotes}
\end{threeparttable}
  
\end{table}
\end{landscape}

Aiming towards a meta-analytically pooled correlation table, correlations of the individual studies had to be grouped, which presented the first challenge, because no uniform set of variables, taxonomy and operationalization was used in the studies. It was decided to cluster conceptually similar variables into groups with a minimum size of 2. For example, whilst Sun et al. assess \q{environmental concern} with three items out of the NEP scale \cite{Sun.2020}, Rai and Beck operationalize the variable with 2 general statements \cite{Rai.2015b}, and Arroyo and Carrete use 9 indicators including an activism and a recycling index, willingness to pay higher prices for water, and a general assessment of the world's environmental situation to determine \q{environmental consciousness} \cite{Arroyo.2019}. Barrier variables range from affordability issues (Perceived behavioral control: \q{A solar system is affordable for my household} \cite{Rai.2015b} and Cost Barrier: \q{The initial costs of installing solar panels would be too high for me.}, \q{I find the initial costs of installing solar panels a financial strain.} \cite{Claudy.2013}) to perceived risks such as \q{Installing solar panels could damage my home.} (Riskiness) \citep{Wolske.2017} or \q{Solar panels would not provide the level of benefits I would be expecting.} (Risk Barrier) \citep{Claudy.2013}. Arroyo and Carrete combine costs, risks, and other barriers such as finding a vendor or lack of government incentives to \q{Perceived barriers} \cite{Arroyo.2019} . 
To increase the number of measures available for pooling per correlation, a parsimonious cluster, meaning a rough classification, in line with the proposed models was chosen. The parsimonious cluster includes intention (8/8), the socio-demographic variables gender (2/8), education (3/8) and income (3/8), the TPB constructs PBC (4/8), benefits (5/8) and social norm (4/8), and the general personal motivations environmental concern (7/8) and novelty seeking (4/8).
In case of conceptually similar measures within one study (e.g. environmental concern and ecological lifestyle \citep{Sun.2020}, clustered under environmental concern), the composite was computed. For composites with one measure in one variable and several measures in the other variable, Formula (\ref{formula1}) was used: 
 
\begin{align}
\label{formula1}
\scriptsize
    r_{xY}=\frac{\sum\nolimits_ {} r_{xy_{i}}}{\sqrt{n+n(n-1)\overline{r}_{y_{i}y_{j}}}},
\end{align}

with $\sum\nolimits_ {} r_{xy_{i}}$ being the sum of correlations between the single and the $y_{i}$ measures, and $\overline{r}_{y_{i}y_{j}}$ the medium correlation between the $y_{i}$ measures. For composites with several measures in both variables, Formula (\ref{formula2}) was applied:

\begin{align}
\label{formula2}
\scriptsize
    r_{XY}=\frac{R_{xy}}{\sqrt{R_{xx}}\sqrt{R_{yy}}},
\end{align}

with $R_{xy}$ being the matrix of cross-correlations between the $x_{i}$ and the $y_{i}$ measures, and $R_{xx}$ and $R_{yy}$ the matrices of cross-correlations between the $x_{i}$ and the $y_{i}$ measures respectively. Note that the two terms in the denominator are $SD_{x}$ and $SD_{y}$ respectively.

Because the variable groups benefits and PBC contain wide ranges of measures, a second and more refined clustering was performed, splitting benefits into personal (5/8) and environmental benefits (2/8), and PBC into hard (2/8) and soft (4/8) barriers. Hard barriers refer to technical constraints (incompatibility barrier \citep{Claudy.2013}, home unsuitable, may move \citep{Wolske.2017}) whereas soft barriers refer to perceptions such as cost and risk barrier \citep{Claudy.2013}, riskiness, expense concerns, PV may improve and trialability \citep{Wolske.2017}, and perceived cost maintenance \citep{AzizN.S..2017}. 
Table \ref{tab:input_parsi} and Table \ref{tab:input_refi} in the Appendix provide the original variable names from primary studies and the (composed) correlations that served as input for the meta-analysis. Focal point of the analysis was the parsimonious cluster.

\subsection{Performing meta-analysis}
\label{S:3.2}
The meta-analysis was performed in Stata 16 using the meta package\footnote{Online-documentation of the package here: https://www.stata.com/manuals/meta.pdf}. Study level artifacts have attenuating effects on correlations, and statistical methods to correct correlations for artifacts are available. For example, measurement error has a systematic multiplicative effect of the square root of both measurements' reliabilities on the study correlation, and differences in spread in independent variables also have systematic, typically alleviating multiplicative effects on study correlations (c.f. \citep{Hunter.2004}).
However, due to lack of reporting, only sampling error could be corrected, and a so-called bare-bones meta-analysis was performed. The prepared initial measures were converted into a standard normal metric by a Fisher \textit{r}-to-\textit{Z}-transformation. Standard errors were calculated. Thereafter, data for meta-analysis was specified using \texttt{meta set}, and the calculations were performed and presented using \texttt{meta summarize}. A random-effect model with inverse variance weighting using the REML (restricted maximum likelihood) method \citep{StataCorpLLC.2019} was applied. The random-effect model was selected because studies were performed independently and are thus not functionally equivalent. It is therefore unlikely that their true effect sizes are exactly the same (c.f. \citep{Brocke.2009}). The Q-Statistics for the analysis confirmed this decision (c.f. \citep{StataCorp.., Klockner.2013}). The resulting pooled correlations and lower and upper confidence intervals were transformed back into the r metric by reversing the initial transformation. Results for the parsimonious cluster are presented in Table \ref{tab:corr_parsi}, the pooled correlation table for the refined cluster is provided in Table \ref{tab:cor_refi} in the Appendix. For interpreting correlations, we refer to Cohen's guidelines for Pearson correlation coefficients, stating that correlations between .1 and .3 are small, between .3 and .5 are medium, and above .5 are large \cite{cohen2013statistical}.

\subsection{Meta-analytical structural equation modeling}
\label{S:3.3}
In the last step, we computed four structural equation models (see Figure \ref{fig:sems_structure}), using the meta-analytically pooled correlation table as input. Due to a lack of overlap in coverage of variables, pooled correlations were computed based on different sample sizes. For fitting the SEM, in line with \cite{Klockner.2013, Bamberg.2007, Bergh.2016}, we used the harmonic mean instead of the arithmetic mean or the median. According to Bergh et al., \q{the harmonic mean is the preferred option because it limits the influence of very large values and also increases the influence of smaller values, in addition to being in most if not all cases smaller than the arithmetic mean} \cite{Bergh.2016}. The harmonic mean from Model 1 (n=1640) deviates from Models 2-4 (n=1714) because fewer variables are included. 
By default, the \texttt{sem}-command of Stata 16 uses a maximum likelihood estimator. Furthermore, it is assumed that exogenous variables correlate with each other, whereas residuals do not correlate with each other or with exogenous variables \citep{StataCorp..}. Indirect effects are computed using the post-estimation command \texttt{estat teffects}. Results are presented as standardized values.

To test the models for overall significance, we performed the Wald test, an equation-level test assessing whether all coefficients, not the intercept, are zero (\texttt{estat eqtest}). To assess the suitability of the models, we compare determination coefficients $R²$, summarizing the explained variance (\texttt{estat eqgof}). Furthermore, following the recommendation of Bergh et al. \cite{Bergh.2016}, we apply multiple fit measures to assess the goodness of fit of the models (\texttt{estat gof, stats(all)}), including $\mathcal{X}$², CFI, TLI, RMSEA, SRMR, AIC and BIC. All measures, including the Wald test are interpreted based on guidelines summarized in Aichholzer \cite{Aichholzer.2017}.
\section{Results}
\label{S:4}

\subsection{Meta-analysis}
\label{S:4.1}

\begin{table}[htbp]
    \scriptsize
    \centering
    \renewcommand{\arraystretch}{1.2}
    \setlength{\tabcolsep}{2pt}
\begin{threeparttable}
    \caption{Meta-analytically pooled parsimonious correlation table with values transformed back to r-metric. Given are the correlation coefficients with significance levels, the confidence intervals, and the sample size for each correlation.}
    \label{tab:corr_parsi}
\begin{tabular}{lllllllll}
    \toprule
          & \textbf{INT} & \textbf{EC} & \textbf{NS} & \textbf{PBC} & \textbf{BE} & \textbf{SN} & \textbf{GEN} & \textbf{EDU} \\
    \midrule
    \textbf{EC} & .34 (p=.001) &     &     &     &     &     &     &  \\
        & [.14; .52], 7 &     &     &     &     &     &     &  \\
    \textbf{NS} & .47 (p\textless.001) & .44 (p\textless.001) &     &     &     &     &     &  \\
        & [.21; .67], 4 & [.35; .53], 4 &     &     &     &     &     &  \\
    \textbf{PBC} & -.11 (p=.165) & -.17 (p=.1)  & -.01 (p=.652) &     &     &     &     &  \\
        & [-.26; .04], 4 & [-.36; .03], 3 & [-.08; .05], 1 &     &     &     &     &  \\
    \textbf{BE} & .53 (p\textless.001) & .69 (p\textless.001) & .64 (p\textless.001) & -.18 (p\textless.001) &     &     &     &  \\
        & [.34; .68], 5 & [.47; .83], 4 & [.35; .81], 3 & [-.24; -.13], 3 &     &     &     &  \\
    \textbf{SN} & .33 (p\textless.001) & .28 (p\textless.001) & .50 (p=.035) & -.10 (p=.455) & .49 (p\textless.001) &     &     &  \\
        & [.17; .46], 4 & [.13; .42], 4 & [.04; .79], 2 & [-.36; .17], 2 & [.25; .68], 3 &     &     &  \\
    \textbf{GEN} & -.01 (p=.645) & .05 (p=.019) & 0 (p\textless.001) & -.04 (p=.752) & 0 (p\textless.001) & -.06 (p=.007) &     &  \\
        & [-.05; .03], 2 & [.01; .09], 2 & [0; 0], 0 & [-.27; .20], 1 & [0; 0], 0 & [-.10; -.02], 1 &     &  \\
    \textbf{EDU} & .05 (p=.532) & .05 (p=.14)  & 0 (p\textless.001) & -.03 (p=.565) & -.01 (p=.931) & .07 (p=.001) & -.09 (p\textless.001) &  \\
        & [-.10; .19], 3 & [-.02; .11], 3 & [0; 0], 0 & [-.15; .08], 2 & [-.14; .13], 1 & [.03; .11], 2 & [-.13; -.04], 2 &  \\
    \textbf{INC} & .18 (p=.17)  & .15 (p=.008) & 0 (p\textless.001) & 0 (p=.979) & .08 (p=.219) & .04 (p=.443) & -.10 (p\textless.001) & .19 (p=.152) \\
        & [-.08; .42], 3 & [.04; .26], 3 & [0; 0], 0 & [-.16; .17], 2 & [-.47; .22], 1 & [-.06; .13], 2 & [-.14; -.05], 2 & [-.07; .44], 3 \\
    \bottomrule

\end{tabular}%
\begin{tablenotes}
    \scriptsize
        \item Upper numbers: Pearson's r (significance level)
        \item Lower numbers: [lower and upper 95\% CI], number of studies
        \item INT Intention; EC Environmental concern; NS Novelty Seeking; PBC Perceived Behavioral Control; BE Benefits; SN subjective norm; GEN Gender; EDU Education; INC Income
\end{tablenotes}
\end{threeparttable}
\end{table}%

Table \ref{tab:corr_parsi} presents the pooled correlation table for the parsimonious cluster, Table \ref{tab:cor_refi} in the Appendix presents the results for the refined cluster. Both tables summarize Pearson's r with significance value, lower and upper confidence interval, and the number of studies. Input data and results of both clusters including theta, lower and upper confidence interval and relevant statistics ($tau²$, $I²$, $H²$, z-statistics with significance values, Q-statistic with significance values) are provided in two separate databases in the Supplementary Material.
It should be noted that significance levels resulting from the Z-statistic should be handled with caution in case of between-study heterogeneity which is present in most correlations, as shown by p-values below .1 in the Q-statistic.

In this analysis, only small portions of the heterogeneity among effect-size estimates can be explained by sampling error (see the I² statistic, typically showing values around 8\%). Some of the between-study variance might be caused by study level artifacts that could not be corrected due to missing information. On the other hand, large between-study variance might suggest the presence of moderating variables.\\

Concerning socio-demographic adoption determinants, our results show non-significant correlations of gender, education and income with intention. Gender and education are unrelated to intention, and between-study variance is low, pointing towards generalizability of the results. Divergently, income shows a slight positive correlation and higher between-study variance, of which only 8\% can be explained by sampling error. Thus, to draw more generalizable conclusions, a greater number of studies is needed.

Variables relating to the TPB have been assessed in 4 (PBC, subjective norm) and 5 (benefits) studies. 
The determinant benefits shows a strong (\textit{r=.53. p\textless.001}), and subjective norm a medium (\textit{r=.33, p\textless.001}), PBC a small negative, yet non-significant correlation (\textit{r=-.11, p=.165}) with intention. 

Single study effects of benefits were consistently positive (see Table \ref{tab:input_parsi}), and ranged from medium to strong in the parsimonious cluster. In the refined cluster, benefits were separated into personal (\textit{r=.54, p\textless.001}) and environmental (\textit{r=.32, p\textless.001}) benefits, suggesting a stronger relationship between personal benefits and intention as compared to environmental benefits. Because the refined clustering did not reduce between-study variance, and personal and environmental benefits were strongly correlated (\textit{r=.742, p\textless.001}), the merit of the separation within this study is limited.

Subjective norms were consistently positively correlated with intention, ranging from small to medium correlations. The between-study variance was relatively low, pointing towards generalizability of results. Additionally, subjective norm is correlated strongly with benefits and novelty seeking (\textit{r=.49, p\textless.001} and \textit{r=.50, p=.035}), suggesting that individuals who perceive a stronger subjective norm to act also tend to evaluate PV systems more favorably, and that subjective norm is perceived stronger by individuals who score high on novelty seeking. Because all correlations between novelty seeking, subjective norm and benefits are particularly strong, more meaningful insights are expected from the SEMs, as they work with partial correlations to determine the unique variance explained by the independent variables.

PBC shows a small negative correlation with intention to adopt, yet, the correlation is non-significant. Similar to subjective norms, between-study variance is rather low. Also, a small negative correlation between PBC and benefits (\textit{r=.-18, p\textless.001}) is shown, suggesting that an increased perception of benefits comes along with a decreased perception of barriers and vice-versa. In the refined cluster, PBC was separated into hard and soft barriers. The deviating correlations suggest a small correlation of hard barriers with intention (\textit{r=-.18, p\textless.001}), whereas soft barriers are unrelated to intention (\textit{r=-.08, p=.284}). In this case, the refined cluster also reduced between-study variance. Hard and soft barriers have a medium relationship (\textit{r=.34, p\textless.001}). Similar to benefits, it should be noted that there appears to be a distinction between largely immutable technical barriers and barriers whose perception could change over time. \\

Regarding the general personal motivations, environmental concern and novelty seeking both show medium correlations with intention (\textit{r=.34, p=.001} and \textit{r=.47, p=.001}), and large correlations with benefits (\textit{r=.69, p\textless.001} and \textit{r=.64, p\textless.001}). In the refined cluster, it becomes apparent that environmental concern is particularly strongly related with environmental benefits, and novelty seeking particularly strongly with personal benefits (\textit{r=.77, p\textless.001} and \textit{r=.63, p\textless.001}).

\subsection{Meta-analytical structural equation modeling}
\label{S:4.2}
\begin{figure}[ht]
    \centering
    \includegraphics[width=\textwidth]{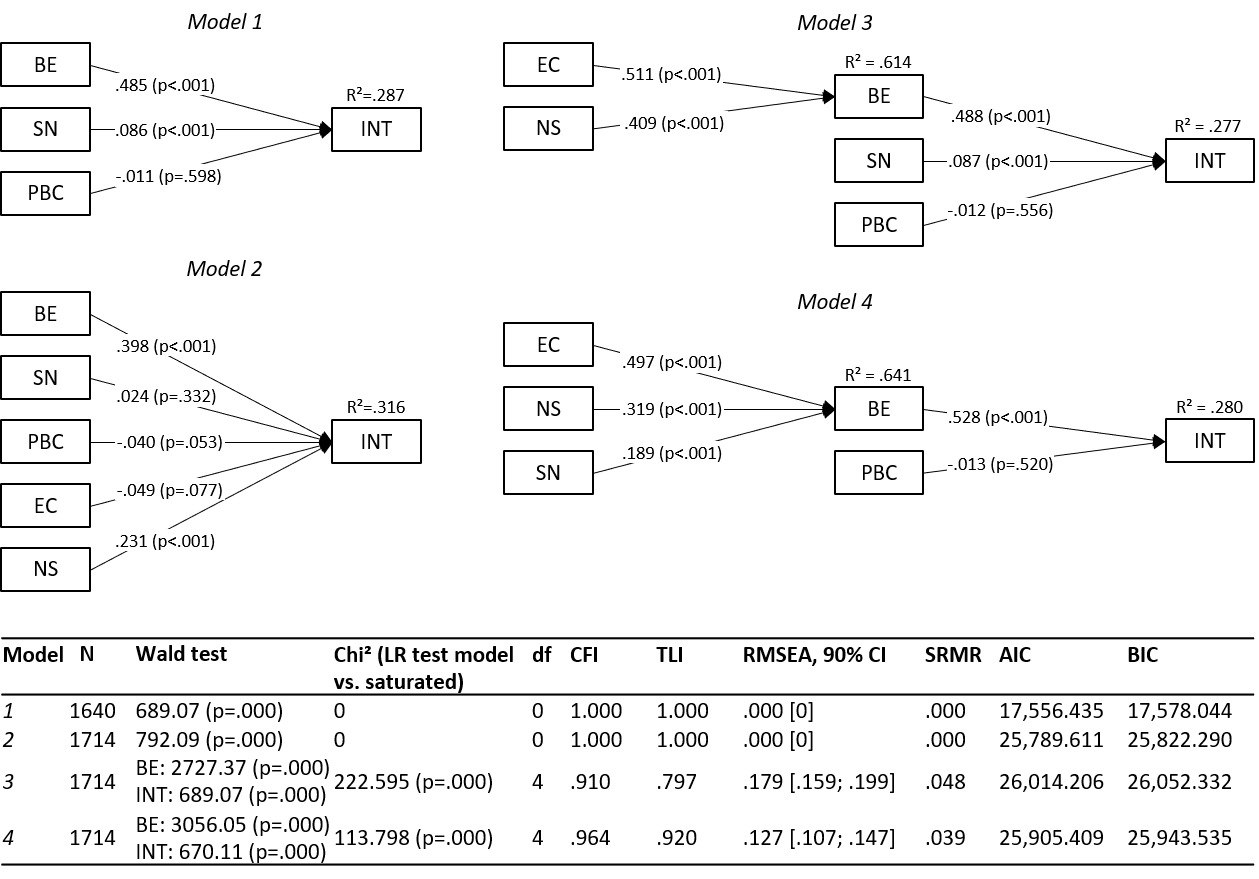}
    \caption{Results of structural equation modeling with meta-analytically summarized input data. Results are presented as standardized values.
    INT Intention; EC Environmental concern; NS Novelty Seeking; PBC Perceived behavioral control; BE Benefits; SN subjective norm}
    \label{fig:sems}
\end{figure}

Figure \ref{fig:sems} presents the final results of the MASEM analysis. For each model, path coefficients and their significance, $R²$, N, and the results for Wald test and fit indices are presented. The four models explain 27.7\% (Model 3) to 31.6\% (Model 2) of the variance in intention to adopt PV.
The Wald-Test confirms overall statistical significance for all four models (p\textless.001), meaning that the Null-Hypothesis that all path coefficients (not the intercept) are zero can be rejected.

Model 1 and Model 2 are saturated models, meaning that the number of data points and estimated parameters is equal (df=0). By definition, this leads to a perfect fit and model fit indices cannot be interpreted.
Concerning Models 3 and 4, model fit indices indicate superiority of Model 4. For Model 4, CFI and SRMR are in a good range (\textgreater.96 and \textless.1,), pointing towards good model fit (\cite{Aichholzer.2017}). Moreover, lower AIC and BIC values indicate that Model 4 has a better fit to the data compared to Model 3.\\

In Model 1, we tested whether the results of the meta-analysis fit the general model of the TPB. Whereas the standardized path coefficient for the effect of benefits on intention is highly significant with a change of one standard deviation in benefits causing a .485 change in intention, the effects of subjective norm and PBC are small, and non-significant in the case of PBC (Figure \ref{fig:sems}). Adding the two variables environmental concern and novelty seeking in Model 2 slightly improves the coefficient of determination ($R²=.316$) compared to the initial model ($R²=.287$). The reduction of the effects of benefits, subjective norm and PBC on intention represents the extent of interference between the three initial and the newly added variables, as coefficients represent direct effects adjusted for the covariances among all variables. Whereas the direct effect of environmental concern on intention is negligible, novelty seeking shows a highly significant effect of $.231$ (\textit{p\textless.001}).

In Model 3, we took into account the distinction between general personal motivations and beliefs about PV and its attributes, and the reasoning \q{that broad dispositions may influence how an innovation is perceived} \citep{Wolske.2017}. Whilst the coefficient of determination of intention was slightly reduced to $R²=.277$, a high proportion of the variance in benefits ($R²=.614$) could be explained by the two general personal motivations with both showing highly significant coefficients. Both general motivations have highly significant indirect effects on intention (EC -\textgreater BE -\textgreater Intent: $\beta$=.248 (p\textless.001); NS -\textgreater BE -\textgreater Intent: $\beta$=.198 (p\textless.001)), suggesting that these general personal motivations are indirectly related to intention via perceived benefits of PV adoption. Model 3 thus provides valuable insights into how people may come to perceive certain benefits of PV adoption based on their broader personal motivations. The results imply that with stronger environmental concern and a stronger propensity to innovation, the perception of benefits of PV systems increases. 
Whereas the direct effect of subjective norm on intention was rather small in the three previous models, Model 4 shows that subjective norms have a highly significant effect on benefits ($\beta$=.189 (p\textless.001)). The indirect effect of subjective norms on intention via perceived benefits amounts to $\beta$=.10 (p\textless.001), suggesting that subjective norms may influence intention to adopt indirectly via increasing perceived benefits of PV adoption, rather than directly. Both the explained variance of benefits and that of intention increase slightly as compared to Model 3.

\section{Discussion}
\label{S:5}
\subsection{Discussion of results and methodology}
\label{S:5.1}
The discussion first puts the results of the MASEM analysis in perspective with the literature body on PV adoption and comparable studies, beginning with the correlational analysis and moving on to the results of the structural models. Thereafter, the chosen methodology is briefly discussed.\\

Other than indicated by Alipour et al. that socio-demographic variables are the most commonly used predictors of PV adoption \cite{Alipour.2020}, they appear not to be the most frequent predictor variables among the included studies in this meta-analysis. They have been accounted for in only two (gender, \citep{Arroyo.2019, Parkins.2018}) and three studies (education, income, \citep{Arroyo.2019, Parkins.2018, AzizN.S..2017}). The correlations between intention and the socio-demographic variables gender and education are negligible, and the small correlation between income and intention is non-significant. 
Concerning gender, Scheller et al. state in a recent analysis that cost-intensive technologies like PV are typically acquired mutually in a partnership \citep{Scheller.2021}, in line with the finding that gender is not important for PV adoption. 

Oppositely, the result for income is surprising, as the influence of income on PV adoption has been shown multiple times, e.g. by \citep{Best.2019, Jan.2020, Nair.2010}. On the one hand, one could argue that intentions are governed more by abstract and general considerations, whereas behavior is affected more strongly by specific, concrete and context-dependent considerations \citep{Arts.2011}, giving rise to the idea that the costs of an innovation might not play a large role when stating an intention. This, however, is not supported by Best et al. and Jan et al. who both find effects of income on intention\cite{Best.2019, Jan.2020}. On the other hand, the impeding role of income can be overcome by incentives tailored to LMI households \cite{Wolske.2020, OShaughnessy.2020}. In this meta-analysis, the insignificance of the correlation between income and intention is related to the strong differences between the included studies, including a slight negative relation in Canada \cite{Parkins.2018}. Aziz et al. reveal a small correlation in Malaysia \cite{AzizN.S..2017}, and Arroyo and Carrete find a medium correlation in Mexico \cite{Arroyo.2019}. The sample of Aziz et al. is biased towards higher incomes \cite{AzizN.S..2017}, potentially reducing the relationship between income and intention, as restrictions in range of a variable in general attenuate effect sizes \citep{Hunter.2004}. Due to the potential role of different policy schemes in the relationship between income and intention, future studies should investigate the conditions under which income plays a bigger or smaller role in PV adoption. Moreover, potential restrictions in range should be carefully documented.

Turning to the TPB constructs benefits, PBC and subjective norm, size and significance of the correlations with intention are broadly in line with Kloeckner and Bamberg and Moeser, who meta-analytically assess the determinants of pro-environmental behavior more generally, but are overall slightly lower in our analysis \cite{Klockner.2013, Bamberg.2007}. This is surprising, especially in the case of PBC, as the adoption of PV systems is subject to a particularly large number of barriers compared to other pro-environmental behaviors such as recycling, water-saving or meat consumption (c.f. \citep{Klockner.2013}), ranging from high investment costs \citep{Jacksohn.2019} to information barriers \citep{Jager.2006}. The direct influence of PBC on PV adoption intention has, amongst others, been shown in \cite{Claudy.2013}. In our analysis, the insignificance and size of the meta-analytically pooled correlation is related to the particularly small and positive effect size of \q{perceived cost maintenance} in \citep{AzizN.S..2017}. Due to a lack of information about the operationalization of the construct, potential reasons for the effect size cannot be further investigated. Yet, looking at the refined cluster, our analysis demonstrates a highly significant effect between such barriers that technically hinder the adoption of PV systems, pointing towards a meaningful distinction between subjectively perceived barriers such as risk and effort, and technical barriers such as unsuitable homes or lack of decision power. 

General personal motivations appear to play an important role in PV adoption intention according to the meta-analysis.
Although PV systems entered the market 40 years ago, they are still in the process of diffusion and far from reaching market saturation \citep{Ford.2017}. This could be a reason why the general personal motivation novelty seeking shows such a large correlation both with intention and benefits. According to DOI, innovativeness particularly characterizes innovators and early adopters, who make up around 16\% of all adopters, and its importance decreases with increasing adoption levels \citep{Rogers.2003}. Studies in Finland \cite{Karjalainen.2019} and the US \cite{Wolske.2020} have shown that early adopters of PV systems correspond to the early adopters of DOI. Accordingly, the level of diffusion within a study's population could moderate the influence of general personal motivations on adoption intention. This could be a starting point to explain the particularly large between-study variance in the correlations between novelty seeking and intention. Whereas the studies of Rai and Beck and Wolske et al. take place in the U.S., showing a similar correlation between novelty seeking and intention (\textit{r=.32}) \citep{Rai.2015b, Wolske.2017}, the correlations of studies in Taiwan are higher (\textit{r=.75} \cite{Sun.2020} and \textit{r=.40} \cite{Chen.2014}), perhaps due to lower diffusion levels. However, without precise information about the diffusion stage of PV within the populations of the studies at the time of the survey, this is only a cautious suggestion. 

Whereas novelty seeking did not appear in Kloeckner \cite{Klockner.2013} and Bamberg and Moeser \cite{Bamberg.2007}, the New Environmental Paradigm scale, a measure related to environmental concern, is included in \cite{Klockner.2013}. The correlations with intention are of comparable size (\textit{r=.32} \citep{Klockner.2013} versus \textit{r=.34}), and support the finding that environmental motivations drive PV adoption \cite{Karytsas.2019, Palm.2018}.
Another interesting finding arises from the refined cluster which suggests that novelty seeking is related more strongly to personal benefits, and environmental concern is more strongly related to environmental benefits. This suggests that the motivation to adopt PV systems can be based on both the concern to protect the environment and the concern from being innovative; there appears to be no single explanation to PV adoption, and no single description for a typical PV adopter.

Overall, our results in the refined cluster suggest a meaningful distinction between hard technical and soft perceived barriers, and personal and environmental benefits. Yet, as the analysis was not able to provide reliable estimates due to the small number of included studies and lacking overlap thereof, more research addressing the relations between such perceived variables and general personal motivations and adoption intention is needed.\\

In analyzing the four models, we found that all models are able to explain adoption intention to a roughly similar extent, showing no large differences in the coefficient of determination (around 30\%). In all four models, the variable benefits is the strongest predictor of intention. Moreover, the suggested extensions in Model 3 and 4 provide important insights into how people may come to perceive certain benefits of PV adoption in the first place, as variance in benefits could be explained to 61\% (Model 3) and 64\% (Model 4). Both, the coefficients of determination for benefits and intention, and the model fit indices suggest superiority of Model 4 over Model 3.
Therefore, our results suggest that the general personal motivations environmental concern and novelty seeking, combined with the perceived subjective norm to adopt a PV system are useful predictors of the perception of benefits, which in turn is the best predictor for adoption intention.
Due to the small correlation between PBC and intention resulting from the meta-analysis, no effect of PBC on intention could be elicited in the models. However, as discussed above, negative effects of soft, perceived barriers such as uncertainty and risks, and of hard technical barriers related to unsuitable homes should be analyzed more precisely in the future.

Compared to the MASEM analyses on antecedents of pro-environmental behavior in Kloeckner and Bamberg and Moeser in which coefficients of determination of around 50\% for intention were reached \cite{Klockner.2013, Bamberg.2007}, the explained variance for intention in our models is lower (around 30\%). One potential reason might be that our analysis explains an investment decision, whereas the other studies address pro-environmental behavior, overwhelmingly including curtailment behaviors. Compared to curtailment behavior, investment decisions have been found more difficult to explain \cite{Kastner.2015}. Furthermore, personal norms have been identified as important explanatory variables for curtailment behavior in the past (c.f. \cite{Kastner.2015}), and are accounted for in the models of \cite{Klockner.2013, Bamberg.2007} with path coefficients of around .25. In our analysis, such predictors were not included, resulting in a more parsimonious model. 

Relating our analysis to the studies of Korcaj et al. and Engelken et al., who investigated the origins of PV adoption intention within TPB frameworks, the finding that benefits can be explained to a greater extent than intentions is supported \cite{Korcaj.2015, Engelken.2018}. 
Their results suggest that whilst the individual attitude towards PV systems measured with simple, global statements is largely explainable through perceived benefits \citep{Korcaj.2015, Engelken.2018} and general personal motivations \citep{Engelken.2018}, the role of attitudes to explain intention is limited with subjective norms having a stronger effect on intention. Indirect effects of the antecedents of attitudes are not reported. Also Abreu et al. find a stronger path coefficient for subjective norms than for attitudes \cite{Abreu.2019}. Oppositely, our comparison of the four Models suggests that subjective norms are a good predictor for perceived benefits, whilst their direct effect on intention is limited. Because the three studies do not provide correlation matrices - which is also why they could not be included in the present study - no investigation of the correlations among intention, attitudes and subjective norms, neither an assessment of the suitability of Model 4 with summary statistics of the three primary studies can be performed.\\

Methodologically, one could discuss whether the variables that have been accumulated are sufficiently similar to pool effect sizes, and whether the accumulation of studies from obviously different contexts is meaningful. In principle, meta-analyses are designed to determine true effect sizes by statistically combining empirical results of different studies. Thereby, random effects of sampling error can be eliminated, and effects of single study artifacts related to e.g. reliability and validity can be accounted for if sufficient information is provided. In case of large, non-artifactual between-study variation, moderator analyses could be performed to determine whether differences in operationalization or context lead to systematic differences in study results. Moderator analyses work by grouping the sample and comparing effect sizes and heterogeneity statistics among the complete sample and the groups \citep{Schulze.2004, Borenstein.2009, Hunter.2004}. However, in the present analysis, only 8 suitable papers could be identified by the systematic literature search, which does not allow a meaningful analysis of moderating variables (c.f. \cite{Bamberg.2007}). Nonetheless, as has been indicated in the above elaborations, a number of contextual moderators, e.g. the nature of financing options and policy schemes, and the local levels of diffusion could systematically affect effect sizes. Moreover, methodological differences related to sampling procedure and operationalization of dependent and independent variables could moderate effect sizes. Studies could be separated into such studies only including homeowners and others, they could be grouped depending on the degree of specificity of the intention, or the measurement of environmental attitudes by previous pro-environmental behavior or value-related statements and the like. An impressive example of moderator analysis is the meta-analysis of van Zomeren et al. \cite{vanZomeren.2008}. Coherent study designs of future studies and reporting standards could allow the verification of the above suggestions.

\subsection{Obstacles to a successful meta-analysis and implications for future research}
\label{S:5.2}
In this study, we attempted to make sense of the vast amount of data that is already available on residential adoption of PV. However, we had to realize that the empirical evidence is in large parts far from being similar enough to combine results meaningfully, and that a lack of comprehensive reporting of results further aggravates the problem.

From the 205 papers identified based on their title, only 24 papers were homogeneous enough in methodology (quantitative survey with residential decision-makers, no off-grid, no choice experiment) and operationalization of PV adoption (adoption intention) to be included in the meta-analysis (for documentation, see Table \ref{tab:fulltext}). Of those, only 5 report bivariate correlations, which are needed to conduct a meta-analysis of this kind. Though the identified 24 studies appear homogeneous at first sight, the approaches vary significantly, as the use and operationalization of constructs vary. Also, scale reliability statistics such as Cronbach’s alpha are not consistently reported. In addition, different requirements were defined for participant selection, potentially resulting in range restrictions and effects of a number of barriers being excluded from single studies. 
Furthermore, contextual factors have not been reported consistently, even though these can affect adoption decisions. For example, technical conditions of the respondents' house (rooftop inclination, orientation and statics, latitude, shading) \citep{Galvin.2020}, current investment costs for the system and expected revenues \citep{Jacksohn.2019}, diffusion stage \citep{Simpson.2017, Palm.2017}, and the presence of LMI-specific incentives \citep{OShaughnessy.2020, Wolske.2020} are factors that undoubtedly affect a household's decision, let alone the administrative workload that must be managed \citep{Jager.2006}.

The authors were aware of heterogeneity in the literature body prior to the analysis. However, the magnitude of these issues was surprising,m given that prior studies had already recognized this problem and proposed a more standardized approach and more detailed reports of statistical parameters \cite{Kastner.2015} to pave the way for future meta-analyses in the field of energy-relevant investment decisions.\\

In order to reduce obstacles to a successful meta-analysis, we suggest the following guidelines for future empirical studies:
\begin{enumerate}
    \item \textit{Future studies should systematically use key variables.}
Key variables could include personal and environmental benefits of adoption, hard barriers in the form of technical conditions of the respondents' house (rooftop inclination, orientation and statics, latitude, shading), soft barriers in the form of perceived risks and effort, the perceived subjective norm, including descriptive and injunctive norm, standard socio-demographic measures and the general personal motivators environmental concern and novelty seeking. For latent variables, established measurement instruments such as the NEP scale \citep{Dunlap.2008, Hawcroft.2010, Amburgey.2012} and the scales of \citep{Manning.1995} could be used. 
    \item \textit{Overarching contextual factors describing the setting of the study should be reported.}
Researchers should indicate the current investment costs for the system, expected revenues, the diffusion stage within the population, the administrative workload related to adoption, the presence of incentives, and information about whether the incentives are LMI-specific or not.
    \item \textit{Reporting standards should be followed and established.}
Reporting standards as existing in e.g. Nature Journals \citep{Editorial.2018} or as recommended by \citep{Cooper.2020} could in the case of residential PV adoption include survey year, sample size, sampling procedure, deviation of study sample from population, construct operationalization (items) and reliability (e.g. Cronbachs alpha), bivariate correlations and confidence intervals for all constructs included.
\end{enumerate}
\section{Conclusion and Implications}
\label{S:6}
To advance theory development concerning residential PV adoption, a MASEM analysis was performed using eight empirical studies on residential PV adoption intention. Four alternative models were tested.
The analysis suggests that perceived benefits are the strongest determinant of adoption intention, with the four models resulting in coefficients of determination of around 30\% for intention. Benefits can be explained with R² of 64\% by the decision-makers' environmental concern, novelty seeking and perceived subjective norm. Thus, Model 4 appears to be the most appropriate model to explain residential intention to adopt a PV system. In the model, benefits and PBC are predictors of intention, and environmental concern, novelty seeking, and subjective norms are predictors of benefits.
Furthermore, results suggest the presence of a variety of moderator variables that could be linked to technical conditions of the respondents' house, current investment costs for the system and expected revenues, diffusion stage, and the presence of LMI-specific incentives.

Following \cite{Kastner.2015}, we would like to emphasize that developing a comprehensive theoretical framework explaining specific behaviors requires the accumulation of quantitative, scientific evidence through meta-analyses. Yet, scattered and uncoordinated studies hamper such analyses, as methodological and content-wise overlap of primary studies is required. To improve the aggregability of scientific evidence in the future, we therefore recommend the use of a comprehensive list of predictors, the collection of contextual variables, and compliance with reporting standards. Compliance of such guidelines could also enhance the integration of empirical evidence in agent-based models, that are increasingly developed to simulate residential investment decisions in the energy domain (c.f. \citep{johanning2020modular, Scheller.2019}). Future research could address questions regarding the role of income under the presence of different policy schemes, and systematically investigate effects of personal and environmental benefits, and hard technical and soft perceived barriers on adoption.\\

Policy-makers should consider the heterogeneity of decision-makers and their context when designing policies to accelerate residential uptake. Overall, our results imply that measures should primarily focus on enhancing the perception of benefits. Measures aimed at increasing the financial benefits of PV installations could include options to reduce initial costs, alleviating the constraining effect of income on adoption, and shifting deployment to low-income households. Promotion strategies for the measures can focus on environmental benefits or the innovativeness of PV systems. Whereas environmental benefits drive adoption independently of diffusion stage, the innovativeness of the product seems to drive adoption particularly in low-diffusion markets, suggesting a focus on the latter aspect in regions with low diffusion stages. As the perception and importance of personal and environmental benefits depends on the characteristics of the decision-maker, promotion strategies could additionally be tailored to consumer segments representing groups of like-minded people rather than socio-demographic groups, to systematically account for the interchangeable or complementary effects of general personal motivations.\\

\section*{Acknowledgement}
Emily Schulte and Fabian Scheller receive funding from the project SUSIC (Smart Utilities and Sustainable Infrastructure Change) with the project number 100378087. The project is financed by the Saxon State government out of the State budget approved by the Saxon State Parliament. Fabian Scheller also kindly acknowledges the financial support of the European Union's Horizon 2020 research and innovation programme under the Marie Sklodowska-Curie grant agreement no. 713683 (COFUNDfellowsDTU).

\newpage

\bibliography{ModelDescription.bib}

\begin{thebibliography}{100}
\expandafter\ifx\csname url\endcsname\relax
  \def\url#1{\texttt{#1}}\fi
\expandafter\ifx\csname urlprefix\endcsname\relax\def\urlprefix{URL }\fi
\expandafter\ifx\csname href\endcsname\relax
  \def\href#1#2{#2} \def\path#1{#1}\fi

\bibitem{IEA.2020}
IEA, {World Energy Outlook 2020}, Paris, 2020.

\bibitem{BundesministeriumfurUmwelt.}
{Bundesministerium f{\"u}r Umwelt}, {Naturschutz und nukleare Sicherheit},
  www.bmu.de, {Naturbewusstsein 2019 -- Bev{\"o}lkerungsumfrage zu Natur und
  biologischer Vielfalt}.

\bibitem{Mercure.2014}
J.-F. Mercure, H.~Pollitt, U.~Chewpreecha, P.~Salas, A.~M. Foley, P.~B. Holden,
  N.~R. Edwards, {The dynamics of technology diffusion and the impacts of
  climate policy instruments in the decarbonisation of the global electricity
  sector}, {Energy Policy} 73 (2014) 686--700.
\newblock \href {https://doi.org/10.1016/j.enpol.2014.06.029}
  {\path{doi:10.1016/j.enpol.2014.06.029}}.

\bibitem{Geels.2018}
F.~W. Geels, T.~Schwanen, S.~Sorrell, K.~Jenkins, B.~K. Sovacool, {Reducing
  energy demand through low carbon innovation: A sociotechnical transitions
  perspective and thirteen research debates}, {Energy Research {\&} Social
  Science} 40 (2018) 23--35.
\newblock \href {https://doi.org/10.1016/j.erss.2017.11.003}
  {\path{doi:10.1016/j.erss.2017.11.003}}.

\bibitem{Axsen.2012}
J.~Axsen, K.~S. Kurani, {Social Influence, Consumer Behavior, and Low-Carbon
  Energy Transitions}, {Annual Review of Environment and Resources} 37~(1)
  (2012) 311--340.
\newblock \href {https://doi.org/10.1146/annurev-environ-062111-145049}
  {\path{doi:10.1146/annurev-environ-062111-145049}}.

\bibitem{Alipour.2020}
M.~Alipour, H.~Salim, R.~A. Stewart, O.~Sahin, {Predictors, taxonomy of
  predictors, and correlations of predictors with the decision behaviour of
  residential solar photovoltaics adoption: A review}, {Renewable and
  Sustainable Energy Reviews} 123 (2020) 109749.
\newblock \href {https://doi.org/10.1016/j.rser.2020.109749}
  {\path{doi:10.1016/j.rser.2020.109749}}.

\bibitem{Karakaya.2015b}
E.~Karakaya, P.~Sriwannawit, {Barriers to the adoption of photovoltaic systems:
  The state of the art}, {Renewable and Sustainable Energy Reviews} 49 (2015)
  60--66.
\newblock \href {https://doi.org/10.1016/j.rser.2015.04.058}
  {\path{doi:10.1016/j.rser.2015.04.058}}.

\bibitem{Rogers.2003}
E.~M. Rogers, {Diffusion of innovations}, 5th Edition, {Free Press}, New York,
  2003.

\bibitem{Labay.1981}
D.~G. Labay, T.~C. Kinnear, {Exploring the Consumer Decision Process in the
  Adoption of Solar Energy Systems}, {The Journal of Consumer Research} 8~(8)
  (1981) 271--278.
\newblock \href {https://doi.org/10.1086/208865} {\path{doi:10.1086/208865}}.

\bibitem{Sawyer.1982}
S.~W. Sawyer, {Leaders in change: solar energy owners and the implications for
  future adoption rates}, {Technological Forecasting and Social Change} 21
  (1982).
\newblock \href {https://doi.org/10.1016/0040-1625(82)90050-6}
  {\path{doi:10.1016/0040-1625(82)90050-6}}.

\bibitem{Faiers.2006}
A.~Faiers, C.~Neame, {Consumer attitudes towards domestic solar power systems},
  {Energy Policy} 34~(14) (2006) 1797--1806.
\newblock \href {https://doi.org/10.1016/j.enpol.2005.01.001}
  {\path{doi:10.1016/j.enpol.2005.01.001}}.

\bibitem{Stern.2000}
P.~C. Stern, {New Environmental Theories: Toward a Coherent Theory of
  Environmentally Significant Behavior}, {Journal of Social Issues} 56~(3)
  (2000) 407--424.
\newblock \href {https://doi.org/10.1111/0022-4537.00175}
  {\path{doi:10.1111/0022-4537.00175}}.

\bibitem{Wolske.2017}
K.~S. Wolske, P.~C. Stern, T.~Dietz, {Explaining interest in adopting
  residential solar photovoltaic systems in the United States: Toward an
  integration of behavioral theories}, {Energy Research {\&} Social Science} 25
  (2017) 134--151.
\newblock \href {https://doi.org/10.1016/j.erss.2016.12.023}
  {\path{doi:10.1016/j.erss.2016.12.023}}.

\bibitem{Ajzen.1991}
I.~Ajzen, {The Theory of Planned Behavior}, {Organizational Behavior and Human
  Decision Processes} 50 (1991) 179--211.

\bibitem{Irfan.2020}
M.~Irfan, Z.-Y. Zhao, H.~Li, A.~Rehman, {The influence of consumers' intention
  factors on willingness to pay for renewable energy: a structural equation
  modeling approach}, {Environmental science and pollution research
  international} 27~(17) (2020) 21747--21761.
\newblock \href {https://doi.org/10.1007/s11356-020-08592-9}
  {\path{doi:10.1007/s11356-020-08592-9}}.

\bibitem{Abreu.2019}
J.~Abreu, N.~Wingartz, N.~Hardy, {New trends in solar: A comparative study
  assessing the attitudes towards the adoption of rooftop PV}, {Energy Policy}
  128 (2019) 347--363.
\newblock \href {https://doi.org/10.1016/j.enpol.2018.12.038}
  {\path{doi:10.1016/j.enpol.2018.12.038}}.

\bibitem{Sun.2020}
P.-C. Sun, H.-M. Wang, H.-L. Huang, C.-W. Ho, {Consumer attitude and purchase
  intention toward rooftop photovoltaic installation: The roles of personal
  trait, psychological benefit, and government incentives}, {Energy {\&}
  Environment} 31~(1) (2020) 21--39.
\newblock \href {https://doi.org/10.1177/0958305X17754278}
  {\path{doi:10.1177/0958305X17754278}}.

\bibitem{Si.2019}
H.~Si, J.-G. Shi, D.~Tang, S.~Wen, W.~Miao, K.~Duan, {Application of the Theory
  of Planned Behavior in Environmental Science: A Comprehensive Bibliometric
  Analysis}, {International journal of environmental research and public
  health} 16~(15) (2019).
\newblock \href {https://doi.org/10.3390/ijerph16152788}
  {\path{doi:10.3390/ijerph16152788}}.

\bibitem{Tan.2017}
C.-S. Tan, H.-Y. Ooi, Y.-N. Goh, {A moral extension of the theory of planned
  behavior to predict consumers' purchase intention for energy-efficient
  household appliances in Malaysia}, {Energy Policy} 107 (2017) 459--471.
\newblock \href {https://doi.org/10.1016/j.enpol.2017.05.027}
  {\path{doi:10.1016/j.enpol.2017.05.027}}.

\bibitem{Korcaj.2015}
L.~Korcaj, U.~J. Hahnel, H.~Spada, {Intentions to adopt photovoltaic systems
  depend on homeowners' expected personal gains and behavior of peers},
  {Renewable Energy} 75 (2015) 407--415.
\newblock \href {https://doi.org/10.1016/j.renene.2014.10.007}
  {\path{doi:10.1016/j.renene.2014.10.007}}.

\bibitem{Litvine.2011}
D.~Litvine, R.~W{\"u}stenhagen, {Helping ``light green'' consumers walk the
  talk: Results of a behavioural intervention survey in the Swiss electricity
  market}, {Ecological Economics} 70~(3) (2011) 462--474.
\newblock \href {https://doi.org/10.1016/j.ecolecon.2010.10.005}
  {\path{doi:10.1016/j.ecolecon.2010.10.005}}.

\bibitem{Ozaki.2011}
R.~Ozaki, {Adopting Sustainable Innovation: What Makes Consumers Sign up to
  Green Electricity?}, {Business Strategy and the Environment} 20~(1) (2011)
  1--17.
\newblock \href {https://doi.org/10.1002/bse.650} {\path{doi:10.1002/bse.650}}.

\bibitem{Gerpott.2010}
T.~J. Gerpott, I.~Mahmudova, {Determinants of green electricity adoption among
  residential customers in Germany}, {International Journal of Consumer
  Studies} 34~(4) (2010) 464--473.
\newblock \href {https://doi.org/10.1111/j.1470-6431.2010.00896.x}
  {\path{doi:10.1111/j.1470-6431.2010.00896.x}}.

\bibitem{Ek.2008}
K.~Ek, P.~S{\"o}derholm, {Norms and economic motivation in the Swedish green
  electricity market}, {Ecological Economics} 68~(1-2) (2008) 169--182.
\newblock \href {https://doi.org/10.1016/j.ecolecon.2008.02.013}
  {\path{doi:10.1016/j.ecolecon.2008.02.013}}.

\bibitem{Bollinger.2012}
B.~Bollinger, K.~Gillingham, {Peer Effects in the Diffusion of Solar
  Photovoltaic Panels}, {Marketing Science} 31~(6) (2012) 900--912.
\newblock \href {https://doi.org/10.1287/mksc.1120.0727}
  {\path{doi:10.1287/mksc.1120.0727}}.

\bibitem{Rai.2013}
V.~Rai, B.~Sigrin, {Diffusion of environmentally-friendly energy technologies:
  buy versus lease differences in residential PV markets}, {Environmental
  Research Letters} 8~(1) (2013) 014022.
\newblock \href {https://doi.org/10.1088/1748-9326/8/1/014022}
  {\path{doi:10.1088/1748-9326/8/1/014022}}.

\bibitem{Palm.2017}
A.~Palm, {Peer effects in residential solar photovoltaics adoption---A mixed
  methods study of Swedish users}, {Energy Research {\&} Social Science} 26
  (2017) 1--10.
\newblock \href {https://doi.org/10.1016/j.erss.2017.01.008}
  {\path{doi:10.1016/j.erss.2017.01.008}}.

\bibitem{Rode.2020}
J.~Rode, S.~M{\"u}ller,
  \href{https://EconPapers.repec.org/RePEc:dar:wpaper:119280}{{I Spot, I Adopt!
  A Discrete Choice Analysis on Peer Effects in Solar Photovoltaic System
  Adoption of Households}} (2019).
\newblock \href {https://doi.org/10.2139/ssrn.3469548}
  {\path{doi:10.2139/ssrn.3469548}}.
\newline\urlprefix\url{https://EconPapers.repec.org/RePEc:dar:wpaper:119280}

\bibitem{scheller2021active}
F.~Scheller, S.~Graupner, J.~Edwards, J.~Weinand, T.~Bruckner, Active peer
  effects in residential photovoltaic adoption: evidence on impact drivers
  among potential and current adopters in germany (2021).
\newblock \href {http://arxiv.org/abs/2105.00796} {\path{arXiv:2105.00796}}.

\bibitem{Dunlap.2008}
R.~E. Dunlap, {The New Environmental Paradigm Scale: From Marginality to
  Worldwide Use}, {The Journal of Environmental Education} 40~(1) (2008) 3--18.
\newblock \href {https://doi.org/10.3200/JOEE.40.1.3-18}
  {\path{doi:10.3200/JOEE.40.1.3-18}}.

\bibitem{Hawcroft.2010}
L.~J. Hawcroft, T.~L. Milfont, {The use (and abuse) of the new environmental
  paradigm scale over the last 30 years: A meta-analysis}, {Journal of
  Environmental Psychology} 30~(2) (2010) 143--158.
\newblock \href {https://doi.org/10.1016/j.jenvp.2009.10.003}
  {\path{doi:10.1016/j.jenvp.2009.10.003}}.

\bibitem{Rai.2015b}
V.~Rai, A.~L. Beck, {Public perceptions and information gaps in solar energy in
  Texas}, {Environmental Research Letters} 10~(7) (2015) 074011.
\newblock \href {https://doi.org/10.1088/1748-9326/10/7/074011}
  {\path{doi:10.1088/1748-9326/10/7/074011}}.

\bibitem{Hunter.2004}
J.~E. Hunter, F.~L. Schmidt, {Methods of meta-analysis: Correcting error and
  bias in research findings}, 2nd Edition, Sage, Thousand Oaks Calif., 2004.

\bibitem{Parkins.2018}
J.~R. Parkins, C.~Rollins, S.~Anders, L.~Comeau, {Predicting intention to adopt
  solar technology in Canada: The role of knowledge, public engagement, and
  visibility}, {Energy Policy} 114 (2018) 114--122.
\newblock \href {https://doi.org/10.1016/j.enpol.2017.11.050}
  {\path{doi:10.1016/j.enpol.2017.11.050}}.

\bibitem{Arroyo.2019}
P.~Arroyo, L.~Carrete, {Motivational drivers for the adoption of green energy},
  {Management Research Review} 42~(5) (2019) 542--567.
\newblock \href {https://doi.org/10.1108/MRR-02-2018-0070}
  {\path{doi:10.1108/MRR-02-2018-0070}}.

\bibitem{Galvin.2020}
R.~Galvin, {I'll follow the sun: Geo-sociotechnical constraints on prosumer
  households in Germany}, {Energy Research {\&} Social Science} 65 (2020)
  101455.
\newblock \href {https://doi.org/10.1016/j.erss.2020.101455}
  {\path{doi:10.1016/j.erss.2020.101455}}.

\bibitem{Cherry.2020}
T.~L. Cherry, H.~S{\ae}le, {Residential Photovoltaic Systems in Norway:
  Household Knowledge, Preferences and Willingness to Pay}, {Challenges in
  Sustainability} 8~(1) (2020).
\newblock \href {https://doi.org/10.12924/cis2020.08010001}
  {\path{doi:10.12924/cis2020.08010001}}.

\bibitem{Bashiri.2018}
A.~Bashiri, S.~H. Alizadeh, {The analysis of demographics, environmental and
  knowledge factors affecting prospective residential PV system adoption: A
  study in Tehran}, {Renewable and Sustainable Energy Reviews} 81 (2018)
  3131--3139.
\newblock \href {https://doi.org/10.1016/j.rser.2017.08.093}
  {\path{doi:10.1016/j.rser.2017.08.093}}.

\bibitem{Mukai.2011}
T.~Mukai, S.~Kawamoto, Y.~Ueda, M.~Saijo, N.~Abe, {Residential PV system users'
  perception of profitability, reliability, and failure risk: An empirical
  survey in a local Japanese municipality}, {Energy Policy} 39~(9) (2011)
  5440--5448.
\newblock \href {https://doi.org/10.1016/j.enpol.2011.05.019}
  {\path{doi:10.1016/j.enpol.2011.05.019}}.

\bibitem{Scheller.2020}
F.~Scheller, I.~Doser, D.~Sloot, R.~McKenna, T.~Bruckner, {Exploring the Role
  of Stakeholder Dynamics in Residential Photovoltaic Adoption Decisions: A
  Synthesis of the Literature}, Energies 13~(23) (2020) 6283.
\newblock \href {https://doi.org/doi.org/10.3390/en13236283}
  {\path{doi:doi.org/10.3390/en13236283}}.

\bibitem{Schulze.2004}
R.~Schulze, {Meta-analysis: A comparison of approaches}, {Hogrefe {\&} Huber},
  Toronto, 2004.

\bibitem{Borenstein.2009}
M.~Borenstein, {Introduction to meta-analysis}, {John Wiley {\&} Sons},
  Chichester U.K., 2009.

\bibitem{Bergh.2016}
D.~D. Bergh, H.~Aguinis, C.~Heavey, D.~J. Ketchen, B.~K. Boyd, P.~Su, C.~L.~L.
  Lau, H.~Joo, {Using meta-analytic structural equation modeling to advance
  strategic management research: Guidelines and an empirical illustration via
  the strategic leadership-performance relationship}, {Strategic Management
  Journal} 37~(3) (2016) 477--497.
\newblock \href {https://doi.org/10.1002/smj.2338}
  {\path{doi:10.1002/smj.2338}}.

\bibitem{Klockner.2013}
C.~A. Kl{\"o}ckner, {A comprehensive model of the psychology of environmental
  behaviour---A meta-analysis}, {Global Environmental Change} 23~(5) (2013)
  1028--1038.
\newblock \href {https://doi.org/10.1016/j.gloenvcha.2013.05.014}
  {\path{doi:10.1016/j.gloenvcha.2013.05.014}}.

\bibitem{Bamberg.2007}
S.~Bamberg, G.~M{\"o}ser, {Twenty years after Hines, Hungerford, and Tomera: A
  new meta-analysis of psycho-social determinants of pro-environmental
  behaviour}, {Journal of Environmental Psychology} 27~(1) (2007) 14--25.
\newblock \href {https://doi.org/10.1016/j.jenvp.2006.12.002}
  {\path{doi:10.1016/j.jenvp.2006.12.002}}.

\bibitem{Kastner.2015}
I.~Kastner, P.~C. Stern, {Examining the decision-making processes behind
  household energy investments: A review}, {Energy Research {\&} Social
  Science} 10 (2015) 72--89.
\newblock \href {https://doi.org/10.1016/j.erss.2015.07.008}
  {\path{doi:10.1016/j.erss.2015.07.008}}.

\bibitem{Bamberg.2003}
S.~Bamberg, {How does environmental concern influence specific environmentally
  related behaviors? A new answer to an old question}, {Journal of
  Environmental Psychology} 23~(1) (2003) 21--32.
\newblock \href {https://doi.org/10.1016/S0272-4944(02)00078-6}
  {\path{doi:10.1016/S0272-4944(02)00078-6}}.

\bibitem{Stamm.1995}
H.~Stamm, T.~M. Schwarb, {Metaanalyse. Eine Einf{\"u}hrung}, {German Journal of
  Human Resource Management} (1995).

\bibitem{Palm.2018}
J.~Palm, {Household installation of solar panels -- Motives and barriers in a
  10-year perspective}, {Energy Policy} 113 (2018) 1--8.
\newblock \href {https://doi.org/10.1016/j.enpol.2017.10.047}
  {\path{doi:10.1016/j.enpol.2017.10.047}}.

\bibitem{Koch.2018}
J.~Koch, O.~Christ, {Household participation in an urban photovoltaic project
  in Switzerland: Exploration of triggers and barriers}, {Sustainable Cities
  and Society} 37 (2018) 420--426.
\newblock \href {https://doi.org/10.1016/j.scs.2017.10.028}
  {\path{doi:10.1016/j.scs.2017.10.028}}.

\bibitem{Karjalainen.2019}
S.~Karjalainen, H.~Ahvenniemi, {Pleasure is the profit - The adoption of solar
  PV systems by households in Finland}, {Renewable Energy} 133 (2019) 44--52.
\newblock \href {https://doi.org/10.1016/j.renene.2018.10.011}
  {\path{doi:10.1016/j.renene.2018.10.011}}.

\bibitem{Scheller.2021}
F.~Scheller, I.~Doser, E.~Schulte, S.~Johanning, R.~McKenna, T.~Bruckner,
  {Stakeholder dynamics in residential solar energy adoption: findings from
  focus group discussions in Germany}, {Energy Research \& Social Science} 76
  (2021) 102065.
\newblock \href {https://doi.org/10.1016/j.erss.2021.102065}
  {\path{doi:10.1016/j.erss.2021.102065}}.

\bibitem{Engelken.2018}
M.~Engelken, B.~R{\"o}mer, M.~Drescher, I.~Welpe, {Why homeowners strive for
  energy self-supply and how policy makers can influence them}, {Energy Policy}
  117 (2018) 423--433.
\newblock \href {https://doi.org/10.1016/j.enpol.2018.02.026}
  {\path{doi:10.1016/j.enpol.2018.02.026}}.

\bibitem{Zander.2019}
K.~K. Zander, G.~Simpson, S.~Mathew, R.~Nepal, S.~T. Garnett, {Preferences for
  and potential impacts of financial incentives to install residential rooftop
  solar photovoltaic systems in Australia}, {Journal of Cleaner Production} 230
  (2019) 328--338.
\newblock \href {https://doi.org/10.1016/j.jclepro.2019.05.133}
  {\path{doi:10.1016/j.jclepro.2019.05.133}}.

\bibitem{Galassi.2014}
V.~Galassi, R.~Madlener, {Identifying Business Models for Photovoltaic Systems
  with Storage in the Italian Market: A Discrete Choice Experiment} (2014).

\bibitem{Scarpa.2010}
R.~Scarpa, K.~Willis, {Willingness-to-pay for renewable energy: Primary and
  discretionary choice of British households' for micro-generation
  technologies}, {Energy Economics} 32~(1) (2010) 129--136.
\newblock \href {https://doi.org/10.1016/j.eneco.2009.06.004}
  {\path{doi:10.1016/j.eneco.2009.06.004}}.

\bibitem{Willis.2011}
K.~Willis, R.~Scarpa, R.~Gilroy, N.~Hamza, {Renewable energy adoption in an
  ageing population: Heterogeneity in preferences for micro-generation
  technology adoption}, {Energy Policy} 39~(10) (2011) 6021--6029.
\newblock \href {https://doi.org/10.1016/j.enpol.2011.06.066}
  {\path{doi:10.1016/j.enpol.2011.06.066}}.

\bibitem{Best.2019}
R.~Best, P.~J. Burke, S.~Nishitateno, {Understanding the determinants of
  rooftop solar installation: evidence from household surveys in Australia},
  {Australian Journal of Agricultural and Resource Economics} 63~(4) (2019)
  922--939.
\newblock \href {https://doi.org/10.1111/1467-8489.12319}
  {\path{doi:10.1111/1467-8489.12319}}.

\bibitem{Nair.2010}
G.~Nair, L.~Gustavsson, K.~Mahapatra, {Factors influencing energy efficiency
  investments in existing Swedish residential buildings}, {Energy Policy}
  38~(6) (2010) 2956--2963.
\newblock \href {https://doi.org/10.1016/j.enpol.2010.01.033}
  {\path{doi:10.1016/j.enpol.2010.01.033}}.

\bibitem{Dharshing.2017}
S.~Dharshing, {Household dynamics of technology adoption: A spatial econometric
  analysis of residential solar photovoltaic (PV) systems in Germany}, {Energy
  Research {\&} Social Science} 23 (2017) 113--124.
\newblock \href {https://doi.org/10.1016/j.erss.2016.10.012}
  {\path{doi:10.1016/j.erss.2016.10.012}}.

\bibitem{Groote.2016}
O.~de~Groote, G.~Pepermans, F.~Verboven, {Heterogeneity in the adoption of
  photovoltaic systems in Flanders}, {Energy Economics} 59 (2016) 45--57.
\newblock \href {https://doi.org/10.1016/j.eneco.2016.07.008}
  {\path{doi:10.1016/j.eneco.2016.07.008}}.

\bibitem{BaltaOzkan.2015}
N.~Balta-Ozkan, J.~Yildirim, P.~M. Connor, {Regional distribution of
  photovoltaic deployment in the UK and its determinants: A spatial econometric
  approach}, {Energy Economics} 51 (2015) 417--429.
\newblock \href {https://doi.org/10.1016/j.eneco.2015.08.003}
  {\path{doi:10.1016/j.eneco.2015.08.003}}.

\bibitem{Setyawati.2020}
D.~Setyawati, {Analysis of perceptions towards the rooftop photovoltaic solar
  system policy in Indonesia}, {Energy Policy} 144 (2020) 111569.
\newblock \href {https://doi.org/10.1016/j.enpol.2020.111569}
  {\path{doi:10.1016/j.enpol.2020.111569}}.

\bibitem{Leenheer.2011}
J.~Leenheer, M.~de~Nooij, O.~Sheikh, {Own power: Motives of having electricity
  without the energy company}, {Energy Policy} 39~(9) (2011) 5621--5629.
\newblock \href {https://doi.org/10.1016/j.enpol.2011.04.037}
  {\path{doi:10.1016/j.enpol.2011.04.037}}.

\bibitem{Aggarwal.2019}
A.~K. Aggarwal, A.~A. Syed, S.~Garg, {Diffusion of residential RT solar -- is
  lack of funds the real issue?}, {International Journal of Energy Sector
  Management} 14~(2) (2019) 316--334.
\newblock \href {https://doi.org/10.1108/IJESM-02-2019-0004}
  {\path{doi:10.1108/IJESM-02-2019-0004}}.

\bibitem{Abdullah.2017}
Abdullah, D.~Zhou, T.~Shah, K.~Jebran, S.~Ali, A.~Ali, {Acceptance and
  willingness to pay for solar home system: Survey evidence from northern area
  of Pakistan}, {Energy Reports} 3 (2017) 54--60.
\newblock \href {https://doi.org/10.1016/j.egyr.2017.03.002}
  {\path{doi:10.1016/j.egyr.2017.03.002}}.

\bibitem{Simpson.2015}
G.~Simpson, J.~Clifton, {The emperor and the cowboys: The role of government
  policy and industry in the adoption of domestic solar microgeneration
  systems}, {Energy Policy} 81 (2015) 141--151.
\newblock \href {https://doi.org/10.1016/j.enpol.2015.02.028}
  {\path{doi:10.1016/j.enpol.2015.02.028}}.

\bibitem{Schelly.2014}
C.~Schelly, {Residential solar electricity adoption: What motivates, and what
  matters? A case study of early adopters}, {Energy Research {\&} Social
  Science} 2 (2014) 183--191.
\newblock \href {https://doi.org/10.1016/j.erss.2014.01.001}
  {\path{doi:10.1016/j.erss.2014.01.001}}.

\bibitem{Rai.2013b}
V.~Rai, S.~A. Robinson, {Effective information channels for reducing costs of
  environmentally- friendly technologies: evidence from residential PV
  markets}, {Environmental Research Letters} 8~(1) (2013) 014044.
\newblock \href {https://doi.org/10.1088/1748-9326/8/1/014044}
  {\path{doi:10.1088/1748-9326/8/1/014044}}.

\bibitem{Palm.2016}
A.~Palm, {Local factors driving the diffusion of solar photovoltaics in Sweden:
  A case study of five municipalities in an early market}, {Energy Research
  {\&} Social Science} 14 (2016) 1--12.
\newblock \href {https://doi.org/10.1016/j.erss.2015.12.027}
  {\path{doi:10.1016/j.erss.2015.12.027}}.

\bibitem{Petrovich.2019}
B.~Petrovich, S.~L. Hille, R.~W{\"u}stenhagen, {Beauty and the budget: A
  segmentation of residential solar adopters}, {Ecological Economics} 164
  (2019) 106353.
\newblock \href {https://doi.org/10.1016/j.ecolecon.2019.106353}
  {\path{doi:10.1016/j.ecolecon.2019.106353}}.

\bibitem{Curtius.2018}
H.~C. Curtius, {Diffusion of Solar Photovoltaics: Consumer Preferences, Peer
  Effects and Implications for Clean Energy Marketing}, {Dissertation},
  {University of St. Gallen}, St. Gallen (2018).

\bibitem{Jan.2020}
I.~Jan, W.~Ullah, M.~Ashfaq, {Social acceptability of solar photovoltaic system
  in Pakistan: Key determinants and policy implications}, {Journal of Cleaner
  Production} 274 (2020) 123140.
\newblock \href {https://doi.org/10.1016/j.jclepro.2020.123140}
  {\path{doi:10.1016/j.jclepro.2020.123140}}.

\bibitem{Parsad.2020}
C.~Parsad, S.~Mittal, R.~Krishnankutty, {A study on the factors affecting
  household solar adoption in Kerala, India}, {International Journal of
  Productivity and Performance Management} 69~(8) (2020) 1695--1720.
\newblock \href {https://doi.org/10.1108/IJPPM-11-2019-0544}
  {\path{doi:10.1108/IJPPM-11-2019-0544}}.

\bibitem{Kapoor.2020}
K.~K. Kapoor, Y.~K. Dwivedi, {Sustainable consumption from the consumer's
  perspective: Antecedents of solar innovation adoption}, {Resources,
  Conservation and Recycling} 152 (2020) 104501.
\newblock \href {https://doi.org/10.1016/j.resconrec.2019.104501}
  {\path{doi:10.1016/j.resconrec.2019.104501}}.

\bibitem{Wolske.2020}
K.~S. Wolske, {More alike than different: Profiles of high-income and
  low-income rooftop solar adopters in the United States}, {Energy Research
  {\&} Social Science} 63 (2020) 101399.
\newblock \href {https://doi.org/10.1016/j.erss.2019.101399}
  {\path{doi:10.1016/j.erss.2019.101399}}.

\bibitem{Alrashoud.2019}
K.~Alrashoud, K.~Tokimatsu, {Factors Influencing Social Perception of
  Residential Solar Photovoltaic Systems in Saudi Arabia}, {Sustainability}
  11~(19) (2019) 5259.
\newblock \href {https://doi.org/10.3390/su11195259}
  {\path{doi:10.3390/su11195259}}.

\bibitem{Arts.2011}
J.~W. Arts, R.~T. Frambach, T.~H. Bijmolt, {Generalizations on consumer
  innovation adoption: A meta-analysis on drivers of intention and behavior},
  {International Journal of Research in Marketing} 28~(2) (2011) 134--144.
\newblock \href {https://doi.org/10.1016/j.ijresmar.2010.11.002}
  {\path{doi:10.1016/j.ijresmar.2010.11.002}}.

\bibitem{Niamir.2020}
L.~Niamir, O.~Ivanova, T.~Filatova, A.~Voinov, H.~Bressers, {Demand-side
  solutions for climate mitigation: Bottom-up drivers of household energy
  behavior change in the Netherlands and Spain}, {Energy Research {\&} Social
  Science} 62 (2020) 101356.
\newblock \href {https://doi.org/10.1016/j.erss.2019.101356}
  {\path{doi:10.1016/j.erss.2019.101356}}.

\bibitem{Simpson.2017}
G.~Simpson, J.~Clifton, {Testing Diffusion of Innovations Theory with data:
  Financial incentives, early adopters, and distributed solar energy in
  Australia}, {Energy Research {\&} Social Science} 29 (2017) 12--22.
\newblock \href {https://doi.org/10.1016/j.erss.2017.04.005}
  {\path{doi:10.1016/j.erss.2017.04.005}}.

\bibitem{Karytsas.2019}
S.~Karytsas, I.~Vardopoulos, E.~Theodoropoulou, {Factors Affecting Sustainable
  Market Acceptance of Residential Microgeneration Technologies. A Two Time
  Period Comparative Analysis}, {Energies} 12~(17) (2019) 3298.
\newblock \href {https://doi.org/10.3390/en12173298}
  {\path{doi:10.3390/en12173298}}.

\bibitem{Jacksohn.2019}
A.~Jacksohn, P.~Gr{\"o}sche, K.~Rehdanz, C.~Schr{\"o}der, {Drivers of renewable
  technology adoption in the household sector}, {Energy Economics} 81 (2019)
  216--226.
\newblock \href {https://doi.org/10.1016/j.eneco.2019.04.001}
  {\path{doi:10.1016/j.eneco.2019.04.001}}.

\bibitem{OShaughnessy.2020}
E.~O'Shaughnessy, G.~Barbose, R.~Wiser, S.~Forrester, N.~Darghouth, {The impact
  of policies and business models on income equity in rooftop solar adoption},
  {Nature Energy} (2020).
\newblock \href {https://doi.org/10.1038/s41560-020-00724-2}
  {\path{doi:10.1038/s41560-020-00724-2}}.

\bibitem{Reames.2020}
T.~G. Reames, {Distributional disparities in residential rooftop solar
  potential and penetration in four cities in the United States}, {Energy
  Research {\&} Social Science} 69 (2020) 101612.
\newblock \href {https://doi.org/10.1016/j.erss.2020.101612}
  {\path{doi:10.1016/j.erss.2020.101612}}.

\bibitem{Chapman.2019}
A.~Chapman, S.~Okushima, {Engendering an inclusive low-carbon energy transition
  in Japan: Considering the perspectives and awareness of the energy poor},
  {Energy Policy} 135 (2019) 111017.
\newblock \href {https://doi.org/10.1016/j.enpol.2019.111017}
  {\path{doi:10.1016/j.enpol.2019.111017}}.

\bibitem{Zander.2020}
K.~K. Zander, {Unrealised opportunities for residential solar panels in
  Australia}, {Energy Policy} 142 (2020) 111508.
\newblock \href {https://doi.org/10.1016/j.enpol.2020.111508}
  {\path{doi:10.1016/j.enpol.2020.111508}}.

\bibitem{Anugwom.2020}
E.~E. Anugwom, K.~N. Anugwom, O.~I. Eya, {Clean energy transition in a
  developing society: Perspectives on the socioeconomic determinants of Solar
  Home Systems adoption among urban households in southeastern Nigeria},
  {African Journal of Science, Technology, Innovation and Development} 12~(5)
  (2020) 653--661.
\newblock \href {https://doi.org/10.1080/20421338.2020.1764176}
  {\path{doi:10.1080/20421338.2020.1764176}}.

\bibitem{Alsabbagh.2019}
M.~Alsabbagh, {Public perception toward residential solar panels in Bahrain},
  {Energy Reports} 5 (2019) 253--261.
\newblock \href {https://doi.org/10.1016/j.egyr.2019.02.002}
  {\path{doi:10.1016/j.egyr.2019.02.002}}.

\bibitem{Claudy.2013}
M.~C. Claudy, M.~Peterson, A.~O'Driscoll, {Understanding the Attitude-Behavior
  Gap for Renewable Energy Systems Using Behavioral Reasoning Theory}, {Journal
  of Macromarketing} 33~(4) (2013) 273--287.
\newblock \href {https://doi.org/10.1177/0276146713481605}
  {\path{doi:10.1177/0276146713481605}}.

\bibitem{Mundaca.2020}
L.~Mundaca, M.~Samahita, {What drives home solar PV uptake? Subsidies, peer
  effects and visibility in Sweden}, {Energy Research {\&} Social Science} 60
  (2020) 101319.
\newblock \href {https://doi.org/10.1016/j.erss.2019.101319}
  {\path{doi:10.1016/j.erss.2019.101319}}.

\bibitem{Alrashoud.2020}
K.~Alrashoud, K.~Tokimatsu, {An exploratory study of the public's views on
  residential solar photovoltaic systems in oil-rich Saudi Arabia},
  {Environmental Development} 35 (2020) 100526.
\newblock \href {https://doi.org/10.1016/j.envdev.2020.100526}
  {\path{doi:10.1016/j.envdev.2020.100526}}.

\bibitem{Schelly.2020}
C.~Schelly, J.~C. Letzelter, {Examining the Key Drivers of Residential Solar
  Adoption in Upstate New York}, {Sustainability} 12~(6) (2020) 2552.
\newblock \href {https://doi.org/10.3390/su12062552}
  {\path{doi:10.3390/su12062552}}.

\bibitem{AzizN.S..2017}
N.~S. Aziz, N.~A. Wahid, M.~A. Sallam, S.~K. Ariffin, {Factors Influencing
  Malaysian Consumers' Intention to Purchase Green Energy: The Case of Solar
  Panel}, {Global Business and Management Research: An International Journal}
  9~(4s) (2017) 328--346.

\bibitem{vanZomeren.2008}
M.~{van Zomeren}, T.~Postmes, R.~Spears, {Toward an integrative social identity
  model of collective action: a quantitative research synthesis of three
  socio-psychological perspectives}, {Psychological bulletin} 134~(4) (2008)
  504--535.
\newblock \href {https://doi.org/10.1037/0033-2909.134.4.504}
  {\path{doi:10.1037/0033-2909.134.4.504}}.

\bibitem{Brocke.2009}
J.~von Brocke, A.~Simons, B.~Niehaves, K.~Reimer,
  \href{http://aisel.aisnet.org/ecis2009/161}{{Reconstructing the giant: On the
  importance of rigour in documenting the literature search process}} (2009).
\newline\urlprefix\url{http://aisel.aisnet.org/ecis2009/161}

\bibitem{Stanley.2013}
T.~D. Stanley, H.~Doucouliagos, M.~Giles, J.~H. Heckemeyer, R.~J. Johnston,
  P.~Laroche, J.~P. Nelson, M.~Paldam, J.~Poot, G.~Pugh, R.~S. Rosenberger,
  K.~Rost, {META-ANALYSIS OF ECONOMICS RESEARCH REPORTING GUIDELINES}, {Journal
  of Economic Surveys} 27~(2) (2013) 390--394.
\newblock \href {https://doi.org/10.1111/joes.12008}
  {\path{doi:10.1111/joes.12008}}.

\bibitem{Aytug.2012}
Z.~G. Aytug, H.~R. Rothstein, W.~Zhou, M.~C. Kern, {Revealed or Concealed?
  Transparency of Procedures, Decisions, and Judgment Calls in Meta-Analyses},
  {Organizational Research Methods} 15~(1) (2012) 103--133.
\newblock \href {https://doi.org/10.1177/1094428111403495}
  {\path{doi:10.1177/1094428111403495}}.

\bibitem{Shamseer.2015}
L.~Shamseer, D.~Moher, M.~Clarke, D.~Ghersi, A.~Liberati, M.~Petticrew,
  P.~Shekelle, L.~A. Stewart, {Preferred reporting items for systematic review
  and meta-analysis protocols (PRISMA-P) 2015: elaboration and explanation},
  {BMJ (Clinical research ed.)} 350 (2015) g7647.
\newblock \href {https://doi.org/10.1136/bmj.g7647}
  {\path{doi:10.1136/bmj.g7647}}.

\bibitem{Lay.2013}
J.~Lay, J.~Ondraczek, J.~Stoever, {Renewables in the energy transition:
  Evidence on solar home systems and lighting fuel choice in Kenya}, {Energy
  Economics} 40 (2013) 350--359.
\newblock \href {https://doi.org/10.1016/j.eneco.2013.07.024}
  {\path{doi:10.1016/j.eneco.2013.07.024}}.

\bibitem{Smith.2014}
M.~G. Smith, J.~Urpelainen, {Early Adopters of Solar Panels in Developing
  Countries: Evidence from Tanzania}, {Review of Policy Research} 31~(1) (2014)
  17--37.
\newblock \href {https://doi.org/10.1111/ropr.12061}
  {\path{doi:10.1111/ropr.12061}}.

\bibitem{Rahut.2018}
D.~B. Rahut, K.~A. Mottaleb, A.~Ali, J.~Aryal, {The use and determinants of
  solar energy by Sub-Saharan African households}, {International Journal of
  Sustainable Energy} 37~(8) (2018) 718--735.
\newblock \href {https://doi.org/10.1080/14786451.2017.1323897}
  {\path{doi:10.1080/14786451.2017.1323897}}.

\bibitem{Opiyo.2019}
N.~N. Opiyo, {Impacts of neighbourhood influence on social acceptance of small
  solar home systems in rural western Kenya}, {Energy Research {\&} Social
  Science} 52 (2019) 91--98.
\newblock \href {https://doi.org/10.1016/j.erss.2019.01.013}
  {\path{doi:10.1016/j.erss.2019.01.013}}.

\bibitem{Bao.2020}
Q.~Bao, E.~Sinitskaya, K.~J. Gomez, E.~F. MacDonald, M.~C. Yang, {A
  human-centered design approach to evaluating factors in residential solar PV
  adoption: A survey of homeowners in California and Massachusetts}, {Renewable
  Energy} 151 (2020) 503--513.
\newblock \href {https://doi.org/10.1016/j.renene.2019.11.047}
  {\path{doi:10.1016/j.renene.2019.11.047}}.

\bibitem{Liang.2020}
J.~Liang, P.~Lui, Y.~Qiu, Y.~D. Wang, B.~Xing, {Time-of-Use electricity pricing
  and residential low-carbon energy technology adoption}, {The Energy Journal}
  41~(3) (2020).
\newblock \href {https://doi.org/10.5547/01956574.41.2.jlia}
  {\path{doi:10.5547/01956574.41.2.jlia}}.

\bibitem{Chen.2014}
K.~K. Chen, {Assessing the effects of customer innovativeness, environmental
  value and ecological lifestyles on residential solar power systems install
  intention}, {Energy Policy} 67 (2014) 951--961.
\newblock \href {https://doi.org/10.1016/j.enpol.2013.12.005}
  {\path{doi:10.1016/j.enpol.2013.12.005}}.

\bibitem{Delmas.2013}
M.~A. Delmas, M.~Fischlein, O.~I. Asensio, {Information strategies and energy
  conservation behavior: A meta-analysis of experimental studies from 1975 to
  2012}, {Energy Policy} 61 (2013) 729--739.
\newblock \href {https://doi.org/10.1016/j.enpol.2013.05.109}
  {\path{doi:10.1016/j.enpol.2013.05.109}}.

\bibitem{StataCorpLLC.2019}
{StataCorp LLC}, \href{https://www.stata.com/manuals/meta.pdf}{{Stata
  Meta-Analysis Reference Manual: Release 16}} (2019).
\newline\urlprefix\url{https://www.stata.com/manuals/meta.pdf}

\bibitem{StataCorp..}
{Stata Corp.}, \href{https://www.stata.com/manuals/semintro11.pdf}{{Intro 11
  --- Fitting models with summary statistics data (sem only)}}.
\newline\urlprefix\url{https://www.stata.com/manuals/semintro11.pdf}

\bibitem{cohen2013statistical}
J.~Cohen, Statistical power analysis for the behavioral sciences, Academic
  press, 1977.

\bibitem{Aichholzer.2017}
J.~Aichholzer, {Einf{\"u}hrung in lineare Strukturgleichungsmodelle mit Stata},
  {Springer Fachmedien Wiesbaden}, Wiesbaden, 2017.
\newblock \href {https://doi.org/10.1007/978-3-658-16670-0}
  {\path{doi:10.1007/978-3-658-16670-0}}.

\bibitem{Jager.2006}
W.~Jager, {Stimulating the diffusion of photovoltaic systems: A behavioural
  perspective}, {Energy Policy} 34~(14) (2006) 1935--1943.
\newblock \href {https://doi.org/10.1016/j.enpol.2004.12.022}
  {\path{doi:10.1016/j.enpol.2004.12.022}}.

\bibitem{Ford.2017}
R.~Ford, S.~Walton, J.~Stephenson, D.~Rees, M.~Scott, G.~King, J.~Williams,
  B.~Wooliscroft, {Emerging energy transitions: PV uptake beyond subsidies},
  {Technological Forecasting and Social Change} 117 (2017) 138--150.
\newblock \href {https://doi.org/10.1016/j.techfore.2016.12.007}
  {\path{doi:10.1016/j.techfore.2016.12.007}}.

\bibitem{Amburgey.2012}
J.~W. Amburgey, D.~B. Thoman, {Dimensionality of the New Ecological Paradigm},
  {Environment and Behavior} 44~(2) (2012) 235--256.
\newblock \href {https://doi.org/10.1177/0013916511402064}
  {\path{doi:10.1177/0013916511402064}}.

\bibitem{Manning.1995}
K.~C. Manning, W.~O. Bearden, T.~J. Madden, {Consumer Innovativeness and the
  Adoption Process}, {Journal of Consumer Psychology} 4~(4) (1995) 329--345.
\newblock \href {https://doi.org/10.1207/s15327663jcp0404{\textunderscore }02}
  {\path{doi:10.1207/s15327663jcp0404{\textunderscore }02}}.

\bibitem{Editorial.2018}
Editorial, {Enhancing reporting standards}, {Nature Climate Change} 8~(3)
  (2018) 173.
\newblock \href {https://doi.org/10.1038/s41558-018-0109-x}
  {\path{doi:10.1038/s41558-018-0109-x}}.

\bibitem{Cooper.2020}
H.~Cooper, {Reporting Quantitative Research in Psychology: How to Meet APA
  Style Journal Article Reporting Standards, Second Edition, Revised: Updated
  for the 7th edition of the Publication Manual}, APA Style Series, 2020.

\bibitem{johanning2020modular}
S.~Johanning, F.~Scheller, D.~Abitz, C.~Wehner, T.~Bruckner, A modular
  multi-agent framework for innovation diffusion in changing business
  environments: conceptualization, formalization and implementation, Complex
  Adaptive Systems Modeling 8~(1) (2020) 1--32.
\newblock \href {https://doi.org/10.1186/s40294-020-00074-6}
  {\path{doi:10.1186/s40294-020-00074-6}}.

\bibitem{Scheller.2019}
F.~Scheller, S.~Johanning, T.~Bruckner,
  \href{http://hdl.handle.net/10419/191981}{{Review of designing empirically
  grounded agent-based models of innovation diffusion}} (2019).
\newline\urlprefix\url{http://hdl.handle.net/10419/191981}

\bibitem{Yadav.2020}
P.~Yadav, P.~J. Davies, S.~Khan, {Breaking into the photovoltaic energy
  transition for rural and remote communities: challenging the impact of
  awareness norms and subsidy schemes}, {Clean Technologies and Environmental
  Policy} 22~(4) (2020) 817--834.
\newblock \href {https://doi.org/10.1007/s10098-020-01823-0}
  {\path{doi:10.1007/s10098-020-01823-0}}.

\bibitem{Sinitskaya.2020}
E.~Sinitskaya, K.~J. Gomez, Q.~Bao, M.~C. Yang, E.~F. MacDonald, {Designing
  linked journey maps to understand the complexities of the residential solar
  energy market}, {Renewable Energy} 145 (2020) 1910--1922.
\newblock \href {https://doi.org/10.1016/j.renene.2019.06.018}
  {\path{doi:10.1016/j.renene.2019.06.018}}.

\bibitem{Roodenrijs.2020}
J.~C.~M. Roodenrijs, D.~L.~T. Hegger, H.~L.~P. Mees, P.~Driessen, {Opening up
  the Black Box of Group Decision-Making on Solar Energy: The Case of Strata
  Buildings in Amsterdam, the Netherlands}, {Sustainability} 12~(5) (2020)
  2097.
\newblock \href {https://doi.org/10.3390/su12052097}
  {\path{doi:10.3390/su12052097}}.

\bibitem{Palm.2020}
A.~Palm, B.~Lantz, {Information dissemination and residential solar PV adoption
  rates: The effect of an information campaign in Sweden}, {Energy Policy} 142
  (2020) 111540.
\newblock \href {https://doi.org/10.1016/j.enpol.2020.111540}
  {\path{doi:10.1016/j.enpol.2020.111540}}.

\bibitem{Makki.2020}
A.~A. Makki, I.~Mosly, {Factors Affecting Public Willingness to Adopt Renewable
  Energy Technologies: An Exploratory Analysis}, {Sustainability} 12~(3) (2020)
  845.
\newblock \href {https://doi.org/10.3390/su12030845}
  {\path{doi:10.3390/su12030845}}.

\bibitem{Lau.2020}
L.-S. Lau, Y.-O. Choong, C.-Y. Wei, A.-N. Seow, C.-K. Choong, A.~Senadjki,
  S.-L. Ching, {Investigating nonusers' behavioural intention towards solar
  photovoltaic technology in Malaysia: The role of knowledge transmission and
  price value}, {Energy Policy} 144 (2020) 111651.
\newblock \href {https://doi.org/10.1016/j.enpol.2020.111651}
  {\path{doi:10.1016/j.enpol.2020.111651}}.

\bibitem{Stikvoort.2020}
B.~Stikvoort, C.~Bartusch, P.~Juslin, {Different strokes for different folks?
  Comparing pro-environmental intentions between electricity consumers and
  solar prosumers in Sweden}, {Energy Research {\&} Social Science} 69 (2020)
  101552.
\newblock \href {https://doi.org/10.1016/j.erss.2020.101552}
  {\path{doi:10.1016/j.erss.2020.101552}}.

\bibitem{DaFigueira.2020}
J.~{Da Figueira}, D.~Coelho, F.~Lopes, {Assessing E3 impacts of RES integration
  using residential consumer's willingness to invest in PV systems}, {EAI
  Endorsed Transactions on Energy Web} 7~(25) (2020) 162403.
\newblock \href {https://doi.org/10.4108/eai.7-1-2020.162403}
  {\path{doi:10.4108/eai.7-1-2020.162403}}.

\bibitem{Kastner.2019}
I.~Kastner, I.~Wittenberg, {How Measurements ``Affect'' the Importance of
  Social Influences on Household's Photovoltaic Adoption---A German Case
  Study}, {Sustainability} 11~(19) (2019) 5175.
\newblock \href {https://doi.org/10.3390/su11195175}
  {\path{doi:10.3390/su11195175}}.

\bibitem{GavaGastaldo.2019}
N.~{Gava Gastaldo}, G.~Rediske, P.~{Donaduzzi Rigo}, C.~{Brum Rosa},
  L.~Michels, J.~C. {Mairesse Siluk}, {What is the Profile of the Investor in
  Household Solar Photovoltaic Energy Systems?}, {Energies} 12~(23) (2019)
  4451.
\newblock \href {https://doi.org/10.3390/en12234451}
  {\path{doi:10.3390/en12234451}}.

\bibitem{Kiprop.2019}
E.~Kiprop, K.~Matsui, N.~Maundu, {The Role of Household Consumers in Adopting
  Renewable Energy Technologies in Kenya}, {Environments} 6~(8) (2019) 95.
\newblock \href {https://doi.org/10.3390/environments6080095}
  {\path{doi:10.3390/environments6080095}}.

\bibitem{HiguerasCastillo.2019}
E.~Higueras-Castillo, F.~Mu{\~n}oz-Leiva, F.~J. Li{\'e}bana-Cabanillas, {An
  examination of attributes and barriers to adopt biomass and solar technology.
  A cross-cultural approach}, {Journal of environmental management} 236 (2019)
  639--648.
\newblock \href {https://doi.org/10.1016/j.jenvman.2019.02.022}
  {\path{doi:10.1016/j.jenvman.2019.02.022}}.

\bibitem{Fachrizal.2019}
M.~Fachrizal, J.~Tang, {Forecasting Annual Solar PV Capacity Installation in
  Thailand Residential Sector: A User Segmentation Approach}, {Engineering
  Journal} 23~(6) (2019) 99--115.
\newblock \href {https://doi.org/10.4186/ej.2019.23.6.99}
  {\path{doi:10.4186/ej.2019.23.6.99}}.

\bibitem{Borghesi.2019}
A.~Borghesi, M.~Milano, {Merging Observed and Self-Reported Behaviour in
  Agent-Based Simulation: A Case Study on Photovoltaic Adoption}, {Applied
  Sciences} 9~(10) (2019) 2098.
\newblock \href {https://doi.org/10.3390/app9102098}
  {\path{doi:10.3390/app9102098}}.

\bibitem{Zainudina.2019}
N.~Zainudina, Z.~M. Jushoa, Z.~Zainalaludina, S.~Osmana, N.~Nordinb, L.~H.
  Paima, A.~Ramasamya, {Climate Change Awareness and Solar Energy Adoption of
  Household}, {International Journal of Advanced Science and Technology}
  28~(8s) (2019) 357--363.

\bibitem{Walters.2018b}
J.~Walters, J.~Kaminsky, L.~Gottschamer, {A Systems Analysis of Factors
  Influencing Household Solar PV Adoption in Santiago, Chile}, {Sustainability}
  10~(4) (2018) 1257.
\newblock \href {https://doi.org/10.3390/su10041257}
  {\path{doi:10.3390/su10041257}}.

\bibitem{Tayal.2018}
D.~Tayal, U.~Evers, {Consumer preferences and electricity pricing reform in
  Western Australia}, {Utilities Policy} 54 (2018) 115--124.
\newblock \href {https://doi.org/10.1016/j.jup.2018.08.008}
  {\path{doi:10.1016/j.jup.2018.08.008}}.

\bibitem{Shakeel.2018}
S.~R. Shakeel, S.~u. Rahman, {Towards the establishment of renewable energy
  technologies' market: An assessment of public acceptance and use in
  Pakistan}, {Journal of Renewable and Sustainable Energy} 10~(4) (2018)
  045907.
\newblock \href {https://doi.org/10.1063/1.5033454}
  {\path{doi:10.1063/1.5033454}}.

\bibitem{Reeves.2018}
D.~C. Reeves, V.~Rai, {Strike while the rebate is hot: Savvy consumers and
  strategic technology adoption timing}, {Energy Policy} 121 (2018) 325--335.
\newblock \href {https://doi.org/10.1016/j.enpol.2018.06.045}
  {\path{doi:10.1016/j.enpol.2018.06.045}}.

\bibitem{Meiklejohn.2018}
D.~Meiklejohn, S.~Bekessy, S.~Moloney, A.~Bezama, {Shifting practices: How the
  rise of rooftop solar PV has changed local government community engagement},
  {Cogent Environmental Science} 4~(1) (2018) 1481584.
\newblock \href {https://doi.org/10.1080/23311843.2018.1481584}
  {\path{doi:10.1080/23311843.2018.1481584}}.

\bibitem{Aklin.2018}
M.~Aklin, C.-y. Cheng, J.~Urpelainen, {Geography, community, household:
  Adoption of distributed solar power across India}, {Energy for Sustainable
  Development} 42 (2018) 54--63.
\newblock \href {https://doi.org/10.1016/j.esd.2017.09.010}
  {\path{doi:10.1016/j.esd.2017.09.010}}.

\bibitem{Stikvoort.2018}
B.~Stikvoort, P.~Juslin, C.~Bartusch, {Good things come in small packages: is
  there a common set of motivators for energy behaviour?}, {Energy Efficiency}
  11~(7) (2018) 1599--1615.
\newblock \href {https://doi.org/10.1007/s12053-017-9537-0}
  {\path{doi:10.1007/s12053-017-9537-0}}.

\bibitem{Klepacka.2018}
A.~M. Klepacka, W.~J. Florkowski, T.~Meng, {Clean, accessible, and cost-saving:
  Reasons for rural household investment in solar panels in Poland},
  {Resources, Conservation and Recycling} 139 (2018) 338--350.
\newblock \href {https://doi.org/10.1016/j.resconrec.2018.09.004}
  {\path{doi:10.1016/j.resconrec.2018.09.004}}.

\bibitem{Curtius.2018b}
H.~C. Curtius, S.~L. Hille, C.~Berger, U.~J.~J. Hahnel, R.~W{\"u}stenhagen,
  {Shotgun or snowball approach? Accelerating the diffusion of rooftop solar
  photovoltaics through peer effects and social norms}, {Energy Policy} 118
  (2018) 596--602.
\newblock \href {https://doi.org/10.1016/j.enpol.2018.04.005}
  {\path{doi:10.1016/j.enpol.2018.04.005}}.

\bibitem{Bondio.2018}
S.~Bondio, M.~Shahnazari, A.~McHugh, {The technology of the middle class:
  Understanding the fulfilment of adoption intentions in Queensland's rapid
  uptake residential solar photovoltaics market}, {Renewable and Sustainable
  Energy Reviews} 93 (2018) 642--651.
\newblock \href {https://doi.org/10.1016/j.rser.2018.05.035}
  {\path{doi:10.1016/j.rser.2018.05.035}}.

\bibitem{Tanaka.2017}
K.~Tanaka, M.~Sekito, S.~Managi, S.~Kaneko, V.~Rai, {Decision-making governance
  for purchases of solar photovoltaic systems in Japan}, {Energy Policy} 111
  (2017) 75--84.
\newblock \href {https://doi.org/10.1016/j.enpol.2017.09.012}
  {\path{doi:10.1016/j.enpol.2017.09.012}}.

\bibitem{Muresan.2017}
I.~C. Muresan, G.~O. Chiciudean, R.~Harun, F.~H. Arion, A.~Porutiu, D.~I.
  Chiciudean, I.~G. Oroian, M.~I. Jitea, {Constraints on Use of Renewable
  Energy Technologies in the Rural Area: a Case Study from the North-West
  Region of Romania}, {Journal of Environmental Protection and Ecology} 18~(4)
  (2017) 1746--1753.

\bibitem{Braito.2017}
M.~Braito, C.~Flint, A.~Muhar, M.~Penker, S.~Vogel, {Individual and collective
  socio-psychological patterns of photovoltaic investment under diverging
  policy regimes of Austria and Italy}, {Energy Policy} 109 (2017) 141--153.
\newblock \href {https://doi.org/10.1016/j.enpol.2017.06.063}
  {\path{doi:10.1016/j.enpol.2017.06.063}}.

\bibitem{Pandey.2017}
S.~Pandey, B.~Kesari, {Study of the factors affecting willingness of rural
  household's to adoption of solar lighting system}, {Journal of Advanced
  Research in Dynamical and Control Systems} 9~(Sp- 17) (2017) 1861--1873.

\bibitem{Bergek.2017}
A.~Bergek, I.~Mignon, {Motives to adopt renewable electricity technologies:
  Evidence from Sweden}, {Energy Policy} 106 (2017) 547--559.
\newblock \href {https://doi.org/10.1016/j.enpol.2017.04.016}
  {\path{doi:10.1016/j.enpol.2017.04.016}}.

\bibitem{Briguglio.2017}
M.~Briguglio, G.~Formosa, {When households go solar: Determinants of uptake of
  a Photovoltaic Scheme and policy insights}, {Energy Policy} 108 (2017)
  154--162.
\newblock \href {https://doi.org/10.1016/j.enpol.2017.05.039}
  {\path{doi:10.1016/j.enpol.2017.05.039}}.

\bibitem{Yuasa.2016}
K.~Yuasa, M.~Yata, {Installation of Residential Energy Systems: Local
  Conditions and Residents$\prime$ Willingness}, {Journal of Asian Architecture
  and Building Engineering} 15~(1) (2016) 127--132.
\newblock \href {https://doi.org/10.3130/jaabe.15.127}
  {\path{doi:10.3130/jaabe.15.127}}.

\bibitem{Rai.2016}
V.~Rai, D.~C. Reeves, R.~Margolis, {Overcoming barriers and uncertainties in
  the adoption of residential solar PV}, {Renewable Energy} 89 (2016) 498--505.
\newblock \href {https://doi.org/10.1016/j.renene.2015.11.080}
  {\path{doi:10.1016/j.renene.2015.11.080}}.

\bibitem{Sigrin.2015}
B.~Sigrin, J.~Pless, E.~Drury, {Diffusion into new markets: evolving customer
  segments in the solar photovoltaics market}, {Environmental Research Letters}
  10~(8) (2015) 084001.
\newblock \href {https://doi.org/10.1088/1748-9326/10/8/084001}
  {\path{doi:10.1088/1748-9326/10/8/084001}}.

\bibitem{Yamamoto.2015}
Y.~Yamamoto, {Opinion leadership and willingness to pay for residential
  photovoltaic systems}, {Energy Policy} 83 (2015) 185--192.
\newblock \href {https://doi.org/10.1016/j.enpol.2015.04.014}
  {\path{doi:10.1016/j.enpol.2015.04.014}}.

\bibitem{Hast.2015}
A.~Hast, B.~Alimohammadisagvand, S.~Syri, {Consumer attitudes towards renewable
  energy in China---The case of Shanghai}, {Sustainable Cities and Society} 17
  (2015) 69--79.
\newblock \href {https://doi.org/10.1016/j.scs.2015.04.003}
  {\path{doi:10.1016/j.scs.2015.04.003}}.

\bibitem{Bigerna.2015}
S.~Bigerna, P.~Polinori, {Assessing the Determinants of Renewable Electricity
  Acceptance Integrating Meta-Analysis Regression and a Local Comprehensive
  Survey}, {Sustainability} 7~(9) (2015) 11909--11932.
\newblock \href {https://doi.org/10.3390/su70911909}
  {\path{doi:10.3390/su70911909}}.

\bibitem{Vasseur.2015b}
V.~Vasseur, R.~Kemp, {A segmentation analysis: the case of photovoltaic in the
  Netherlands}, {Energy Efficiency} 8~(6) (2015) 1105--1123.
\newblock \href {https://doi.org/10.1007/s12053-015-9340-8}
  {\path{doi:10.1007/s12053-015-9340-8}}.

\bibitem{Vasseur.2015}
V.~Vasseur, R.~Kemp, {The adoption of PV in the Netherlands: A statistical
  analysis of adoption factors}, {Renewable and Sustainable Energy Reviews} 41
  (2015) 483--494.
\newblock \href {https://doi.org/10.1016/j.rser.2014.08.020}
  {\path{doi:10.1016/j.rser.2014.08.020}}.

\bibitem{Schaffer.2015}
A.~J. Schaffer, S.~Brun, {Beyond the sun---Socioeconomic drivers of the
  adoption of small-scale photovoltaic installations in Germany}, {Energy
  Research {\&} Social Science} 10 (2015) 220--227.
\newblock \href {https://doi.org/10.1016/j.erss.2015.06.010}
  {\path{doi:10.1016/j.erss.2015.06.010}}.

\bibitem{Karakaya.2015}
E.~Karakaya, A.~Hidalgo, C.~Nuur, {Motivators for adoption of photovoltaic
  systems at grid parity: A case study from Southern Germany}, {Renewable and
  Sustainable Energy Reviews} 43 (2015) 1090--1098.
\newblock \href {https://doi.org/10.1016/j.rser.2014.11.077}
  {\path{doi:10.1016/j.rser.2014.11.077}}.

\bibitem{Tsantopoulos.2014}
G.~Tsantopoulos, G.~Arabatzis, S.~Tampakis, {Public attitudes towards
  photovoltaic developments: Case study from Greece}, {Energy Policy} 71 (2014)
  94--106.
\newblock \href {https://doi.org/10.1016/j.enpol.2014.03.025}
  {\path{doi:10.1016/j.enpol.2014.03.025}}.

\bibitem{Balcombe.2014}
P.~Balcombe, D.~Rigby, A.~Azapagic, {Investigating the importance of
  motivations and barriers related to microgeneration uptake in the UK},
  {Applied Energy} 130 (2014) 403--418.
\newblock \href {https://doi.org/10.1016/j.apenergy.2014.05.047}
  {\path{doi:10.1016/j.apenergy.2014.05.047}}.

\bibitem{Karytsas.2014}
S.~Karytsas, H.~Theodoropoulou, {Socioeconomic and demographic factors that
  influence publics' awareness on the different forms of renewable energy
  sources}, {Renewable Energy} 71 (2014) 480--485.
\newblock \href {https://doi.org/10.1016/j.renene.2014.05.059}
  {\path{doi:10.1016/j.renene.2014.05.059}}.

\bibitem{Alam.2014}
S.~S. Alam, N.~H. {Nik Hashim}, M.~Rashid, N.~A. Omar, N.~Ahsan, M.~D. Ismail,
  {Small-scale households renewable energy usage intention: Theoretical
  development and empirical settings}, {Renewable Energy} 68 (2014) 255--263.
\newblock \href {https://doi.org/10.1016/j.renene.2014.02.010}
  {\path{doi:10.1016/j.renene.2014.02.010}}.

\bibitem{VarelaMargolles.2014}
A.~Varela-Margolles, J.~Onsted, {Do Incentives Work?: An Analysis of
  Residential Solar Energy Adoption in Miami-Dade County, Florida},
  {Southeastern Geographer} 54~(1) (2014) 18--35.
\newblock \href {https://doi.org/10.1353/sgo.2014.0009}
  {\path{doi:10.1353/sgo.2014.0009}}.

\bibitem{Kaldellis.2013}
J.~K. Kaldellis, M.~Kapsali, E.~Kaldelli, E.~Katsanou, {Comparing recent views
  of public attitude on wind energy, photovoltaic and small hydro
  applications}, {Renewable Energy} 52 (2013) 197--208.
\newblock \href {https://doi.org/10.1016/j.renene.2012.10.045}
  {\path{doi:10.1016/j.renene.2012.10.045}}.

\bibitem{Li.2013}
X.~Li, H.~Li, X.~Wang, {Farmers' willingness to convert traditional houses to
  solar houses in rural areas: A survey of 465 households in Chongqing, China},
  {Energy Policy} 63 (2013) 882--886.
\newblock \href {https://doi.org/10.1016/j.enpol.2013.09.004}
  {\path{doi:10.1016/j.enpol.2013.09.004}}.

\bibitem{Muller.2013}
S.~M{\"u}ller, J.~Rode, {The adoption of photovoltaic systems in Wiesbaden,
  Germany}, {Economics of Innovation and New Technology} 22~(5) (2013)
  519--535.
\newblock \href {https://doi.org/10.1080/10438599.2013.804333}
  {\path{doi:10.1080/10438599.2013.804333}}.

\bibitem{Zhai.2012}
P.~Zhai, E.~D. Williams, {Analyzing consumer acceptance of photovoltaics (PV)
  using fuzzy logic model}, {Renewable Energy} 41 (2012) 350--357.
\newblock \href {https://doi.org/10.1016/j.renene.2011.11.041}
  {\path{doi:10.1016/j.renene.2011.11.041}}.

\bibitem{Zhao.2012}
T.~Zhao, L.~Bell, M.~W. Horner, J.~Sulik, J.~Zhang, {Consumer responses towards
  home energy financial incentives: A survey-based study}, {Energy Policy} 47
  (2012) 291--297.
\newblock \href {https://doi.org/10.1016/j.enpol.2012.04.070}
  {\path{doi:10.1016/j.enpol.2012.04.070}}.

\bibitem{Zhang.2012}
X.~Zhang, L.~Shen, S.~Y. Chan, {The diffusion of solar energy use in HK: What
  are the barriers?}, {Energy Policy} 41 (2012) 241--249.
\newblock \href {https://doi.org/10.1016/j.enpol.2011.10.043}
  {\path{doi:10.1016/j.enpol.2011.10.043}}.

\bibitem{Axsen.2012b}
J.~Axsen, J.~TyreeHageman, A.~Lentz, {Lifestyle practices and pro-environmental
  technology}, {Ecological Economics} 82 (2012) 64--74.
\newblock \href {https://doi.org/10.1016/j.ecolecon.2012.07.013}
  {\path{doi:10.1016/j.ecolecon.2012.07.013}}.

\bibitem{Lau.2011}
K.~L. Lau, E.~Ng, Z.~J. He, {Residents' preference of solar access in
  high-density sub-tropical cities}, {Solar Energy} 85~(9) (2011) 1878--1890.
\newblock \href {https://doi.org/10.1016/j.solener.2011.04.026}
  {\path{doi:10.1016/j.solener.2011.04.026}}.

\bibitem{Palm.2011}
J.~Palm, M.~Tengvard, {Motives for and barriers to household adoption of
  small-scale production of electricity: examples from Sweden},
  {Sustainability: Science, Practice and Policy} 7~(1) (2011) 6--15.
\newblock \href {https://doi.org/10.1080/15487733.2011.11908061}
  {\path{doi:10.1080/15487733.2011.11908061}}.

\bibitem{Welsch.2009}
H.~Welsch, J.~K{\"u}hling, {Determinants of pro-environmental consumption: The
  role of reference groups and routine behavior}, {Ecological Economics} 69~(1)
  (2009) 166--176.
\newblock \href {https://doi.org/10.1016/j.ecolecon.2009.08.009}
  {\path{doi:10.1016/j.ecolecon.2009.08.009}}.

\bibitem{Urmee.2009}
T.~Urmee, D.~Harries, {A survey of solar PV program implementers in Asia and
  the Pacific regions}, {Energy for Sustainable Development} 13~(1) (2009)
  24--32.
\newblock \href {https://doi.org/10.1016/j.esd.2009.01.002}
  {\path{doi:10.1016/j.esd.2009.01.002}}.

\bibitem{Caird.2008}
S.~Caird, R.~Roy, H.~Herring, {Improving the energy performance of UK
  households: Results from surveys of consumer adoption and use of low- and
  zero-carbon technologies}, {Energy Efficiency} 1~(2) (2008) 149--166.
\newblock \href {https://doi.org/10.1007/s12053-008-9013-y}
  {\path{doi:10.1007/s12053-008-9013-y}}.

\bibitem{Haw.2008}
L.~C. Haw, M.~F. Zaky, E.~Salleh, {Assessment of Non-Technical Barriers for
  Widespread Adoption of Building Integrated Photovoltaic System (BIPV) in
  Malaysian Urban Residential Sector: Paper No: 153} (2008).

\bibitem{Faiers.2007b}
A.~Faiers, C.~Neame, M.~Cook, {The adoption of domestic solar-power systems: Do
  consumers assess product attributes in a stepwise process?}, {Energy Policy}
  35~(6) (2007) 3418--3423.
\newblock \href {https://doi.org/10.1016/j.enpol.2006.10.029}
  {\path{doi:10.1016/j.enpol.2006.10.029}}.

\end{thebibliography}
\bibliographystyle{elsarticle-num}

\appendix
\newpage
\section{Appendix}
\label{Appendix}

\begin{ThreePartTable}

\tiny
\begin{longtable}[l]{lp{25em}lp{12em}l|lllllll}

\caption{List of 110 papers eligible for full-text screening. The screening procedure consisted of (1) Focal technology is rooftop PV; (2) Quantitative survey with residential decision-makers; (3) Dependent variable is intention; (4) Exclusion criteria (a=choice experiment, b=off-grid, c=not available); (5) Overall suitability; (6) Correlations provided in publication; (7) Correlations received. Y: Yes; N: No}\\
\label{tab:fulltext}\\
\toprule
\multicolumn{5}{p{45em}}{\textit{Description}} & \multicolumn{7}{c}{\textit{Screening procedure}} \\
\midrule
\textbf{No.} & \textbf{Title} & \textbf{Year} & \textbf{Journal} & \textbf{Ref.} & \textbf{1} & \textbf{2} & \textbf{3} & \textbf{4} & \textbf{5} & \textbf{6} & \textbf{7} \\
\midrule
\endfirsthead

\caption*{List of 110 papers (continued). The screening procedure consisted of (1) Focal technology is rooftop PV; (2) Quantitative survey with residential decision-makers; (3) Dependent variable is intention; (4) Exclusion criteria (a=choice experiment, b=off-grid, c=not available); (5) Overall suitability; (6) Correlations provided in publication; (7) Correlations received. Y: Yes; N: No}\\
\toprule
\multicolumn{5}{p{45em}}{\textit{Description}} & \multicolumn{7}{c}{\textit{Screening procedure}} \\
\midrule
\textbf{No.} & \textbf{Title} & \textbf{Year} & \textbf{Journal} & \textbf{Ref.} & \textbf{1} & \textbf{2} & \textbf{3} & \textbf{4} & \textbf{5} & \textbf{6} & \textbf{7} \\
\midrule
\endhead

\endfoot

\midrule
\endlastfoot
    
    1   & Unrealized opportunities for residential solar panels in Australia & 2020 & Energy Policy &   \cite{Zander.2020}  & Y   & Y   & Y   &     & Y   & N & N \\
    2   & Consumer attitude and purchase intention toward rooftop photovoltaic installation: The roles of personal trait, psychological benefit, and government incentives & 2020 & Energy \& Environment &  \cite{Sun.2020}   & Y   & Y   & Y   &     & Y   & Y &  \\
    3   & Examining the Key Drivers of Residential Solar Adoption in Upstate New York & 2020 & Sustainability &   \cite{Schelly.2020}  & Y   & Y   & N   &     &   N  &     &  \\
    4   & Breaking into the photovoltaic energy transition for rural and remote communities: challenging the impact of awareness norms and subsidy schemes & 2020 & Clean Technologies and Environmental Policy &   \cite{Yadav.2020}  & Y   & Y   & N   &     &   N  &     &  \\
    5   & More alike than different: Profiles of high-income and low-income rooftop solar adopters in the United States & 2020 & Energy Research \& Social Science &  \cite{Wolske.2020}  & Y   & Y   & N   &     &  N   &     &  \\
    6   & Designing linked journey maps to understand the complexities of the residential solar energy market & 2020 & Renewable Energy &   \cite{Sinitskaya.2020}  & Y   & N   &     &     &  N   &     &  \\
    7   & Opening up the Black Box of Group Decision-Making on Solar Energy: The Case of Strata Buildings in Amsterdam, the Netherlands & 2020 & Sustainability &     & Y   &     & N   &     &  N   &     &  \\
    8   & A study on the factors affecting household solar adoption in Kerala, India & 2020 & International Journal of Productivity and Performance Management &   \cite{Roodenrijs.2020}  & Y   & Y   & Y   &     & Y   & N &  N \\
    9   & Information dissemination and residential solar PV adoption rates: The effect of an information campaign in Sweden & 2020 & Energy Policy & \cite{Palm.2020}    & Y   & Y   & N   &     &  N   &     &  \\
    10  & Demand-side solutions for climate mitigation: Bottom-up drivers of household energy behavior change in the Netherlands and Spain & 2020 & Energy Research \& Social Science &   \cite{Niamir.2020}  & Y   & Y   & N   &     &  N   &     &  \\
    11  & Factors Affecting Public Willingness to Adopt Renewable Energy Technologies: An Exploratory Analysis & 2020 & Sustainability &  \cite{Makki.2020}   & N   &     &     &     &  N   &     &  \\
    12  & Time-of-Use electricity pricing and residential low-carbon energy technology adoption & 2020 & The Energy Journal &   \cite{Liang.2020}  &     &     &     & c   &  N   &     &  \\
    13  & Sustainable consumption from the consumer’s perspective: Antecedents of solar innovation adoption & 2020 & Resources, Conservation \& Recycling &  \cite{Kapoor.2020}   & N   &     &     &     &   N  &     &  \\
    14  & The influence of consumers’ intention factors on willingness to pay for renewable energy: a structural equation modeling approach & 2020 & Environmental Science and Pollution Research &  \cite{Irfan.2020}   & N   &     &     &     &  N   &     &  \\
    15  & Residential Photovoltaic Systems in Norway: Household Knowledge, Preferences and Willingness to Pay & 2020 & Challenges in Sustainability &  \cite{Cherry.2020}   & Y   & Y   & Y   &     & Y   & N & N \\
    16  & A human-centered design approach to evaluating factors in residential solar PV adoption: A survey of homeowners in California and Massachusetts & 2020 & Renewable Energy &  \cite{Bao.2020}   & Y   & Y   & Y   & a   &  N   &     &  \\
    17  & Analysis of perceptions towards the rooftop photovoltaic solar system policy in Indonesia & 2020 & Energy Policy &   \cite{Setyawati.2020}  & Y   & Y   & N   &     &  N   &     &  \\
    18  & Investigating nonusers’ behavioural intention towards solar photovoltaic technology in Malaysia: The role of knowledge transmission and price value & 2020 & Energy Policy &  \cite{Lau.2020}   & Y   & Y   & Y   &     & Y   & N & N \\
    19  & An exploratory study of the public's views on residential solar photovoltaic systems in oil-rich Saudi Arabia & 2020 & Environmental Development & \cite{Alrashoud.2020}    & Y   & Y   & N   &     &  N   &     &  \\
    20  & Different strokes for different folks? Comparing pro-environmental intentions between electricity consumers and solar prosumers in Sweden & 2020 & Energy Research and Social Science &  \cite{Stikvoort.2020}   & N   &     &     &     &  N   &     &  \\
    21  & Social acceptability of solar photovoltaic system in Pakistan: Key determinants and policy implications & 2020 & Journal of Cleaner Production &  \cite{Jan.2020}   & Y   & Y   & N   &     &  N   &     &  \\
    22  & Assessing E3 impacts of RES integration using residential consumer’s willingness to invest in PV systems & 2020 & EAI Endorsed Transactions on Energy Web &  \cite{DaFigueira.2020}   & Y   & Y   & N   &     &   N  &     &  \\
    23  & Clean energy transition in a developing society: Perspectives on the socioeconomic determinants of Solar Home Systems adoption among urban households in southeastern Nigeria & 2020 & African Journal of Science, Technology, Innovation and Development &  \cite{Anugwom.2020}   & Y   & Y   &     & b   &   N  &     &  \\
    24  & What drives home solar PV uptake? Subsidies, peer effects and visibility in Sweden & 2020 & Energy Research and Social Science &  \cite{Mundaca.2020}   & Y   & Y   & N   &     &  N   &     &  \\
    25  & How Measurements “Affect” the Importance of Social Influences on Household’s Photovoltaic Adoption— A German Case Study & 2019 & Sustainability &  \cite{Kastner.2019}   & Y   & Y   & N   &     &   N  &     &  \\
    26  & Drivers of renewable technology adoption in the household sector & 2019 & Energy Economics &  \cite{Jacksohn.2019}    & Y   & Y   & N   &     &   N  &     &  \\
    27  & What is the Profile of the Investor in Household Solar Photovoltaic Energy Systems? & 2019 & Energies & \cite{GavaGastaldo.2019}   & Y   & Y   & N   &     &   N  &     &  \\
    28  & Understanding the determinants of rooftop solar installation: evidence from household surveys in Australia & 2019 & Australian Journal of Agricultural and Resource Economics &  \cite{Best.2019}   & Y   & Y   & Y   &     & Y   & N & N \\
    29  & Factors Influencing Social Perception of Residential Solar Photovoltaic Systems in Saudi Arabia & 2019 & Sustainability &  \cite{Alrashoud.2019}   & Y   & Y   & N   &     &   N  &     &  \\
    30  & Diffusion of residential RT solar – is lack of funds the real issue? & 2019 & International Journal of Energy Sector Management &  \cite{Aggarwal.2019}   & Y   & Y   & Y   &     & Y   & N & N \\
    31  & New trends in solar: A comparative study assessing the attitudes towards the adoption of rooftop PV & 2019 & Energy Policy & \cite{Abreu.2019}    & Y   & Y   & Y   &     & Y   & N & N \\
    32  & Preferences for and potential impacts of financial incentives to install residential rooftop solar photovoltaic systems in Australia & 2019 & Journal of Cleaner Production &  \cite{Zander.2019}   & Y   & Y   &     & a   &  N   &     &  \\
    33  & Impacts of neighbourhood influence on social acceptance of small solar home systems in rural western Kenya & 2019 & Energy Research \& Social Science &   \cite{Opiyo.2019}  & Y   &     &     & b   &   N  &     &  \\
    34  & The Role of Household Consumers in Adopting Renewable Energy Technologies in Kenya & 2019 & Sustainability &  \cite{Kiprop.2019}   & N   &     &     &     &  N   &     &  \\
    35  & Factors Affecting Sustainable Market Acceptance of Residential Microgeneration Technologies. A Two Time Period Comparative Analysis & 2019 & Energies &  \cite{Karytsas.2019}   & N   &     &     &     &   N  &     &  \\
    36  & An examination of attributes and barriers to adopt biomass and solar technology. A cross-cultural approach & 2019 & Journal of Environmental Management &  \cite{HiguerasCastillo.2019}   & Y   & Y   & Y   &     & Y   & N & N \\
    37  & Forecasting Annual Solar PV Capacity Installation in Thailand Residential Sector: A User Segmentation Approach & 2019 & Engineering Journal &   \cite{Fachrizal.2019}  & Y   & Y   & N   &     &  N   &     &  \\
    38  & Merging Observed and Self-Reported Behaviour in Agent-Based Simulation: A Case Study on Photovoltaic Adoption & 2019 & Applied Sciences &  \cite{Borghesi.2019}   & Y   & N   &     &     &  N   &     &  \\
    39  & Public perception toward residential solar panels in Bahrain & 2019 & Energy Reports &   \cite{Alsabbagh.2019}  & Y   & Y   & N   &     &   N  &     &  \\
    40  & Climate Change Awareness and Solar Energy Adoption of Household & 2019 & International Journal of Advanced Science and Technology &  \cite{Zainudina.2019}   & Y   & Y   & Y   &     & Y   & N & N \\
    41  & Engendering an inclusive low-carbon energy transition in Japan: Considering the perspectives and awareness of the energy poor & 2019 & Energy Policy &   \cite{Chapman.2019}  & Y   & Y   & N   &     &  N   &     &  \\
    42  & The use and determinants of solar energy by Sub-Saharan African households & 2018 & International Journal of Sustainable Energy &   \cite{Rahut.2018}  & Y   & Y   &     & b   &   N  &     &  \\
    43  & Predicting intention to adopt solar technology in Canada: The role of knowledge, public engagement, and visibility & 2018 & Energy Policy & \cite{Parkins.2018}    & Y   & Y   & Y   &     & Y   & N & Y \\
    44  & Motivational drivers for the adoption of green energy: The case of purchasing photovoltaic systems & 2018 & Management Research Review &   \cite{Arroyo.2019}  & Y   & Y   & Y   &     & Y   & N & Y \\
    45  & A Systems Analysis of Factors Influencing Household Solar PV Adoption in Santiago, Chile & 2018 & Sustainability &   \cite{Walters.2018b}  & Y   & N   &     &     &   N  &     &  \\
    46  & Consumer preferences and electricity pricing reform in Western Australia & 2018 & Utilities Policy &  \cite{Tayal.2018}   & Y   & Y   & N   &     &   N  &     &  \\
    47  & Towards the establishment of renewable energy technologies' market: An assessment of public acceptance and use in Pakistan & 2018 & Journal of Renewable and Sustainable Energy &   \cite{Shakeel.2018}  & N   &     &     &     &  N   &     &  \\
    48  & Strike while the rebate is hot: Savvy consumers and strategic technology adoption timing & 2018 & Energy Policy &   \cite{Reeves.2018}  & Y   & Y   & N   &     &   N  &     &  \\
    49  & Shifting practices: How the rise of rooftop solar PV has changed local government community engagement & 2018 & Cogent Environmental Science &   \cite{Meiklejohn.2018}  & Y   & N   &     &     &  N   &     &  \\
    50  & Why homeowners strive for energy self-supply and how policy makers can influence them & 2018 & Energy Policy &  \cite{Engelken.2018}   & N   &     &     &     &   N  &     &  \\
    51  & Geography, community, household: Adoption of distributed solar power across India & 2018 & Energy for Sustainable Development &  \cite{Aklin.2018}   & Y   & Y   & N   &     &   N  &     &  \\
    52  & Good things come in small packages: is there a common set of motivators for energy behaviour? & 2018 & Energy Efficiency &   \cite{Stikvoort.2018}  & Y   & Y   & Y   &     & Y   & N &  N \\
    53  & Clean, accessible, and cost-saving: Reasons for rural household investment in solar panels in Poland & 2018 & Resources, Conservation \& Recycling &  \cite{Klepacka.2018}   & Y   & Y   & N   &     &   N  &     &  \\
    54  & Shotgun or snowball approach? Accelerating the diffusion of rooftop solar photovoltaics through peer effects and social norms & 2018 & Energy Policy &  \cite{Curtius.2018b}   & Y   & Y   & Y   &     & Y   & N &  N \\
    55  & The technology of the middle class: Understanding the fulfillment of adoption intentions in Queensland's rapid uptake residential solar photovoltaics market & 2018 & Renewable and Sustainable Energy Reviews & \cite{Bondio.2018}    & Y   & Y   & Y   &     & Y   & N &  N \\
    56  & The analysis of demographics, environmental and knowledge factors affecting prospective residential PV system adoption: A study in Tehran & 2018 & Renewable and Sustainable Energy Reviews &  \cite{Bashiri.2018}   & Y   & Y   & Y   &     & Y   & N & N \\
    57  & Explaining interest in adopting residential solar photovoltaic systems in the United States: Toward an integration of behavioral theories & 2017 & Energy Research \& Social Science &  \cite{Wolske.2017}   & Y   & Y   & Y   &     & Y   & Y &  \\
    58  & Decision-making governance for purchases of solar photovoltaic systems in Japan & 2017 & Energy Policy &  \cite{Tanaka.2017}   & Y   & Y   & N   &     &  N   &     &  \\
    59  & Testing Diffusion of Innovations Theory with data: Financial incentives, early adopters, and distributed solar energy in Australia & 2017 & Energy Research \& Social Science &  \cite{Simpson.2017}   & Y   & Y   & N   &     &   N  &     &  \\
    60  & Constraints on Use of Renewable Energie Technologies in the Rural Area: A Case STudy from the North-West Region of Romania & 2017 & Journal of Environmental Protection and Ecology &   \cite{Muresan.2017}  & N   &     &     &     &   N  &     &  \\
    61  & Individual and collective socio-psychological patterns of photovoltaic investment under diverging policy regimes of Austria and Italy & 2017 & Energy Policy &   \cite{Braito.2017}  & Y   & Y   & N   &     &   N  &     &  \\
    62  & Study on the factors affecting willingness of rural household's to adoption of solar lighting systems & 2017 & Journal of Advanced Research in Dynamical and Control Systems &  \cite{Pandey.2017}   & N   &     &     &     &   N  &     &  \\
    63  & Motives to adopt renewable electricity technologies: Evidence from Sweden & 2017 & Energy Policy &  \cite{Bergek.2017}   & Y   & Y   & N   &     &  N   &     &  \\
    64  & When households go solar: Determinants of uptake of a Photovoltaic Scheme and policy insights & 2017 & Energy Policy &   \cite{Briguglio.2017}  & Y   & N   &     &     & N    &     &  \\
    65  & Factors Influencing Malaysian Consumers’ Intention to Purchase Green Energy: The Case of Solar Panel & 2017 & GLobal Business and Management Research: An International Journal &  \cite{AzizN.S..2017}   & Y   & Y   & Y   &     & Y   & Y &  \\
    66  & Installation of Residential Energy Systems: Local Conditions and Residents' Willingness & 2016 & Journal of Asian Architecture and Building Engineering &  \cite{Yuasa.2016}   & N   &     &     &     &   N  &     &  \\
    67  & Overcoming barriers and uncertainties in the adoption of residential solar PV & 2016 & Renewable Energy &  \cite{Rai.2016}   & Y   & Y   & N   &     &  N   &     &  \\
    68  & Heterogeneity in the adoption of photovoltaic systems in Flanders & 2016 & Energy Economics &  \cite{Groote.2016}   & Y   & N   &     &     &   N  &     &  \\
    69  & Diffusion into new markets: evolving customer segments in the solar photovoltaics market & 2015 & Environmental Research Letters &   \cite{Sigrin.2015}  & Y   & Y   & N   &     &   N  &     &  \\
    70  & Public perceptions and information gaps in solar energy in Texas & 2015 & Environmental Research Letters &  \cite{Rai.2015b}   & Y   & Y   & Y   &     & Y   & N & Y \\
    71  & Intentions to adopt photovoltaic systems depend on homeowners' expected personal gains and behavior of peers & 2015 & Renewable Energy &   \cite{Korcaj.2015}  & Y   & Y   & Y   &     & Y   & N & N \\
    72  & Opinion leadership and willingness to pay for residential photovoltaic systems & 2015 & Energy Policy &  \cite{Yamamoto.2015}   & Y   & Y   & N   &     &  N   &     &  \\
    73  & The emperor and the cowboys: The role of government policy and industry in the adoption of domestic solar microgeneration systems & 2015 & Energy Policy &   \cite{Simpson.2015}  & Y   & Y   & N   &     &   N  &     &  \\
    74  & Consumer attitudes towards renewable energy in China—The case of Shanghai & 2015 & Sustainable Cities and Society &   \cite{Hast.2015}  & N   &     &     &     &   N  &     &  \\
    75  & Assessing the Determinants of Renewable Electricity Acceptance Integrating Meta-Analysis Regression and a Local Comprehensive Survey & 2015 & sustainability &  \cite{Bigerna.2015}   & N   &     &     &     &   N  &     &  \\
    76  & A segmentation analysis: the case of photovoltaic in the Netherlands & 2015 & Energy Efficiency &  \cite{Vasseur.2015b}   & Y   & Y   & N   &     &   N  &     &  \\
    77  & Regional distribution of photovoltaic deployment in the UK and its determinants: A spatial econometric approach & 2015 & Energy Economics & \cite{BaltaOzkan.2015}    & Y   & N   &     &     &  N   &     &  \\
    78  & The adoption of PV in the Netherlands: A statistical analysis of adoption factors & 2015 & Renewable and Sustainable Energy Reviews &   \cite{Vasseur.2015}  & Y   & Y   & Y   &     & Y   & N &  N \\
    79  & Beyond the sun—Socioeconomic drivers of the adoption of small-scale photovoltaic installations in Germany & 2015 & Energy Research and Social Science &   \cite{Schaffer.2015}  & Y   & N   &     &     &  N   &     &  \\
    80  & Motivators for adoption of photovoltaic systems at grid parity: A case study from Southern Germany & 2015 & Renewable and Sustainable Energy Reviews &  \cite{Karakaya.2015}   & Y   & N   &     &     &   N  &     &  \\
    81  & Public attitudes towards photovoltaic developments: Case study from Greece & 2014 & Energy Policy &  \cite{Tsantopoulos.2014}   & Y   & Y   & Y   &     & Y   & N & N \\
    82  & Assessing the effects of customer innovativeness, environmental value and ecological lifestyles on residential solar power systems install intention & 2014 & Energy Policy &  \cite{Chen.2014}   & Y   & Y   & Y   &     & Y   & Y &  \\
    83  & Investigating the importance of motivations and barriers related to microgeneration uptake in the UK & 2014 & Applied Energy &  \cite{Balcombe.2014}   & N   &     &     &     &  N   &     &  \\
    84  & Socioeconomic and demographic factors that influence publics' awareness on the different forms of renewable energy sources & 2014 & Renewable Energy &  \cite{Karytsas.2014}   & N   &     &     &     &  N   &     &  \\
    85  & Early Adopters of Solar Panels in Developing Countries: Evidence from Tanzania & 2014 & Review of Policy Research &   \cite{Smith.2014}  & Y   &     &     & b   &   N  &     &  \\
    86  & Residential solar electricity adoption: What motivates, and what matters? A case study of early adopters & 2014 & Energy Research and Social Science &   \cite{Schelly.2014}  & Y   & N   &     &     &   N  &     &  \\
    87  & Small-scale households renewable energy usage intention: Theoretical development and empirical settings & 2014 & Renewable Energy &   \cite{Alam.2014}  & N   &     &     &     &   N  &     &  \\
    88  & Do Incentives Work? An Analysis of Residential Solar Energy Adoption in Miami-Dade County, Florida & 2014 & Southeastern Geographer &  \cite{VarelaMargolles.2014}   & N   &     &     &     &   N  &     &  \\
    89  & Effective information channels for reducing costs of environmentally- friendly technologies: evidence from residential PV markets & 2013 & Environmental Research Letters &   \cite{Rai.2013b}  & Y   & Y   & N   &     &  N   &     &  \\
    90  & Comparing recent views of public attitude on wind energy, photovoltaic and small hydro applications & 2013 & Renewable Energy &   \cite{Kaldellis.2013}  & N   &     &     &     &   N  &     &  \\
    91  & Farmers' willingness to convert traditional houses to solar houses in rural areas: A survey of 465 households in Chongqing, China & 2013 & Energy Policy &  \cite{Li.2013}   & N   &     &     &     &  N   &     &  \\
    92  & Renewables in the energy transition: Evidence on solar home systems and lighting fuel choice in Kenya & 2013 & Energy Economics &  \cite{Lay.2013}   &     &     &     & b   &  N   &     &  \\
    93  & The adoption of photovoltaic systems in Wiesbaden, Germany & 2013 & Economics of Innovation and New Technology &   \cite{Muller.2013}  & Y   & N   &     &     &   N  &     &  \\
    94  & Understanding the Attitude-Behavior Gap for Renewable Energy Systems Using Behavioral Reasoning Theory & 2013 & Journal of Macromarketing &   \cite{Claudy.2013}  & Y   & Y   & Y   &     & Y   & Y &  \\
    95  & Analyzing consumer acceptance of photovoltaics (PV) using fuzzy logic model & 2012 & Renewable Energy &  \cite{Zhai.2012}   & Y   & Y   & N   &     &   N  &     &  \\
    96  & Consumer responses towards home energy financial incentives: A survey-based study & 2012 & Energy Policy &   \cite{Zhao.2012}  & Y   & Y   & N   &     &  N   &     &  \\
    97  & The diffusion of solar energy use in HK: What are the barriers? & 2012 & Energy Policy &   \cite{Zhang.2012}  & Y   & Y   & N   &     &   N  &     &  \\
    98  & Lifestyle practices and pro-environmental technology & 2012 & Ecological Economics &   \cite{Axsen.2012b}  & N   &     &     &     &  N   &     &  \\
    99  & Own power: Motives of having electricity without the energy company & 2011 & Energy Policy & \cite{Leenheer.2011}    & N   &     &     &     &  N   &     &  \\
    100 & Residents’ preference of solar access in high-density sub-tropical cities & 2011 & Solar Energy &   \cite{Lau.2011}  & N   &     &     &     &   N  &     &  \\
    101 & Motives for and barriers to household adoption of small-scale production of electricity: examples from Sweden & 2011 & Sustainability: Science, Practice and Policy &  \cite{Palm.2011}   & N   &     &     &     &   N  &     &  \\
    102 & Factors influencing energy efficiency investments in existing Swedish residential buildings & 2010 & Energy Policy &  \cite{Nair.2010}   & N   &     &     &     &  N   &     &  \\
    103 & Determinants of pro-environmental consumption: The role of reference groups and routine behavior & 2009 & Ecological Economics &  \cite{Welsch.2009}   & N   &     &     &     &   N  &     &  \\
    104 & A survey of solar PV program implementers in Asia and the Pacific regions & 2009 & Energy for Sustainable Development &   \cite{Urmee.2009}  & Y   & Y   & N   &     &  N   &     &  \\
    105 & Improving the energy performance of UK households: Results from surveys of consumer adoption and use of low- and zero-carbon technologies & 2008 & Energy Efficiency &  \cite{Caird.2008}   & N   &     &     &     &    N &     &  \\
    106 & Assessment of Non-Technical Barriers for Widespread Adoption of Building Integrated Photovoltaic System (BIPV) in Malaysian Urban Residential Sector & 2008 & Conference on Passive and Low Energy Architecture &   \cite{Haw.2008}  & Y   & Y   & N   &     &  N   &     &  \\
    107 & The adoption of domestic solar-power systems: Do consumers assess product attributes in a stepwise process? & 2007 & Energy Policy &  \cite{Faiers.2007b}   & N   &     &     &     &  N   &     &  \\
    108 & Stimulating the diffusion of photovoltaic systems: A behavioural perspective & 2006 & Energy Policy &   \cite{Jager.2006}  & Y   & Y   & N   &     &  N   &     &  \\
    109 & Consumer attitudes towards domestic solar power systems & 2006 & Energy Policy &   \cite{Faiers.2006}  & Y   & Y   & N   &     &  N   &     &  \\
    110 & Exploring the Consumer Decision Process in the Adoption of Solar Energy Systems & 1981 & Journal of Consumer Research &  \cite{Labay.1981}   & N   &     &     &     &  N   &     &  \\

\end{longtable}
\end{ThreePartTable}

\newpage
\begin{landscape}
\begin{ThreePartTable}
    \tiny
    \renewcommand{\arraystretch}{1.0} 
    \setlength{\tabcolsep}{4pt}


\begin{longtable}{p{0.1\linewidth} p{0.03\linewidth} p{0.15\linewidth} p{0.1\linewidth} p{0.065\linewidth} p{0.02\linewidth} p{0.32\linewidth} p{0.13\linewidth} p{0.05\linewidth}}

    \caption{List of 24 studies identified as relevant for meta-analysis in the screening procedure. Due to lacking reporting of data, only 8 studies could be included in the final analysis.}\\
    \label{tab:lit_select}\\
    \toprule
    Authors & Year & Title & Journal & Region & N & Main Objective & Dep. variable & data source \\
    \midrule
    \endfirsthead
    
    \caption*{List of 24 studies identified as relevant for meta-analysis (continued)}\\
    \toprule
    Authors & Year & Title & Journal & Region & N & Main Objective & Dep. variable & data source\\
    \midrule
    \endhead

    \endfoot

    \bottomrule
    \endlastfoot
    
    \multicolumn{9}{l}{\textbf{Included in meta-analysis}} \\
    \midrule
    Pi-Chuan Sun, Hsueh-Mei Wang, Hsien-Long Huang, Chien-Wei Ho & 2020  & Consumer attitude and purchase intention toward rooftop photovoltaic installation: The roles of personal trait, psychological benefit, and government incentives & Energy \& Environment & Taiwan & 300   & "examine the relationship among personal traits (including environmental concern, an ecological lifestyle, and consumer innovativeness), psychological benefits (including a warm glow and a “nature experience”), attitudes toward rooftop photovoltaic, government incentives, and intentions to install rooftop photovoltaic" & rooftop photovoltaic installation intention & paper   \\
    \hline
    Pilar Arroyo, Lorena Carrete & 2018  & Motivational drivers for the adoption of green energy: The case of purchasing photovoltaic systems & Management Research Review & Mexico & 72    & "propose and empirically test a model where different motivational drivers are used to stimulate the intention of individuals to purchase green energy" & intention to purchase a photovoltaic system in short and medium terms       & request \\
    \hline
    John R. Parkins, Curtis Rollins, Sven Anders, Louise Comeau & 2018  & Predicting intention to adopt solar technology in Canada: The role of knowledge, public engagement, and visibility & Energy Policy & Canada & 2065  & provide insights if the combined experience of  perceived knowledge, public engagement, and visual exposure to renewable energy technology influences adoption intention of solar technologies & solar adoption intention   & request \\
    \hline
    Nik Salehah Nik Abdul Aziz, Nabsiah Abdul Wahid, Methaq Ahmed Sallam, Shaizatulaqma Kamarul Ariffin & 2017  & Factors Influencing Malaysian Consumers’ Intention to Purchase Green Energy: The Case of Solar Panel & Global Business and Management Research: An International Jounal & Malaysia & 211   & "study investigates whether Malaysian consumers’ intention to purchase solar panel\newline{}are influenced by perceived government policy, perceived costs and maintenance, product knowledge and experience, solar panel aesthetics, social influence, environmental concern, product benefits and demographic factors (education level, income)" & consumer’s solar panel purchase intention & paper  \\
    \hline
    Kimberly S. Wolske, Paul C. Stern, Thomas Dietz & 2017  & Explaining interest in adopting residential solar photovoltaic systems in the United States: Toward an integration of behavioral theories & Energy Research \& Social Science & United States (Arizona, California, New Yersey, New York) & 904   & "offer a better understanding of what factors influence interest in contacting an RPV [residential photovoltaics] installer, a key precursor to actual adoption" & interest in pursuing residential solar & paper       \\
    \hline
    Varun Rai, Ariane L Beck & 2015  & Public perceptions and information gaps in solar energy in Texas & Environmental Research Letters & United States (Texas) & 522   & "investigates the behavioral, normative, and control factors affecting intentions and behavior related to residential solar PV" & intentions to consider installing solar,  intentions to call a solar installer          & request \\
    \hline
    Kee Kuo Chen & 2014  & Assessing the effects of customer innovativeness, environmental value and ecological lifestyles on residential solar power systems install intention & Energy Policy & Taiwan & 203   & "proposes a model integrating customer innovativeness in the relationships between ecological lifestyle, environmental value and SPS install intention" & customer intention to install solar power system (SPS) in their private houses & paper      \\
    \hline
    Marius C. Claudy, Mark Peterson, Aidan O’Driscoll & 2013  & Understanding the Attitude-Behavior Gap for Renewable Energy Systems Using Behavioral Reasoning Theory & Journal of Macromarketing & Ireland & 254   & "examine[…] a proposed model in which reasons both for and against adopting solar panels mediate the relationship between consumers’ attitudes, values and adoption intentions" & adoption intention of solar energy panels & paper      \\
    \midrule
    \multicolumn{9}{l}{\textbf{Excluded from meta-analysis due to lacking target data}}\\
    \midrule
    Todd L. Cherry, Hanne Sæle & 2020  & Residential Photovoltaic Systems in Norway: Household Knowledge, Preferences and Willingness to Pay & Challenges in Sustainability & Norway & 1000  & "This study seeks to contribute to this need by investigating consumer preferences related to PV prosumer adoption; thereby offering new insights that can inform more effective and efficient PV programs and policies." & consideration of a solar system, WTP &         \\
    \hline
    Lin-Sea Lau, Yuen-Onn Choong, Chooi-Yi Wei, Ai-Na Seow , Chee-Keong Choong, Abdelhak Senadjki, Suet-Ling Ching & 2020  & Investigating nonusers’ behavioural intention towards solar photovoltaic technology in Malaysia: The role of knowledge transmission and price value & Energy Policy & Malaysia & 392   & "in an attempt to understand nonusers’ behavioural intention in adopting solar PV technology. This study aimed to examine how knowledge, price value, social influence, and facilitating conditions contribute to the behavioural intention of using solar PV systems" & intention to use solar energy &         \\
    \hline
    Chandan Parsad, Shashank Mittal and Raveesh Krishnankutty & 2020  & A study on the factors affecting household solar adoption in Kerala, India & International Journal of Productivity and Performance Management & India & (1) 201; (2) 331 & "Building upon behavioural finance and institutional theory, we posit that, in addition to a rational evaluation of the economics of the investment opportunities, various nonfinancial factors affect the household’s decision to invest in renewables. We analyse the investment decisions of a large sample of household investors, to identify the main determinants of their choices." & investment in rooftop PV for households adoption intention &         \\
    \hline
    Kerstin K. Zander & 2020  & Unrealised opportunities for residential solar panels in Australia & Energy Policy & Australia & 1126  & "This study therefore aims to 1) assess the relative importance of different motivations for the adoption of residential rooftop solar panels, 2) compare these motivations between those who have already adopted them and those who intend to do so, and 3) assess the relative importance of barriers for not adopting solar panels." & intention to install solar panels &         \\
    \hline
    Joana Abreu, Nathalie Wingartz, Natasha Hardy & 2019  & New trends in solar: A comparative study assessing the attitudes towards the adoption of rooftop PV  & Energy Policy & United States & 400   & "The study has two primary research objectives: a) to detect early in the development of the product whether the survey participants reveal more concern toward the adhesive technology, and b) to observe if the adhesive technology can address some concerns traditionally raised by solar adopters (e.g., reliability, attractiveness), and increase purchase intentions." & intentions to purchase solar PV       &  \\
    \hline
    Ashwini Kumar Aggarwal, Asif Ali Syed, Sandeep Garg & 2019  & Diffusion of residential RT solar – is lack of funds the real issue? & International Journal of Energy Sector Management & India & 405   & "The purpose of this paper is to examine the triggers of the PI of the residential RT solar buyer and to develop an empirically valid, robust behavioral model covering these relationships. As a by-product, in the analysis, the role of financial self-efficacy is also discussed." & purchase intention (PI) for residential rooftop solar PV (SPV)        &  \\
    \hline
    Rohan Best , Paul J. Burke, Shuhei Nishitateno & 2019  & Understanding the determinants of rooftop solar installation: evidence from household surveys in Australia & Australian Journal of Agricultural and Resource Economics & Australia & 11978; 17768 & "we use household-level data to identify economic, social, and environmental factors that influence actual uptake and the intention to install." & households’ intention to install solar PV &         \\
    \hline
    Elena Higueras-Castillo, Francisco Muñoz-Leiva, Francisco José Liébana-Cabanillas & 2019  & An examination of attributes and barriers to adopt biomass and solar technology. A cross-cultural approach & Journal of Environmental Management & Spain, Germany, Mexico. & 489   & "This research assesses consumers' perceptions of the main characteristics of renewable energy sources (biomass and solar) that influence their adoption. The study is focused on cultural differences in the adoption of clean energies between three regions with a different level of collectivism: Spain, Germany and Mexico. Furthermore, it examines pro-environmental behavior and New Ecological Paradigm (NEP) scale." & intention to adopt renewable energy (biomass, solar panels) &         \\
    \hline
    Norzalina Zainudina, Zuroni Md Jusoha, Zumilah Zainalaludina, Syuhaili Osmana, Nurnaddia Nordinb, Laily Hj Paima and Abinasan Ramasamya & 2019  & Climate Change Awareness and Solar Energy Adoption of Household & International Journal of Advanced Science and Technology & Malaysia & 120   & "this paper was designed to have better understanding on what motives consumers to make buy solar photovoltaic products and what discourages them from doing so as the first step in promoting green buying. Therefore study aim to have a better understanding on consumer acceptance on solar photovoltaic products" & consumer's intention to buy (BI) solar photovoltaic        &  \\
    \hline
    Ali Bashiri, Sasan H. Alizadeh & 2018  & The analysis of demographics, environmental and knowledge factors affecting prospective residential PV system adoption: A study in Tehran & Renewable and Sustainable Energy Reviews & Iran  & 345   & "This study aims to evaluate those factors which affect adoption of photovoltaic systems by taking into consideration Tehran's unique circumstances, air pollution and high-density of population, low price of energy and governmental financial supports. Moreover, families who are more likely to adopt these systems are identified." & tendency to adopt photovoltaic systems &         \\
    \hline
    Steven Bondio, Mahdi Shahnazari, Adam McHugh & 2018  & The technology of the middle class: Understanding the fulfilment of adoption intentions in Queensland's rapid uptake residential solar photovoltaics market & Renewable and Sustainable Energy Reviews & Australia & 647   & "The overarching research objective that led to the production of this paper was to empirically examine the behavioural drivers of fulfilling an intention to adopt residential PV by households as decision-making units. The analysis investigated the differences in demographics and motivational factors among households who intended to purchase solar PV and those who had already adopted the technology." & intention to purchase PV by households &         \\
    \hline
    Hans Christoph Curtius, Stefanie Lena Hille, Christian Berger, Ulf Joachim Jonas Hahnel, Rolf Wüstenhagen & 2018  & Shotgun or snowball approach? Accelerating the diffusion of rooftop solar photovoltaics through peer effects and social norms & Energy Policy & Switzerland & 410   & In our analysis of behavioral factors driving PV diffusion, we confirm the existence of peer effects and draw on the social psychology literature to investigate how two types of social norms shape homeowners’ intentions to install solar panels. We find that both descriptive norms [...] and injunctive norms [...] play a role in potential PV adopters’ decision-making, and discuss how post-grid parity PV policy can be designed to facilitate an accelerated diffusion of solar photovoltaics." & homeowners' intention to install PV &        \\
    \hline
    Britt Stikvoort, Peter Juslin, Cajsa Bartusch & 2018  & Good things come in small packages: is there a common set of motivators for energy behaviour? & Energy Efficiency & Sweden & 83    & "The specific aim of this current survey study was thus to investigate the (dis)similarity of the role of a specific set of determinants (defined later on) when looking at different energy behaviours: investing in solar panels, turning off appliances that are on standby, reducing the time spent showering and replacing old household appliances with new energy-efficient ones." & Investing in photovoltaic (PV) panels &         \\
    \hline
    Liridon Korcaj, Ulf J.J. Hahnel, Hans Spada & 2015  & Intentions to adopt photovoltaic systems depend on homeowners' expected personal gains and behavior of peers & Renewable Energy & Germany & 200   & "To promote adoption in the future absence of this tariff, we explored further motives of homeowners relevant to PV system purchase intention." & purchase intention of a PV system &        \\
    \hline
    Véronique Vasseur, René Kemp & 2015  & The adoption of PV in the Netherlands: A statistical analysis of adoption factors & Renewable and Sustainable Energy Reviews & Netherlands & 779   & "This paper tests theoretical propositions about adoption variables for the case of solar PV in the Netherlands and offers a novel contribution to the literature through the use of a segmentation model." & intention to adopt solar PV &        \\
    \hline
    Georgios Tsantopoulos, Garyfallos Arabatzis, Stilianos Tampakis & 2014  & Public attitudes towards photovoltaic developments: Case study from Greece & Energy Policy & Greece & 1068  & "The purpose of this study is to outline the views of Greek citizens on a series of issues related to investments in photovoltaic systems. The data, collected through a structured questionnaire, are important due to the fact that they depict the attitude of Greek citizens before governmental decisions were taken to suspend applications for new licenses." & willingness to invest in residential photovoltaic systems &        \\
\end{longtable}%
\end{ThreePartTable}
\end{landscape}

\newpage

\begin{table}[ht]
    \tiny
    \centering
    \renewcommand{\arraystretch}{1.2}
    \setlength{\tabcolsep}{3pt}
\begin{threeparttable}
    \caption{Meta-analytically pooled refined correlation table with values transformed back to r-metric}
    \label{tab:cor_refi}
\begin{tabular}{lllllllllll}
    \toprule
          & \textbf{INT} & \textbf{EC} & \textbf{NS} & \textbf{HBA} & \textbf{SBA} & \textbf{PBEN} & \textbf{EBEN} & \textbf{SN} & \textbf{GEN} & \textbf{EDU} \\
    \midrule
    \textbf{EC} & .34 (p=.001) &     &     &     &     &     &     &     &     &  \\
        & [.14; .52], 7 &     &     &     &     &     &     &     &     &  \\
    \textbf{NS} & .47 (p\textless.001) & .44 (p\textless.001) &     &     &     &     &     &     &     &  \\
        & [.21; .67], 4 & [.35; .53], 4 &     &     &     &     &     &     &     &  \\
    \textbf{HBA} & -.18 (p\textless.001) & -.09 (p=.009) & -.13 (p\textless.001) &     &     &     &     &     &     &  \\
        & [-.24; -.13], 2 & [-.15; -.02], 1 & [-.19; -.06], 1 &     &     &     &     &     &     &  \\
    \textbf{SBA} & -.08 (p=.284) & -.17 (p=.126) & .05 (p=.122) & .34 (p\textless.001)  &     &     &     &     &     &  \\
        & [-.21; .06], 4 & [-.36; .05], 3 & [-.01; .12], 1 & [.27; .41], 2 &     &     &     &     &     &  \\
    PBEN & .54 (p\textless.001)  & .65 (p\textless.001)  & .63 (p\textless.001)  & -.18 (p\textless.001) & -.12 (p\textless.001) &     &     &     &     &  \\
        & [.32; .71], 5 & [.41; .81], 4 & [.33; .82], 3 & [-.24; -.13], 2 & [-.17; -.06], 3 &     &     &     &     &  \\
    EBEN & .32 (p=.03) & .77 (p\textless.001)  & .35 (p\textless.001)  & -.24 (p\textless.001) & -.15 (p\textless.001) & .74 (p\textless.001)  &     &     &     &  \\
        & [.03; .55], 2 & [.75; .8], 1 & [.29; .4], 1 & [-.33; -.15], 2 & [-.21; -.1], 2 & [.71; .77], 2 &     &     &     &  \\
    \textbf{SN} & .33 (p\textless.001)  & .28 (p\textless.001)  & .5 (p=.035) & -.33 (p\textless.001) & -.1 (p=.455) & .49 (p\textless.001)  & .37 (p\textless.001)  &     &     &  \\
        & [.17; .46], 4 & [.13; .42], 4 & [.04; .79], 2 & [-.39; -.27], 1 & [-.36; .17], 2 & [.25; .68], 3 & [.31; .43], 1 &     &     &  \\
    \textbf{GEN} & -.01 (p=.645) & .05 (p=.019) & 0 (p\textless.001) & 0 (p\textless.001) & -.04 (p=.752) & 0 (p\textless.001) & 0 (p\textless.001) & -.06 (p=.007) &     &  \\
        & [-.05; .03], 2 & [.01; .09], 2 & [0; 0], 0 & [0; 0], 0 & [-.27; .2], 1 & [0; 0], 0 & [0; 0], 0 & [-.1; -.02], 1 &     &  \\
    \textbf{EDU} & .05 (p=.532) & .05 (p=.14)  & 0 (p\textless.001) & 0 (p\textless.001) & -.03 (p=.565) & -.01 (p=.931) & 0 (p\textless.001) & .07 (p=.001) & -.09 (p\textless.001) &  \\
        & [-.1; .19], 3 & [-.02; .11], 3 & [0; 0], 0 & [0; 0], 0 & [-.15; .08], 2 & [-.14; .13], 1 & [0; 0], 0 & [.03; .11], 2 & [-.13; -.04], 2 &  \\
    \textbf{INC} & .18 (p=.17)  & .15 (p=.008) & 0 (p\textless.001) & 0 (p\textless.001) & 0 (p=.979) & .08 (p=.219) & 0 (p\textless.001) & .04 (p=.443) & -.1 (p\textless.001)  & .19 (p=.152) \\
        & [-.08; .42], 3 & [.04; .26], 3 & [0; 0], 0 & [0; 0], 0 & [-.16; .17], 2 & [-.05; .22], 1 & [0; 0], 0 & [-.06; .13], 2 & [-.14; -.05], 2 & [-.07; .44], 3 \\
    \bottomrule
\end{tabular}%
\begin{tablenotes}
    \item Upper numbers: Pearson's r (significance level)
    \item Lower numbers: [lower and upper 95\% CI], number of studies
    \item INT Intention; EC Environmental concern; NS Novelty Seeking; HBA Hard Barriers; SBA Soft Barriers; PBE Personal Benefits; EBE Environmental Benefits; SN Subjective Norm; GEN Gender; EDU Education; INC Income
\end{tablenotes}
\end{threeparttable}
\end{table}%

\newpage
\begin{table}[h]
    \tiny
    \centering
    \renewcommand{\arraystretch}{1}
    \setlength{\tabcolsep}{2.5pt}
    \caption{Parsimonious input correlation table with original and composed correlations, correlations provided in r-metric}
\begin{tabularx}{\textwidth}{p{0.18\textwidth} p{0.22\textwidth} p{0.065\textwidth} p{0.065\textwidth} p{0.065\textwidth} p{0.065\textwidth} p{0.065\textwidth} p{0.065\textwidth} p{0.065\textwidth} p{0.065\textwidth} p{0.065\textwidth} p{0.065\textwidth}}
          &       & \textbf{INT} & \textbf{EC} & \textbf{NS} & \textbf{BA} & \textbf{BE} & \textbf{SN} & \textbf{GEN} & \textbf{EDU} \\
    \midrule
    \multicolumn{10}{l}{\textbf{Environmental concern}} \\
    \midrule
    \citep{Sun.2020} & Environmental concern & \multirow{2}[2]{*}{.632} & \multicolumn{7}{c}{\multirow{2}[2]{*}{}} \\
    \citep{Sun.2020} & Ecological lifestyle &       & \multicolumn{7}{c}{} \\
    \midrule
    \citep{Rai.2015b} & Environmental Concern & .187 &       &       &       &       &       &       &  \\
    \midrule
    \citep{Chen.2014} & Environmental value & \multirow{2}[2]{*}{.640} & \multicolumn{7}{c}{\multirow{2}[2]{*}{}} \\
    \citep{Chen.2014} & Ecological lifestyle &       & \multicolumn{7}{c}{} \\
    \midrule
    \citep{Arroyo.2019} & Environmental conciousness & .046 &       &       &       &       &       &       &  \\
    \midrule
    \citep{Parkins.2018} & Environmental Values & .048 &       &       &       &       &       &       &  \\
    \midrule
    \citep{AzizN.S..2017} & Environmental concern & .354 &       &       &       &       &       &       &  \\
    \midrule
    \citep{Wolske.2017} & Awareness of consequences (AC) & \multirow{2}[2]{*}{.324} & \multicolumn{7}{c}{\multirow{2}[2]{*}{}} \\
    \citep{Wolske.2017} & Personan norm to act (PN) &       & \multicolumn{7}{c}{} \\
    \midrule
    \multicolumn{10}{l}{\textbf{Novelty seeking}} \\
    \midrule
    \citep{Sun.2020} & Consumer innovativeness & .75  & .581 &       &       &       &       &       &  \\
    \midrule
    \citep{Rai.2015b} & Personal Norm & .324 & .412 &       &       &       &       &       &  \\
    \midrule
    \citep{Chen.2014} & Novelty seeking & .4   & .379 &       &       &       &       &       &  \\
    \midrule
    \citep{Wolske.2017} & Consumer novelty seeking (CNS) & \multirow{2}[2]{*}{.324} & \multirow{2}[2]{*}{.395} & \multicolumn{6}{c}{\multirow{2}[2]{*}{}} \\
    \citep{Wolske.2017} & Openness &       &       & \multicolumn{6}{c}{} \\
    \midrule
    \multicolumn{10}{l}{\textbf{Barriers}} \\
    \midrule
    \citep{Claudy.2013} & Cost Barrier & \multirow{3}[2]{*}{-.267} & \multirow{3}[2]{*}{/} & \multirow{3}[2]{*}{/} & \multicolumn{5}{c}{\multirow{3}[2]{*}{}} \\
    \citep{Claudy.2013} & Incompatibility Barrier &       &       &       & \multicolumn{5}{c}{} \\
    \citep{Claudy.2013} & Risk Barrier &       &       &       & \multicolumn{5}{c}{} \\
    \midrule
    \citep{Arroyo.2019} & Perceived barriers & -.153 & -.014 & /     &       &       &       &       &  \\
    \midrule
    \citep{AzizN.S..2017} & Perceived cost maintenance & .097 & -.359 & /     &       &       &       &       &  \\
    \midrule
    \citep{Wolske.2017} & Riskiness & \multirow{6}[2]{*}{-.117} & \multirow{6}[2]{*}{-.104} & \multirow{6}[2]{*}{-.015} & \multicolumn{5}{c}{\multirow{6}[2]{*}{}} \\
    \citep{Wolske.2017} & Expense Concerns &       &       &       & \multicolumn{5}{c}{} \\
    \citep{Wolske.2017} & Home unsuitable &       &       &       & \multicolumn{5}{c}{} \\
    \citep{Wolske.2017} & May move &       &       &       & \multicolumn{5}{c}{} \\
    \citep{Wolske.2017} & PV may improve &       &       &       & \multicolumn{5}{c}{} \\
    \citep{Wolske.2017} & Trialability &       &       &       & \multicolumn{5}{c}{} \\
    \midrule
    \multicolumn{10}{l}{\textbf{Benefits}} \\
    \midrule
    \citep{Sun.2020} & Attitude towards rooftop PV & \multirow{2}[2]{*}{.765} & \multirow{2}[2]{*}{.864} & \multirow{2}[2]{*}{.666} & \multirow{2}[2]{*}{/} & \multicolumn{4}{c}{\multirow{2}[2]{*}{}} \\
    \citep{Sun.2020} & Warm glow &       &       &       &       & \multicolumn{4}{c}{} \\
    \midrule
    \citep{Claudy.2013} & Environmental Benefit & \multirow{4}[2]{*}{.340} & \multirow{4}[2]{*}{/} & \multirow{4}[2]{*}{/} & \multirow{4}[2]{*}{-.165} & \multicolumn{4}{c}{\multirow{4}[2]{*}{}} \\
    \citep{Claudy.2013} & Independence Benefit &       &       &       &       & \multicolumn{4}{c}{} \\
    \citep{Claudy.2013} & Economic Benefit &       &       &       &       & \multicolumn{4}{c}{} \\
    \citep{Claudy.2013} & Attitude &       &       &       &       & \multicolumn{4}{c}{} \\
    \midrule
    \citep{Rai.2015b} & Attitude & .349 & .422 & .786 & /     &       &       &       &  \\
    \midrule
    \citep{AzizN.S..2017} & Product benefit & .562 & .657 & /     & -.199 &       &       &       &  \\
    \midrule
    \citep{Wolske.2017} & Personal Benefits & \multirow{2}[2]{*}{.531} & \multirow{2}[2]{*}{.704} & \multirow{2}[2]{*}{.376} & \multirow{2}[2]{*}{-.187} & \multicolumn{4}{c}{\multirow{2}[2]{*}{}} \\
    \citep{Wolske.2017} & Environmental Benefits &       &       &       &       & \multicolumn{4}{c}{} \\
    \midrule
    \multicolumn{10}{l}{\textbf{Subjective norm}} \\
    \midrule
    \citep{Rai.2015b} & Subjective Norm & \multirow{2}[2]{*}{.416} & \multirow{2}[2]{*}{.436} & \multirow{2}[2]{*}{.675} & \multirow{2}[2]{*}{/} & \multirow{2}[2]{*}{.677} & \multicolumn{3}{c}{\multirow{2}[2]{*}{}} \\
    \citep{Rai.2015b} & Descriptive Norm &       &       &       &       &       & \multicolumn{3}{c}{} \\
    \midrule
    \citep{Parkins.2018} & Regularly sees PV & .101 & .085 & /     & /     & /     &       &       &  \\
    \midrule
    \citep{AzizN.S..2017} & Social norms & .397 & .29  & /     & .042 & .341 &       &       &  \\
    \midrule
    \citep{Wolske.2017} & Observability & \multirow{2}[2]{*}{.384} & \multirow{2}[2]{*}{.315} & \multirow{2}[2]{*}{.284} & \multirow{2}[2]{*}{-.233} & \multirow{2}[2]{*}{.405} & \multicolumn{3}{c}{\multirow{2}[2]{*}{}} \\
    \citep{Wolske.2017} & Social support &       &       &       &       &       & \multicolumn{3}{c}{} \\
    \midrule
    \multicolumn{10}{l}{\textbf{Gender}} \\
    \midrule
    \citep{Arroyo.2019} & Gender & -.028 & .046 & /     & -.038 & /     & /     &       &  \\
    \midrule
    \citep{Parkins.2018} & Gender & -.009 & .081 & /     & /     & /     & -.059 &       &  \\
    \midrule
    \multicolumn{10}{l}{\textbf{Education}} \\
    \midrule
    \citep{Arroyo.2019} & Academic level & .265 & .008 & /     & .003 & /     & /     & -.031 &  \\
    \midrule
    \citep{Parkins.2018} & Education & \multirow{2}[2]{*}{.012} & \multirow{2}[2]{*}{.058} & \multirow{2}[2]{*}{/} & \multirow{2}[2]{*}{/} & \multirow{2}[2]{*}{/} & \multirow{2}[2]{*}{.071} & \multirow{2}[2]{*}{-.088} & \multirow{2}[2]{*}{} \\
    \citep{Parkins.2018} & Recode for University Education &       &       &       &       &       &       &       &  \\
    \midrule
    \citep{AzizN.S..2017} & Education & -.048 & .093 & /     & -.047 & -.006 & .036 & /     &  \\
    \midrule
    \multicolumn{10}{l}{\textbf{Income}} \\
    \midrule
    \citep{Arroyo.2019} & Socioeconomic level & .442 & .2   & /     & -.114 & /     & /     & 0     & .376 \\
    \midrule
    \citep{Parkins.2018} & Household income before taxes in 2013 & -.002 & .003 & /     & /     & /     & .069 & -.099 & .279 \\
    \midrule
    \citep{AzizN.S..2017} & Income & .142 & .199 & /     & .065 & .085 & -.036 & /     & -.068 \\
    \bottomrule
\end{tabularx}%
  \label{tab:input_parsi}%
\end{table}%

\newpage
\begin{table}
    \tiny
    \centering
    \renewcommand{\arraystretch}{1}
    \setlength{\tabcolsep}{1pt}
    \caption{Refined input correlation table with original and composed correlations, correlations provided in r-metric}
\begin{tabularx}{\textwidth}{p{0.18\textwidth} p{0.22\textwidth} p{0.055\textwidth} p{0.055\textwidth} p{0.055\textwidth} p{0.055\textwidth} p{0.055\textwidth} p{0.055\textwidth} p{0.055\textwidth} p{0.055\textwidth} p{0.055\textwidth} p{0.055\textwidth}}
          &       & \textbf{INT} & \textbf{EC} & \textbf{NS} & \textbf{HBA} & \textbf{SBA} & \textbf{PBEN} & \textbf{EBEN} & \textbf{SN} & \textbf{GEN} & \textbf{EDU} \\
    \midrule
    \multicolumn{12}{l}{\textbf{Environmental concern}} \\
    \midrule
    \citep{Sun.2020} & Environmental concern & \multirow{2}[2]{*}{.632} & \multicolumn{9}{l}{\multirow{2}[2]{*}{}} \\
    \citep{Sun.2020} & Ecological lifestyle &       & \multicolumn{9}{l}{} \\
    \midrule
    \citep{Rai.2015b} & Environmental Concern & .187 &       &       &       &       &       &       &       &       &  \\
    \midrule
    \citep{Chen.2014} & Environmental value & \multirow{2}[2]{*}{.640} & \multicolumn{9}{l}{\multirow{2}[2]{*}{}} \\
    \citep{Chen.2014} & Ecological lifestyle &       & \multicolumn{9}{l}{} \\
    \midrule
    \citep{Arroyo.2019} & Environmental conciousness & .046 &       &       &       &       &       &       &       &       &  \\
    \midrule
    \citep{Parkins.2018} & Environmental Values & .048 &       &       &       &       &       &       &       &       &  \\
    \midrule
    \citep{AzizN.S..2017} & Environmental concern & .354 &       &       &       &       &       &       &       &       &  \\
    \midrule
    \citep{Wolske.2017} & Awareness of consequences & \multirow{2}[2]{*}{.324} & \multicolumn{9}{l}{\multirow{2}[2]{*}{}} \\
    \citep{Wolske.2017} & Personan norm to act &       & \multicolumn{9}{l}{} \\
    \midrule
    \multicolumn{12}{l}{\textbf{Novelty seeking}} \\
    \midrule
    \citep{Sun.2020} & Consumer innovativeness & .75  & .581 &       &       &       &       &       &       &       &  \\
    \midrule
    \citep{Rai.2015b} & Personal Norm & .324 & .412 &       &       &       &       &       &       &       &  \\
    \midrule
    \citep{Chen.2014} & Novelty seeking & .4   & .379 &       &       &       &       &       &       &       &  \\
    \midrule
    \citep{Wolske.2017} & Consumer novelty seeking (CNS) & \multirow{2}[2]{*}{.324} & \multirow{2}[2]{*}{.395} & \multicolumn{8}{l}{\multirow{2}[2]{*}{}} \\
    \citep{Wolske.2017} & Openness &       &       & \multicolumn{8}{l}{} \\
    \midrule
    \multicolumn{12}{l}{\textbf{Hard barriers}} \\
    \midrule
    \citep{Claudy.2013} & Incompatibility Barrier & -.23 & /     & /     &       &       &       &       &       &       &  \\
    \midrule
    \citep{Wolske.2017} & Home unsuitable & \multirow{2}[2]{*}{-.169} & \multirow{2}[2]{*}{-.087} & \multirow{2}[2]{*}{-.129} & \multicolumn{7}{l}{\multirow{2}[2]{*}{}} \\
    \citep{Wolske.2017} & May move &       &       &       & \multicolumn{7}{l}{} \\
    \midrule
    \multicolumn{12}{l}{\textbf{Soft Barriers}} \\
    \midrule
    \citep{Claudy.2013} & Cost Barrier & \multirow{2}[2]{*}{-.222} & \multirow{2}[2]{*}{/} & \multirow{2}[2]{*}{/} & \multirow{2}[2]{*}{.394} & \multicolumn{6}{l}{\multirow{2}[2]{*}{}} \\
    \citep{Claudy.2013} & Risk Barrier &       &       &       &       & \multicolumn{6}{l}{} \\
    \midrule
    \citep{Arroyo.2019} & Perceived barriers & -.153 & -.014 & /     & /     &       &       &       &       &       &  \\
    \midrule
    \citep{AzizN.S..2017} & Perceived cost maintenance & .097 & -.359 & /     & /     &       &       &       &       &       &  \\
    \midrule
    \citep{Wolske.2017} & Riskiness & \multirow{4}[2]{*}{-.048} & \multirow{4}[2]{*}{-.086} & \multirow{4}[2]{*}{.052} & \multirow{4}[2]{*}{.320} & \multicolumn{6}{l}{\multirow{4}[2]{*}{}} \\
    \citep{Wolske.2017} & Expense Concerns &       &       &       &       & \multicolumn{6}{l}{} \\
    \citep{Wolske.2017} & PV may improve &       &       &       &       & \multicolumn{6}{l}{} \\
    \citep{Wolske.2017} & Trialability &       &       &       &       & \multicolumn{6}{l}{} \\
    \midrule
    \multicolumn{12}{l}{\textbf{Personal Benefits}} \\
    \midrule
    \citep{Sun.2020} & Attitude towards rooftop PV & .8   & .850 & .67  & /     & /     &       &       &       &       &  \\
    \midrule
    \citep{Claudy.2013} & Independence Benefit & \multirow{3}[2]{*}{.305} & \multirow{3}[2]{*}{/} & \multirow{3}[2]{*}{/} & \multirow{3}[2]{*}{-.182} & \multirow{3}[2]{*}{-.101} & \multicolumn{5}{l}{\multirow{3}[2]{*}{}} \\
    \citep{Claudy.2013} & Economic Benefit &       &       &       &       &       & \multicolumn{5}{l}{} \\
    \citep{Claudy.2013} & Attitude &       &       &       &       &       & \multicolumn{5}{l}{} \\
    \midrule
    \citep{Rai.2015b} & Attitude & .349 & .422 & .786 & /     & /     &       &       &       &       &  \\
    \midrule
    \citep{AzizN.S..2017} & Product benefit & .562 & .657 & /     & /     & -.199 &       &       &       &       &  \\
    \midrule
    \citep{Wolske.2017} & Personal Benefits & .55  & .541 & .353 & -.276 & -.104 &       &       &       &       &  \\
    \midrule
    \multicolumn{12}{l}{\textbf{Environmental Benefits}} \\
    \midrule
    \citep{Claudy.2013} & Environmental Benefit & .17  & /     & /     & -.23 & -.165 & .748 &       &       &       &  \\
    \midrule
    \citep{Wolske.2017} & Environmental Benefits & .44  & .773 & .348 & -.169 & -.150 & .74  &       &       &       &  \\
    \midrule
    \multicolumn{12}{l}{\textbf{Subjective norm}} \\
    \midrule
    \citep{Rai.2015b} & Subjective Norm & \multirow{2}[2]{*}{.416} & \multirow{2}[2]{*}{.436} & \multirow{2}[2]{*}{.675} & \multirow{2}[2]{*}{/} & \multirow{2}[2]{*}{/} & \multirow{2}[2]{*}{.677} & \multirow{2}[2]{*}{/} & \multicolumn{3}{l}{\multirow{2}[2]{*}{}} \\
    \citep{Rai.2015b} & Descriptive Norm &       &       &       &       &       &       &       & \multicolumn{3}{l}{} \\
    \midrule
    \citep{Parkins.2018} & Regularly sees PV & .101 & .085 & /     & /     & /     & /     & /     &       &       &  \\
    \midrule
    \citep{AzizN.S..2017} & Social norms & .397 & .29  & /     & /     & .042 & .341 & /     &       &       &  \\
    \midrule
    \citep{Wolske.2017} & Observability & \multirow{2}[2]{*}{.384} & \multirow{2}[2]{*}{.315} & \multirow{2}[2]{*}{.284} & \multirow{2}[2]{*}{-.330} & \multirow{2}[2]{*}{-.233} & \multirow{2}[2]{*}{.405} & \multirow{2}[2]{*}{.371} & \multicolumn{3}{l}{\multirow{2}[2]{*}{}} \\
    \citep{Wolske.2017} & Social support &       &       &       &       &       &       &       & \multicolumn{3}{l}{} \\
    \midrule
    \multicolumn{12}{l}{\textbf{Gender}} \\
    \midrule
    \citep{Arroyo.2019} & Gender & -.028 & .046 & /     & /     & -.038 & /     & /     & /     &       &  \\
    \midrule
    \citep{Parkins.2018} & Gender & -.009 & .081 & /     & /     & /     & /     & /     & -.059 &       &  \\
    \midrule
    \multicolumn{12}{l}{\textbf{Education}} \\
    \midrule
    \citep{Arroyo.2019} & Academic level & .265 & .008 & /     & /     & .003 & /     & /     & /     & -.031 &  \\
    \midrule
    \citep{Parkins.2018} & Education & \multirow{2}[2]{*}{.012} & \multirow{2}[2]{*}{.058} & \multirow{2}[2]{*}{/} & \multirow{2}[2]{*}{/} & \multirow{2}[2]{*}{/} & \multirow{2}[2]{*}{/} & \multirow{2}[2]{*}{/} & \multirow{2}[2]{*}{.071} & \multirow{2}[2]{*}{-.088} & \multirow{2}[2]{*}{} \\
    \citep{Parkins.2018} & Recode for University Education &       &       &       &       &       &       &       &       &       &  \\
    \midrule
    \citep{AzizN.S..2017} & Education & -.048 & .093 & /     & /     & -.047 & -.006 & /     & .036 & /     &  \\
    \midrule
    \multicolumn{12}{l}{\textbf{Income}} \\
    \midrule
    \citep{Arroyo.2019} & Socioeconomic level & .442 & .2   & /     & /     & -.114 & /     & /     & /     & 0     & .376 \\
    \midrule
    \citep{Parkins.2018} & Household income before taxes in 2013 & -.002 & .003 & /     & /     & /     & /     & /     & .069 & -.099 & .279 \\
    \midrule
    \citep{AzizN.S..2017} & Income & .142 & .199 & /     & /     & .065 & .085 & /     & -.036 & /     & -.068 \\
    \bottomrule
\end{tabularx}%
  \label{tab:input_refi}%
\end{table}%








\end{document}